\def\ispreprint{1}  

\if \ispreprint1
\documentclass[preprint,authoryear,5p]{elsarticle}  
\else
\documentclass[preprint,authoryear,12pt]{elsarticle}  
\fi

\usepackage[utf8]{inputenc}
\usepackage{lineno} 
\usepackage{amssymb}
\usepackage{amsmath,bm}
\usepackage{graphicx}
\usepackage[draft]{hyperref} 
\usepackage{appendix}
\usepackage{changepage,geometry}  
\usepackage{multicol}
\usepackage{natbib} 

\begin{document}

\title{Wind erosion and transport on planetesimals}

\author[a1]{Alice C. Quillen\corref{cor1}}
\ead{alice.quillen@rochester.edu}
\author[a1]{Stephen Luniewski}
\ead{sluniews@ur.rochester.edu}
\author[a1]{Adam E. Rubinstein}
\ead{arubinst@ur.rochester.edu}
\author[a2,a1]{J\'er\'emy Couturier}
\ead{jcouturi@ur.rochester.edu}
\author[a2]{Rachel Glade}
\ead{rglade@rochester.edu}
\author[a2,a1]{Miki Nakajima}
\ead{mnakajima@rochester.edu} 
\address[a1]{Department of Physics and Astronomy, University of Rochester, Rochester, NY 14627, USA}
\address[a2]{Department of Earth and Environmental Science, University of Rochester, Rochester, NY 14627, USA}

\cortext[cor1]{Corresponding author}


\begin{abstract}

We consider the possibility that aeolian (wind blown) processes occur on small, 1 to 100~km diameter, planetesimals when they were embedded in the protosolar nebula. Drag from a headwind within a protostellar disk is sufficiently large to loft cm and smaller sized particles off the surface of a 10 km diameter asteroid in the inner solar system (at a few AU),  and micron sized particles off the surface of a 10 km diameter object in the Transneptunian region.  The headwind is sufficiently strong to overcome surface cohesion in the inner solar system, but not in the outer solar system.  However, in the outer solar system, surface particles can be redistributed or escape due to impacts from particles that are in the protosolar disk's wind.  Based on scaling crater ejecta, we estimate that impacts from particles in the headwind will lead to erosion of mass rather than accretion for planetesimals below about 6 km in diameter.   The erosion limit is independent of material strength but proportional to the wind velocity.  We explore the sensitivity of splash particle trajectories to particle size,  headwind velocity and Reynolds number.  Winds from a protostellar disk could account for Kuiper Belt Object (486958) Arrokoth's smooth undulating terrain but only during an epoch of high particle flux and low wind velocity.  These conditions could have been present during and just after coalescence of Arrokoth's building blocks.   
\end{abstract}

\maketitle

\if\ispreprint1
\else 
\linenumbers 
\fi 

\section{Introduction}
\label{sec:intro}

Formation of planetesimals from small grains within a protosolar disk encompasses
cohesion between particles, erosion or accretion due to impacts, settling, and angular momentum exchange between particles and gas; for a recent review, see \citet{Simon_2022}. 
The lifetime of a protoplanetary disk is estimated to 
be a few Myr \citep{Williams_2011}, implying that planetesimals could interact with a gas disk for a prolonged period.  
Within a protostellar disk, the gaseous component feels a pressure force,  which lowers the gas velocity with respect to planetesimals.  After it has formed, 
a planetesimal embedded in the disk feels a wind known as a `headwind'. 
The drag force from this wind can cause the planetesimal to steadily drift inwards \citep{Whipple_1972}.   For meter sized bodies, the radial drift rate is high enough  that it presents a problem for planetesimal growth scenarios, known as the `meter-sized' barrier \citep{Adachi_1976,Weidenschilling_1977}.
Depending upon the size and cohesion strength of particles on the surface of a planetesimal, the gas flow from a headwind can cause erosion, 
with small planetesimals on eccentric orbits most susceptible to mass loss 
\citep{Rozner_2020,Demirci_2020,Demirci_2020b,Cedenblad_2021}. 
The disk can also contain particles that can be accreted onto planetesimals or planetary embryos
\citep{Ormel_2010,Lambrechts_2012,Johansen_2015,Ormel_2017}.
We consider how this headwind, comprised of both gas and particles, can modify the surface of a planetesimal.  We refer to an object that is composed of solids, and that has diameter substantially larger than a meter but is smaller than a few hundred km, as a planetesimal. 

Pebbles and small particles, comprised of ices and dust,  can co-exist in a gaseous protostellar disk that contains planetesimals. 
We refer to a conglomerate of solids that is well-coupled to the gas disk as a `pebble', following
\citet{Lambrechts_2012}.  
Low porosity (fluffy) dust particles in comet 67P/Churyumov-Gerasimenko are interpreted to be remnants from the pre-pebble growth stage of the protoplanetary disk, which were captured and preserved between pebbles within the comet during formation \citep{Fulle_2016,Mannel_2019,Guttler_2019}.   
Planetesimal formation via streaming instability \citep{Youdin_2005}, in turbulent vortices \citep{Cuzzi_2008} or via gravitational instability, may be inefficient \citep{Birnstiel_2012,Robinson_2020,Nesvorny_2021,Lorek_2022}, with only 10 to 50\% of the total dust mass incorporated into planetesimals \citep{Rucska_2021}, and smaller (cm-sized and below) particles predominantly left behind in the gas disk \citep{Rucska_2023,Johansen_2007,Nesvorny_2021}.  


Because particles could be present within the protosolar nebula during late states of planetesimal formation, and prior to evaporation of the disk, they can hit a planetesimal's surface \citep{Ormel_2010,Lambrechts_2012,Johansen_2015}.   For example, 
\citet{Johansen_2015} explored drag assisted accretion of chondrules from the protostellar disk.  


Aeolian (wind-driven) particle transport has occurred on many bodies in the Solar system including Earth, Mars, Venus, Triton, Titan  (e.g., \citealt{Kok_2012,Tefler_2018,Gunn_2022}), Pluto  \citep{Tefler_2018},  Io \citep{McDonald_2022}, and comet 67P/Churyumov-Gerasimenko \citep{ElMaarry_2017}. 
The ubiquity of aeolian processes 
in the Solar system suggests that planetesimal surfaces can be modified by protostellar-disk headwinds
and the particles within them.  

On Earth, if the wind is strong enough, particles loft off the surface, and are transported
by the wind by hopping across the surface, a process called 
`saltation' \citep{Bagnold_1941}, from the Latin word `salto', which means to leap or spring.
Particles entrained in the wind can cause additional grains to be splashed off the surface via impacts \citep{Bagnold_1941},  which can lead to erosion.   We consider analogies of these processes, 
but in the rarefied and low-g environments of planetesimal surfaces. 
As is true on Earth, 
impacts by particles in a wind on a planetesimal can overcome cohesion,  causing particles to be splashed off the surface.   However, on a planetesimal surface gravity is low, so particles
lofted off the surface could escape the planetesimal altogether causing erosion.  
Particles that do not escape are transported across the surface and could fill in low lying topography. 
As aeolian processes on Earth and other planets are sensitive to particle size, these processes
could lead to size segregation on planetesimal surfaces. 


In section \ref{sec:winds}, we give estimates for the physical properties of the protosolar disk.    Using the disk model, we estimate the size of a particle on the surface of a planetesimal that can be pushed by a disk headwind.  We estimate what size particles would be prevented from lofting off the surface due to cohesion.   
In section \ref{sec:impacts},  we discuss impacts on a planetesimal from particles that are in the disk wind.  We delineate a regime where impacts from particles in a wind cause erosion rather than accretion.  
In section \ref{sec:traj},  we discuss trajectories of particles that are splashed off the surface due to impacts from particles that are part of a headwind. 
In section \ref{sec:apps} we consider winds as a process for surface alteration in the 
outer regions of the solar system, with particular application to the surface of Cold Classical Kuiper Belt Object (CCKBO)  (486958) Arrokoth. 

\section{Disk winds on planetesimals}
\label{sec:winds}

\subsection{Winds on planetesimals from the protosolar disk}

We estimate the relative velocity between a planetesimal, 
and gas flowing within the protosolar disk, prior to its evaporation and soon after
the planetesimal formed.  

Because the gas in the disk feels pressure, its mean tangential circular velocity
is expected to be below that of a circular Keplerian orbit.  
The velocity of a planetesimal in a circular orbit about the Sun 
differs from the mean gas rotational velocity which can be estimated using the radial component
of Euler's equation. The planetesimal feels a headwind given by this velocity difference; 
\begin{equation}
u_{hw}   \approx \eta_g v_K,  \label{eqn:headwind}
\end{equation}
(e.g., \citealt{Nakagawa_1986,Armitage_2020}), 
where pressure gradient parameter
\begin{equation}
\eta_g \equiv -\frac{1}{2} \frac{P_g}{\rho_g v_K^2}  \frac{\partial \ln P_g}{\partial \ln r} 
. \label{eqn:eta_g}
\end{equation}
Here the gas density and pressure in the disk midplane are $\rho_g$ and  $P_g$, respectively. 
The Keplerian orbital velocity for a circular orbit
about the Sun at radius $r$ is 
$v_K = \sqrt{GM_\odot/r}$, where  the gravitational constant is $G$ and $M_\odot$ is the
mass of the Sun.   

It is convenient to define a quantity  
\begin{equation}
 c_g = \frac{P_g}{\rho_g} = \sqrt{\frac{ k_B T_g}{\bar m}},    \label{eqn:c_g}
\end{equation}
which is called the isothermal sound speed (e.g., \citealt{Dullemond_2004})
and is similar in size to the sound speed (which depends on an adiabatic index).  Here  
 $k_B$ is Boltzmann's constant, $T_g$ is the disk temperature and $\bar m$ is the mean
 molecular mass.  
 In terms of $c_g$,  equation \ref{eqn:eta_g} becomes 
$\eta_g = - \frac{1}{2} \left(\frac{c_g}{v_K}\right)^2  \frac{\partial \ln P_g}{\partial \ln r} $. 

The protosolar disk could be turbulent 
due to the magneto-rotational (MRI) instability \citep{Balbus_2011} or other processes \citep{Frank_2002}. 
The dimensionless parameter $\alpha_g$ is used characterize the strength of the turbulence and the associated turbulent viscosity. 
The largest turbulent eddies have velocity with respect to the mean rotational velocity 
\begin{equation}
  u_{turb} \sim \sqrt{\alpha_g} c_g 
  \label{eqn:vt}
\end{equation}
\citep{Pringle_1981,SS_1978,Shakura_1973}.
Turbulent eddies are conventionally described in terms of the sound speed \citep{SS_1978} rather than
the isothermal sound speed but as equation \ref{eqn:vt} is approximate, the difference between these
descriptions is not significant. 
   
\subsection{Protosolar disk model}
\label{sec:disk}

Following \citet{Weidenschilling_1977}, we adopt power law forms for the disk's 
gas surface density and temperature, 
\begin{align}
\Sigma_{g}(r) & =  g_\Sigma 
\left( \frac{r}{ 10 \ {\rm AU} } \right)^{-\alpha_\Sigma}   \label{eqn:Sig_g} \\
T_{g}(r) & = g_T   \left( \frac{r}{ 1 \ {\rm AU} } \right)^{-\alpha_T}.    \label{eqn:T_g}  
\end{align}
The coefficients $g_\Sigma$ and $g_T$ give the surface density at 10 AU and temperature at 1 AU. 
For the minimum mass solar nebula \citep{Hayashi_1981,Nakagawa_1986}, 
the gas surface density at 10 AU is 
 $538$ kg~m$^{-2}$ and the exponents for equations \ref{eqn:Sig_g} and \ref{eqn:T_g}   are 
 \begin{align}
 \alpha_\Sigma = \frac{3}{2} \qquad {\rm and} \qquad \alpha_T = \frac{1}{2} .  \label{eqn:alpha_exps}
 \end{align}
An estimate for the gas surface density and temperature in the protosolar nebula that takes into account a history of planet migration \citep{Desch_2007} has similar but slightly higher exponents and 
 a surface density that is about 6 times higher than the minimum mass solar nebula at 10 AU.    
 The surface density for the model by \citet{Desch_2007} is similar to the protosolar disk evolution models by \citet{Lenz_2020} at ages less than two million years.
To facilitate estimating physical quantities, we adopt a model 
that has the exponents  
of the minimum mass solar nebula by \citet{Hayashi_1981}, and as shown in equations \ref{eqn:Sig_g}, 
 \ref{eqn:T_g} and \ref{eqn:alpha_exps}, but  
 has the surface density at 10 AU and temperature at 1 AU of the disk model by \citet{Desch_2007}, giving coefficient $g_\Sigma$ equal to 
\begin{align}
 g_{\Sigma, {\rm adopted}} &\equiv 3430\ {\rm kg~m}^{-2}  .\label{eqn:gSigma}
\end{align} 

There is a range of estimates for the 
disk midplane temperature at 1 AU.  At 1 AU,  the passive radiative models by   
by \citet{Chiang_1997} and \citet{Chiang_2010}, have $T_g \sim 150$ and  120 K respectively. 
The minimum mass solar nebula model \citep{Hayashi_1981,Nakagawa_1986} has 
$T_g \sim 280$ at the same radius, whereas 300 K was adopted by  \citet{Ormel_2017}. 
Following the discussion by \citet{Lenz_2020} on the disk temperature (their section 3.7),  
the Sun was about 5/7 of its current luminosity when the protosolar disk was present.  
Neglecting dissipative heating associated with accretion, 
the disk temperature $T_g \approx  \left(\theta_d L_\star /(4 \pi \sigma_{SB} r^2\right)^\frac{1}{4}$
with $\sigma_{SB}$ the Stephan-Boltzmann constant and $\theta_d$ representing 
 the fraction of radiation absorbed by the disk.  Taking stellar luminosity $L_\star = \frac{5}{7} L_\odot$ and 
 $\theta_d= 0.05$,   we estimate a disk temperature of about 170 K at 1 AU.  
For our disk model we adopt a $g_T$  value  of 
\begin{align}
 g_{T,{\rm adopted}} & \equiv 170\ {\rm K} \label{eqn:gT}
\end{align}
 similar to the models by \citet{Lenz_2020}. 

To estimate the size of the headwind and turbulent eddy velocity, we estimate the gas density $\rho_g(r)$ and pressure $P_g(r)$ in the disk midplane as a function of $r$, the distance from the Sun. 
Following \citet{Armitage_2020}, we approximate the gas density vertical profile with 
an isothermal profile, 
\begin{align}
\rho_{g,rz} (r,z) &= \rho_g(r) \exp \left(- \frac{z^2}{2h_g^2} \right),  \label{eqn:rho_g}
\end{align}
where $z$ is the height above the midplane, and $h_g$ is the vertical scale height which is a function of orbital radius $r$. 
The disk surface mass density 
$\Sigma_g(r) = (2\pi)^{1/2} \rho_g(r) h_g(r) $.  
Neglecting vertical temperature variations,  hydrostatic equilibrium (see for example \citealt{Armitage_2020}) gives scale height 
\begin{align}
h_g \approx \frac{c_g}{\Omega_K}, \label{eqn:h_g}
\end{align}
with Keplerian angular rotation rate $\Omega_K = v_K/r$
and with isothermal sound speed defined in equation \ref{eqn:c_g}. 
The gas pressure in the midplane is related to the midplane density and isothermal sound speed
via the ideal gas law,
\begin{equation}
P_g = \frac{\rho_g k_B T_g}{\bar m} =  \rho_g c_g^2 . \label{eqn:P_g}
\end{equation}
We adopt mean molecular mass
$\bar m \approx 2 m_H$ with $m_H$,  the mass of a hydrogen atom. 


Our adopted disk model has 
$c_g \propto r^{-\frac{1}{4}}$,   
$h_g/r \propto  r^\frac{1}{4} $,
$\rho_g \propto  r^{-\frac{11}{4}}$, 
$P_g \propto r^{-\frac{13}{4}}$, 
$\eta_g \propto r^\frac{1}{2}$,  
$u_{hw} \propto r^0$, 
$u_{turb} \propto r^{-\frac{1}{4}}$.
%
%
An advantage of using the -3/2 and -1/2 exponents (equations \ref{eqn:alpha_exps}) 
for radial scaling of surface density and temperature, respectively, 
is that the headwind velocity is independent of radius. 
In terms of the assumed surface density at 10 AU and the temperature at 1 AU,  
$\Sigma_g \propto g_\Sigma$, $c_g \propto g_T^\frac{1}{2}$, $h_g \propto g_T^\frac{1}{2}$, 
$\rho_g \propto g_\Sigma g_T^{-\frac{1}{2}}$, $P_g \propto g_\Sigma g_T^\frac{1}{2}$,  $\eta_g 
\propto g_T$, $u_{hw} \propto g_T$, $u_{turb} \propto g_T^\frac{1}{2}$. 
We include these factors in our numerical estimates so that quantities can be estimated for a 
disk with a different density or temperature.     Hereafter $\tilde g_T = g_T/g_{T, {\rm adopted}}$ 
and $\tilde g_\Sigma = g_\Sigma/g_{\Sigma, {\rm adopted}}$. 

The gas in the protosolar nebula is rarefied compared to the atmospheres of Earth or Mars. 
The midplane gas density is 
\begin{align}
\rho_g & = 1.8 \times 10^{-8}\ {\rm kg~m}^{-3} \left( \frac{r}{10 \ {\rm AU}} \right)^{-\frac{11}{4}} \tilde g_\Sigma \tilde g_T^{-\frac{1}{2}}.
\end{align}
Over a wide range of temperatures, 
the cross section for collisions between molecular hydrogen molecules is 
$\sigma_{H2} = 2 \times 10^{-19} {\rm\ m}^2$ \citep{Massey_1933}. 
In the kinetic theory of gases, the mean free path $\lambda_g$, is the average distance a particle travels between collisions with other moving particles   
$\lambda_g = (\sqrt{2} n \sigma)^{-1} $, where $n$ is the number density of particles and
$\sigma$ is their collisional cross sectional area \citep{Chapman_1990}.    
With number density of molecules $n_g = \rho_g/\bar m$, 
we estimate the  
the mean free path $\lambda_g$ in the disk midplane; 
\begin{align}
\lambda_g & \sim  \frac{1}{\sqrt{2} n_g \sigma_{H2}} =  \frac{\bar m}{\sqrt{2} \rho_g \sigma_{H2}}  \nonumber \\
 &\sim \ 0.6 \ {\rm m} \left( \frac{r}{10\ {\rm AU}} \right)^\frac{11}{4} \tilde g_\Sigma^{-1} \tilde g_T^\frac{1}{2} \label{eqn:lam_g}
 ,
\end{align}
The kinematic viscosity due to collisions is 
\begin{equation}
\nu_g \approx  \lambda_g v_{\rm th} /2, \label{eqn:nu_g}
\end{equation}
(equations 5.5 and 5.8c by \citealt{Vincenti_1986}).  The  
thermal velocity $ v_{\rm th}$ is defined as the mean of the velocity magnitude computed with a Maxwell-Boltzmann velocity distribution in three dimensions; 
\begin{align}
v_{\rm th} = \sqrt{\frac{ 8}{\pi}} c_g,  \label{eqn:vth}
\end{align}  
in terms of the isothermal sound speed (equation \ref{eqn:c_g}). 
The Reynolds number of wind flow with velocity $u$ about a planetesimal of diameter $D_a$ is 
\begin{align}
Re_a = \frac{u D_a}{\nu_g}. \label{eqn:Re}
\end{align}  

Using equation \ref{eqn:headwind}, the headwind velocity is 
\begin{equation}
u_{hw} \sim 38\ {\rm m/s}\ \left( \frac{r}{10\ {\rm AU}} \right)^{0} \tilde g_T.
\label{eqn:uhw}
\end{equation}
This value is higher than that estimated by 
\citet{Chiang_2010} (their equation 8 gives 25 m/s at 1 AU) but is lower than the $\approx 54$ m/s estimated by \citet{Johansen_2014,Ormel_2017}. 
The difference in headwind speed is most likely to the different disk temperatures as  
$T_g \sim 120 K$ at 1 AU for the disk by \citet{Chiang_2010} but is 300 K at 1 AU in the model by \citet{Ormel_2017}.  

The Mach number of the headwind 
\begin{equation}
M\!a \approx \frac{u_{hw}}{c_g} = 0.08 \left( \frac{r}{10\ {\rm AU}} \right)^{\frac{1}{4}} \tilde g_T^{\frac{1}{2}}.
\label{eqn:Mach}
\end{equation}
The turbulent velocity for the largest eddy (via equation \ref{eqn:vt}) is 
\begin{equation}
u_{turb} = 47\ {\rm m/s} \left( \frac{\alpha_g}{10^{-2} } \right)^\frac{1}{2}
\left( \frac{r}{10\ {\rm AU}} \right)^{-\frac{1}{4}} \tilde g_T^\frac{1}{2}.
\label{eqn:uturb}
\end{equation}

If the wind velocity is due to a headwind with velocity 
in equations \ref{eqn:headwind} and \ref{eqn:uhw}, the Reynolds number for flow about a planetesimal with diameter $D_a$ is 
\begin{align}
Re_{a,hw} \sim & \frac{u_{hw} D_a}{\nu_g } \nonumber  \\
\sim & \ 1600 \left( \frac{r}{10\ {\rm AU}}\right)^{-\frac{5}{2}} 
\left( \frac{D_a}{10\ {\rm km}} \right) \tilde g_\Sigma . \label{eqn:Re_hw}
\end{align}
If the wind velocity is due to turbulent eddies, with velocity in equation \ref{eqn:vt}, the Reynolds number 
\begin{align}
Re_{a,vt} \sim &  \frac{u_{turb} D_a}{\nu_g} \nonumber  \\
\sim & \ 1900 \left( \frac{r}{10\ {\rm AU}}\right)^{-\frac{11}{4}} 
\left( \frac{\alpha_g}{0.01} \right)
\left( \frac{D_a}{10\ {\rm km}} \right) \nonumber \\
& \ \ \ \ \times \tilde g_\Sigma \tilde g_T^{-\frac{1}{2}}
. \label{eqn:Re_vt}
\end{align}
Disk-averaged values of $\alpha_g \sim  10^{-2}$ are inferred from protostar ages and accretion rates \citep{Calvet_2000}.


To characterize the regimes in inner and outer solar system, we list physical quantities for the protostellar disk at 4 different orbital radii in Table \ref{tab:tab}.

\begin{table*}[!htb] \centering 
\if \ispreprint1 
\else
\begin{adjustwidth}{-2.0cm}{-2.0cm} 
\small
\fi
\caption{Physical quantities at different orbital radii \label{tab:tab}}
\begin{tabular}{lllllllll}
\hline
Quantity & Symbol  &   Eqn.  & units  & & & & \\
\hline 
Radius from Sun                          &     $r$       &  & AU &  1      &   3.2     &  10   &     45  \\               
\hline
Disk gas surface density & $ \Sigma_g$ &   \ref{eqn:Sig_g}  & kg m$^{-2}$ &
       1.1 $\times 10^5$  & 1.9 $\times 10^4$  & 3.4$\times 10^3$  & 3.6$\times 10^2$ \\
Disk gas midplane density & $\rho_g$ & \ref{eqn:rho_g} &
   kg m$^{-3}$  & 1.0$\times10^{-5}$  & 4.2$\times10^{-7}$  & 1.8$\times10^{-8}$  & 2.9$\times10^{-10}$ \\
Isothermal sound speed  & $c_g$  & \ref{eqn:c_g} & m/s  & 838  & 626  & 471  & 323 \\   
Disk aspect ratio & $h_g/r$ &\ref{eqn:h_g} & - & 0.028  & 0.038  & 0.050  & 0.073 \\
Midplane gas pressure & $P_g$ & \ref{eqn:P_g} & Pa &  7.2  & 0.16  & 4.1$\times 10^{-3}$  & 3.1$\times 10^{-5}$ \\
Midplane temperature & $T_g$ & \ref{eqn:T_g} & K &   170  & 95  & 54  & 25 \\
Mean free path & $\lambda_g$ & \ref{eqn:lam_g} &  m  
              & 1.1$\times 10^{-3}$  & 2.8$\times 10^{-2}$  & 0.65  & 40 \\
Pressure parameter & $\eta_g$  & \ref{eqn:eta_g} & -   
             & 1.3$\times 10^{-3}$ & 2.3$\times 10^{-3}$  & 4.1$\times 10^{-3}$  & 8.6$\times 10^{-3}$ \\
Kinematic viscosity & $\nu_g$ & \ref{eqn:nu_g} & m$^2$/s   & 0.77  & 14.1  & 243  & $10^{4}$ \\
Headwind velocity & $u_{hw}$ & \ref{eqn:uhw} &   m/s    & 38 & 38 & 38 & 38  \\
Turbulent velocity &  $u_{turb}$ & \ref{eqn:vt}   &   m/s   &  84  & 63  & 47  & 32 \\
Particle radius; $t_{stop}(s)\!=\!3000$~s\!\! &  $s$  & \ref{eqn:tstop} & mm  &
       15  & 2.4  & 0.08  & 9$\times 10^{-4}$ \\
Cohesion threshold (ice) &   $ s_{pull}$ & mm & \ref{eqn:Fc} & 
 		0.065  & 2.2  & 67 & 6200\\
\hline
\multicolumn{2}{l}{For $D_a =10$ km,  $\rho_a = 1000~{\rm kg~m}^{-3}$:} & \\
Reynolds number -headwind &  $Re_{hw} $   & \ref{eqn:Re_hw} & -  
& 5$\times 10^5$  & 3$\times10^4$  & 1600  & 37 \\
Reynolds number -turbulence &  $Re_{vt} $   & \ref{eqn:Re_vt} & 
-  & 1.1$\times 10^6$  & 4.4$\times10^4$  & 1900  & 31 \\
\hline
\end{tabular}
{\\ Notes:  Adopting mean molecular mass $\bar m = 2m_H$ 
and turbulent viscosity parameter $\alpha_g=0.01$. 
The particle radius with stopping time equal to 3000 s
is computed for a particle density  $\rho_s = 500~ {\rm kg~m}^{-3}$. 
The cohesion thresholds are computed using the same particle density, 
$\gamma_c = 0.244~ {\rm J~m}^{-2}$ (for ice; \citealt{Gundlach_2015}),
 and ${\cal B} \sim 8.74$. 
}
\if \ispreprint1 
\else
\end{adjustwidth}
\fi
\end{table*}

\subsection{Lofting particles -- saltation}

We consider small particles that are on the surface of a planetesimal but that interact with the gas in the protostellar disk. 
A particle with radius $s$ has dimensionless Knudsen number 
\begin{equation} 
    K\!n_s = \frac{\lambda_g}{s},
\end{equation}
 where $\lambda_g$ is the mean free path in the gas. 
Saltation can still occur in the rarefied regime with $K\!n_s \gtrsim 1$ (as shown in Figure 2b by \citealt{Gunn_2022}). 
The Reynolds number for the particle 
\begin{equation}
Re_s = \frac{u s}{\nu_g},
\end{equation}
where gas kinematic viscosity $\nu_g$ is given by equation \ref{eqn:nu_g} and $u$ is the relative velocity between particle and gas.  
With relative velocity set by a headwind (equation \ref{eqn:uhw}) 
or a turbulent eddy (equation \ref{eqn:uturb}) we expect the wind to be subsonic with 
Mach number $M\!a <1$ (as given in equation \ref{eqn:Mach}).

The threshold for saltation, or lofting of a particle, depends on lift and drag forces from a wind exceeding gravitational and adhesive or cohesive forces (e.g., \citealt{Shao_2000,Kok_2012,Gunn_2022}).  
In atmospheres, the drag and lift forces on a particle are usually described as proportional to $u^2$, the square of the relative velocity between wind and particle.   The drag and lift forces are proportional to the ram pressure. 
At low Reynolds number, the drag force is said to be in the Stokes regime, 
and at high Knudsen number, the drag force is said to be in the free molecular or 
Epstein regime \citep{Epstein_1924}.  In Stokes and Epstein regimes, the drag force is proportional to
the relative velocity rather than its square. 

At low Reynolds number, $Re_s \lesssim 1$, and low Mach number, the drag force   
$F_D$ on a particle of radius $s$ and mass  $m_s = \frac{4\pi}{3} \rho_s s^3$ gives an acceleration 
\begin{align}
a_D & =  \frac{F_D}{m_s} \nonumber \\
 &\approx \frac{\rho_g v_{\rm th}}{\rho_s s}  u \times 
 \begin{cases}
  1 & {\rm for \ }K\!n_s \gtrsim \frac{4}{9}, Re_s \lesssim 1   \\
\frac{9 }{4} \frac{\lambda_g}{s} &  { \rm for \ } K\!n_s \lesssim \frac{4}{9}, Re_s \lesssim 1  \\
  \end{cases} \label{eqn:FDcases}
\end{align}
where the first case is the Epstein or free-molecular regime \citep{Epstein_1924} 
and the second case is the Stokes regime \citep{Weidenschilling_1977}. 
Here $\rho_g$ is the gas density and $\rho_s$ is the particle density. 
In the Epstein regime $F_D/m_s \propto r^{-3}  g_\Sigma g_T$ for our adopted disk model. 
For $K\!n_s <1 $  and $1 \lesssim Re_s \lesssim 1000$, and low Mach number and in the ram pressure regime, the drag coefficient is of order unity, giving acceleration
\begin{align}
a_D = \frac{F_D}{m_s}  \sim   \frac{\rho_g }{\rho_s }  \frac{u^2}{s} \ \ { \rm for \ } Re_s\gtrsim1, K\!n_s \lesssim 1.  \label{eqn:FDcase3}
\end{align}
This gives a drag force $F_D \sim \rho_g u^2 \pi s^2$ which is approximately equal to that caused by ram pressure. 
An analytical expression for the acceleration from drag, covering a range of Reynolds numbers and low and moderate Knudsen
numbers but not high Mach number, and that approximates expressions by \citet{Loth_2008b,Singh_2022},  
\begin{align}
a_D & \sim \frac{\rho_g v_{\rm th}^2}{\rho_s s}
 \left( \frac{u}{v_{\rm th}}\right)
  \left( \frac {\frac{9}{4} K\!n_s }{1 + \frac{9}{4} K\!n_s}\right)
   \left(1 + \frac{3}{16} Re_s \right) 
   .  \label{eqn:a_D}
\end{align}

In the presence of a velocity gradient, a spherical particle near a hard surface, feels a lower pressure on its side that is further from the hard surface.  
This gives a lift force known as the Saffman lift force \citep{Saffman_1965}.   
The size of the lift force is similar to the size of the drag force \citep{Loth_2008b,Demirci_2020b,Ekanayake_2021}, though in some regimes lift may be negative \citep{Luo_2016}. 
Following \citet{Shao_2000}, 
the relative velocity between a particle near the surface and a wind is similar in size to the difference between the velocity of the surface and the wind velocity at large distances from the planetesimal, and this relative velocity is also similar in size to the quantity called the friction velocity which is computed from the wind's shear stress.   
We assume that the size of the lift force is similar to the size of the drag force and the relative velocity between wind and particle can be estimated from the headwind velocity.  We estimate a threshold for saltation by equating the acceleration due to drag with the gravitational acceleration $g_a$ on a planetesimal's surface 
\begin{align}
a_D(s,u) \sim g_a . \label{eqn:assumption}
\end{align}  
This equation shows explicitly that the acceleration is
sensitive to particle size $s$ and wind velocity $u$.
We solve equation \ref{eqn:assumption} to find $s_{max}$, a maximum radius of a particle that can be lofted by a wind with velocity $u$. The same relation can be used to estimate the critical wind velocity as a function of particle radius $s$ required for saltation.  The thresholds estimated from the drag acceleration are consistent to order of magnitude  
with thresholds estimated in other studies (e.g., \citealt{Shao_2000,Kok_2012,Demirci_2020b,Gunn_2022}). 

Solving equation \ref{eqn:assumption} for a particle size $s_{max}$  gives us an estimate for the maximum particle that can be lofted by the wind;  
\begin{align}
 s_{max} \sim  
 \begin{cases}
    \frac{\rho_g v_{\rm th}^2}{\rho_s g_a} \left( \frac{u}{v_{\rm th}}  \right) 
            & {\rm for  \ } K\!n_s \gtrsim \frac{4}{9}, Re_s \lesssim1 \\
    \left( \frac{\rho_g v_{\rm th}^2}{\rho_s g_a}   \frac{9\lambda_g  }{4} \right)^\frac{1}{2} \!\!
    \left(\frac{u}{ v_{\rm th}}\right)^\frac{1}{2}
       & {\rm for  \ } K\!n_s \lesssim \frac{4}{9}, Re_s \lesssim 1 \\
 \frac{\rho_g v_{\rm th}^2}{\rho_s g_a} \left( \frac{u}{ v_{\rm th}} \right)^2 
      &  {\rm for  \ } K\!n_s \lesssim \frac{4}{9}, Re_s \gtrsim 1 \\
\end{cases}  .
 \label{eqn:smax}
 \end{align}
From top to bottom, equation \ref{eqn:smax} shows the Epstein, Stokes and ram pressure drag regimes.  For our figures, we compute $s_{max}$ using the single drag formula of equation \ref{eqn:a_D} which covers these regimes,  however equation \ref{eqn:smax} shows how $s_{max}$ scales with physical quantities.   

A more accurate estimate for the lift could take into account the sensitivity of the lift force to the velocity shear in the wind flow, the distance of a particle to the surface, the surface roughness, particle spin,  the presence of other particles in the flow,  and in a dilute gas, how gas molecules are reflected from the particle surface upon collisions \citep{Luo_2016}.  

\subsection{Cohesion}

Cohesion due to interparticle forces, such as the van der Waals force, is 
particularly important for small particles \citep{Gundlach_2015}.  On Earth, cohesion prevents saltation of aerosols (particles with diameter less than 1 $\mu$m,  e.g., \citealt{Greeley_1985,Castellanos_2005,Kok_2012}).  
Recent experiments \citep{Gundlach_2015} find a cohesion energy per unit
area for ice of $\gamma_c = 0.244~ {\rm J~m}^{-2}$. 
In comparison to mm-sized silica dust particles \citep{Dominik_1997,Chokshi_1993,Poppe_2000},
 the sticking threshold of water-ice particles is approximately tenfold higher. 
For a recent discussion of experiments
on the cohesion of free particles with different compositions see \cite{Kimura_2020}.

The force required to separate (or pull apart) two particles of radius $s$ is approximately 
\begin{equation}
F_{coh} \sim  \pi \gamma_c s e^{-{\cal B}} ,\label{eqn:Fc}
\end{equation}
where ${\cal B}$ takes into account surficial particle roughness and the interparticle distance across which adhesion forces act (see chap 17 by \citealt{Israelachvili_2011}).
Averaging over a wide range of aeolian settings, 
\citet{Gunn_2022} find ${\cal B} \sim 8.74$ and that this description 
is approximately equivalent to other models for cohesion (e.g., \citealt{Shao_2000,Demirci_2020}).

Small particles are prevented from lofting off the surface if the drag force from the
wind is weaker than the cohesion force.   Using equations \ref{eqn:FDcases} and  \ref{eqn:Fc}, we estimate the threshold particle size allowing a drag from a wind to exceed cohesion ($F_D  = F_{coh}$) 
\begin{equation}
s_{pull} \sim
\begin{cases}
\frac{3}{4} \frac{ \gamma_c e^{- {\cal B}}}{\rho_g v_{\rm th}^2} \left(\frac{v_{\rm th}}{u} \right) & {\rm for \ } K\!n_s \gtrsim \frac{4}{9}, Re_s \lesssim 1  \\
\frac{3}{4}  \frac{ \gamma_c e^{- {\cal B}}}{\rho_g v_{\rm th}^2}  \left( \frac{v_{\rm th}}{u} \right)^2
& {\rm for \ } K\!n_s \lesssim \frac{4}{9}, Re_s \gtrsim 1\\ 
\end{cases} .
 \label{eqn:s_pull}
\end{equation}
Particles with $s \lesssim s_{pull}$ would remain on the surface due to cohesion. 
The Stokes regime is a special case as cohesion and drag scale the same way with particle radius $s$. 
In the ram pressure regime, 
cohesion exceeds the drag force if 
\begin{align}
\frac{ \gamma_c e^{- {\cal B}}}{\rho_g c_g^2}  \left( \frac{v_{\rm th}}{u} \right) \gtrsim 3 \lambda_g \ \ {\rm and\   \ } K\!n_s \lesssim \frac{4}{9}, Re_s \lesssim 1.
\end{align}
We list cohesion particle size thresholds at four different orbital disk radii in Table \ref{tab:tab}.
To compute these, we solve for the particle size
that satisfies $a_D = F_D$ using equations \ref{eqn:a_D} 
and \ref{eqn:Fc} and with a particle density of $\rho_s = 500$ kg~m$^{-3}$.

\subsection{Regime for saltation by disk winds}

\begin{figure*}[!htbp]\centering
\includegraphics[width=4.8 truein, trim = 10 0 0 0,clip]{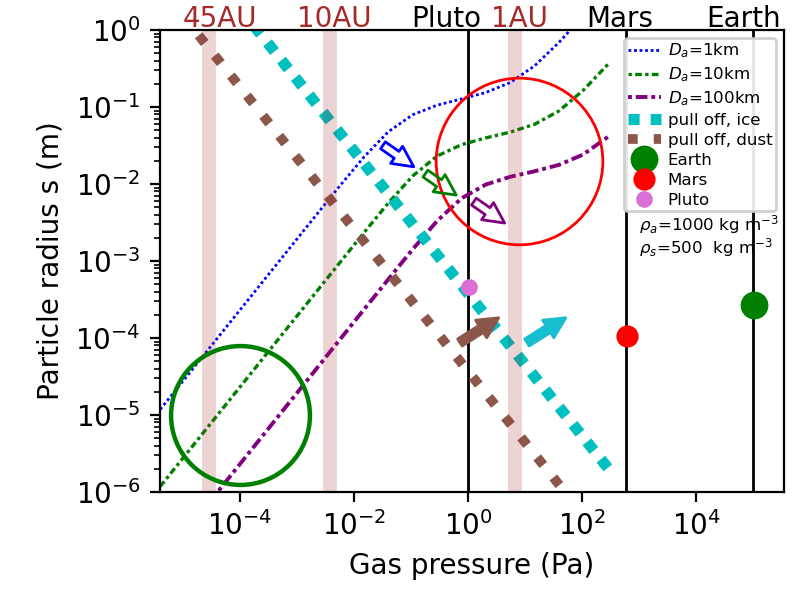}
\caption{Regimes for saltation on different objects in the solar system. 
The $x$-axis shows atmospheric or protosolar disk pressure in Pa. 
The three thin dot-dashed lines show maximum particle size for saltation 
computed by numerically solving for the grain radius which has drag force equal 
to that from gravity.  
These are computed using the disk model of section \ref{sec:disk}  and 
gravitational acceleration for  
 spherical bodies of diameter $1, 10, 100$ km, respectively, mean planetesimal density
$\rho_a = 1000~{\rm kg~m}^{-3}$, wind velocity equal to that of a headwind  and grain density
$\rho_s = 500~{\rm kg~m}^{-3}$. 
For saltation to take place, grains should lie below these lines, as shown by the unfilled colored arrows. 
Midplane pressures in the protosolar disk are shown with wide vertical tan bars
at 1, 10 and 45 AU. 
Atmospheric pressure on Pluto, Earth and Mars are shown with thin vertical black lines. 
The solid dots show the maximum grain radius allowing saltation on Pluto, Mars
and Earth. 
The thick cyan and intermediate thickness brown dotted lines gives the minimum particle radius allowing 
a wind to exceed cohesion for icy particles and dusty particles, respectively.  
The smaller the particle, the harder it is
to pull away from the surface, so for saltation to be initiated, grains should be above these dotted lines,  depending upon their composition, as shown by the large solid cyan and brown arrows.  
The red circle shows that the drag force from a disk headwind within a few AU
is sufficiently strong 
to exceed both cohesion and gravity for cm sized surface particles. 
In the outer solar system,  the drag force from a headwind in a protosolar disk is sufficiently strong to exceed 
gravity for $\mu$m sized particles, but not strong enough to overcome cohesion.  
For saltation to be initiated, another mechanism is required to pull particles
from the surface.  \label{fig:codot}
  }
\end{figure*}


To illustrate the regime where a wind can allow saltation to take place,
we plot in Figure \ref{fig:codot} the maximum radius for a $\rho_s$  = 500 kg~m$^{-3}$ density grain on the surface of 1 km, 10 km and 100 km diameter spherical
planetsimals  to be lofted by a headwind.  
A planetesimal bulk density $\rho_a = 1000~ {\rm kg~m}^{-3}$ is used with 
the planetesimal diameter to compute the surface gravitational acceleration $g_a$.    
We choose a moderately low grain density to represent fractal or porous particles. 
The three thin dotted and dot dashed curves in Figure \ref{fig:codot} show  the maximum grain radius allowing saltation as a function of gas pressure $P_g$ in the disk with disk properties described in section \ref{sec:disk}.  These are computed with a  headwind velocity given by equation \ref{eqn:uhw}.
The limiting grain radius is found by searching for a root of the equation $a_D - g_a =0$  and using equation \ref{eqn:a_D} for the acceleration from drag that covers a range of Reynolds and Knudsen numbers. 
Specifically, to find the root we use python's \texttt{scipy.optimize.brentq} routine that utilizes Brent's method \citep{Brent_2013}. 

In Figure \ref{fig:codot}, 
the thick tan vertical bars show midplane gas pressures for the adopted protosolar disk model at $r=1$, 10 and 45 AU.  
The thick cyan dotted line gives the minimum particle radius allowing 
a wind to exceed the surface cohesion force for ice.  
This is computed by numerically finding a root of $F_D - F_{coh}=0$ with 
equation \ref{eqn:a_D} for the drag force and equation \ref{eqn:Fc} for the cohesion force. 
The brown dashed line is similarly computed assuming a cohesion 10 times lower, representing a dusty rather than icy particle. 

To place disk winds in context with aeolian processes in atmospheres,  in Figure \ref{fig:codot} we show atmospheric pressure for the Earth, Mars and Pluto with thin solid vertical black lines. 
Estimates for the maximum size of particles that can be lofted by winds on these planets are shown with solid dots.  
These are computed using the average properties for the three planets listed in Table S1 by \citet{Gunn_2022} (pressure, temperature, gas and particle densities, gas viscosity and mean free path, and wind speed) and by finding a root of $a_D - g_a = 0$. The saltation limit for Pluto is below the cohesion limit, consistent with the need for additional mechanisms to loft particles  \citep{Tefler_2018}.  


The dot-dashed curves in Figure \ref{fig:codot} are curved. The ram pressure drag  regime $Re_s >1, K\!n_s<1$ is on 
the right.   Particles are in the Epstein regime, with $K\!n_s >1$, on the left side of the plot.   At $P_g \sim 10$ Pa, particles with radius $\sim $ 0.1~m can be in the Stokes regime. 
At a disk pressure $P_g \lesssim 1 $ Pa  
the drag force is about an order of magnitude higher than if it were computed using ram pressure scaling.   
Consequently, the dot-dashed curves are about an order of magnitude higher on the left in Figure \ref{fig:codot} than they would be had we assumed a $u^2$ dependence for the drag force.  For the same reason, the minimum particle size allowing drag to exceed cohesion is about an 
order of magnitude higher on the left than it would have been, had we assumed drag force $F_D \propto u^2$.  
The protosolar disk wind is predominantly lower pressure than planetary atmospheres. 
The disk in the inner solar system has pressure similar to Pluto and Triton, where the pressure is so low that \citet{Gunn_2022} found that their estimate for the drag coefficient was not accurate (see their Figure S1). 
Our estimate for the saltation threshold differs from that used by \citet{Gunn_2022} because it transitions in the high Knudsen number limit to Epstein drag. 

Figure \ref{fig:codot} suggests that disk headwinds can exceed the cohesion force 
and allow cm sized surface particles to be lofted on small bodies in the inner solar system. 
The regime where this might occur is shown with a large red circle in the middle
of Figure \ref{fig:codot}.   

In the outer solar system, shown with the green circle on the lower left in Figure \ref{fig:codot}, headwinds are strong enough to exceed gravity for small ($\mu$m sized) particles and cannot overcome cohesion.  
Following \citet{Tefler_2018} for aeolian processes on Pluto, (also see \citealt{Kok_2012})
if another mechanism can liberate small ($\mu$m sized) particles from the surface, a headwind could allow them to be transported by the wind. 

\subsubsection{Wind density perturbations and gusts}
\label{sec:gusts}

Estimates for disk density perturbations range from about $\delta\rho_g/{\bar\rho_g} \sim 0.2$,  
where $\bar \rho_g$ is the mean value,  
for hydromagnetic turbulence associated with an $\alpha$-disk \citep{Nelson_2010} 
and for compressible hydrodynamic turbulence \citep{Sakurai_2021}.  Spiral density waves, driven by planets \citep{Zhu_2015}, external perturbations or gravitational instability \citep{Kratter_2016}, could also cause high amplitude density perturbations in the disk.  Consequently a planetesimal embedded
in the protosolar nebula could experience gusts from winds at a higher density than the mean value of the disk.  If the density perturbation is caused by a spiral density wave, the amplitude of an associated velocity perturbation would be sensitive to the amplitude of the density perturbation.  While a planetesimal passes through 
the peak density, the headwind velocity could be of order a few percent of
the Keplerian velocity, exceeding the velocity of a headwind in a quiescent disk.  
Together density and velocity perturbations from spiral density waves could give an order of magnitude increase in drag compared to that estimated from a mean disk density.  As a  consequence, 
the regime for saltation could be an order of magnitude higher than estimated  with the three dot-dashed lines in Figure \ref{fig:codot}.  

While spiral density waves and turbulence can cause both density and velocity perturbations in the gas,  the situation is somewhat different when streaming instability is active 
and there are large variations in the ratio of mass surface density in particles compared to gas  (e.g., \citealt{Johansen_2007}). For turbulence driven by streaming instability,  
density perturbations in the gas are small \citep{Johansen_2007}.  However,  
the velocity of the gas with respect to clumps that are forming planetesimals can differ from the headwind velocity,  leading to either higher or lower drag on clumps compared to that estimated from the headwind \citep{Johansen_2007}.  

\section{Particles splashed off the surface by particles in a diskwind}
\label{sec:impacts}

The disk contains pebbles and small particles that can impact a planetesimal surface. 
In subsection \ref{sec:sc}, 
we review criteria allowing these particles to impact a planetesimal.  
In the following subsections we consider the fate of particles that are ejected off the surface of a planetesimal due to these impacts.  

\subsection{Particles in the wind that can impact a small planetesimal}
\label{sec:sc}

Aerodynamics in a protostellar disk affects the size distribution of particles that
can impact a planetesimal
\citep{Ormel_2010,Lambrechts_2012,Ormel_2017}.
The stopping time for a particle of radius $s$ within a gas with relative velocity $u$ depends on the drag force on the particle,  
$t_{stop}(s) \sim m_s u/F_D$ \citep{Weidenschilling_1977}.  
With acceleration $a_D= F_D/m_s$ from equation \ref{eqn:FDcases} 
the stopping time 
\begin{align}
t_{stop}(s) & =\frac{\rho_s s }{\rho_g v_{\rm th}} \times 
	 \begin{cases} 
		 1 & {\rm for\ } s \lesssim 9 \lambda_g/4 \\
		\frac{4 s}{9 \lambda_g} &
		{\rm for\ } s \gtrsim 9 \lambda_g/4   \\
	\end{cases}.
\label{eqn:tstop}
\end{align}
If the particle is small compared to the mean free path, then it is in  
the Epstein regime of free molecular drag, otherwise it is in the Stokes regime corresponding to a small Reynolds number. 
In both cases, the drag force is proportional to the relative velocity between 
particle and gas.   In the Epstein regime,  $t_{stop} \propto r^3  g_\Sigma^{-1}$ for our adopted disk model. 
Here we assume the particle is small and consequently have neglected the ram pressure regime. 

The dimensionless stopping time, often called the Stokes number, is 
\begin{align}
St  = &~ t_{stop}\Omega_K \label{eqn:St} 
\end{align}
where $\Omega_K$ is the Keplerian orbital angular rotation rate. 
The Stokes number (equation \ref{eqn:St}) determines whether a particle is well coupled to the gaseous disk on a time period of about a rotation period about the Sun.    With Stokes number $St<1$, particles orbit with the gas and are slower than a planetesimal that is on a Keplerian orbit.  However, particles that are well-coupled to the disk on a rotation period do not necessarily flow with the gas around a small planetesimal.  This concept is an extension of what is known as `pebble accretion' in the context of accretion of particles, known as pebbles, onto large bodies such as planets \citep{Ormel_2010,Lambrechts_2012,Guillot_2014,Visser_2016,Homann_2016,Ormel_2017}.  Small particles are well-coupled to the gas and don't necessarily accrete onto a planetesimal because they remain coupled to the gas disk, whereas intermediate sized ones, called pebbles, can accrete onto the planet with the aid of gas drag.  

\begin{figure}[ht]
    \centering
    \includegraphics[width=3.5truein, trim = 15 0 0 0, clip]{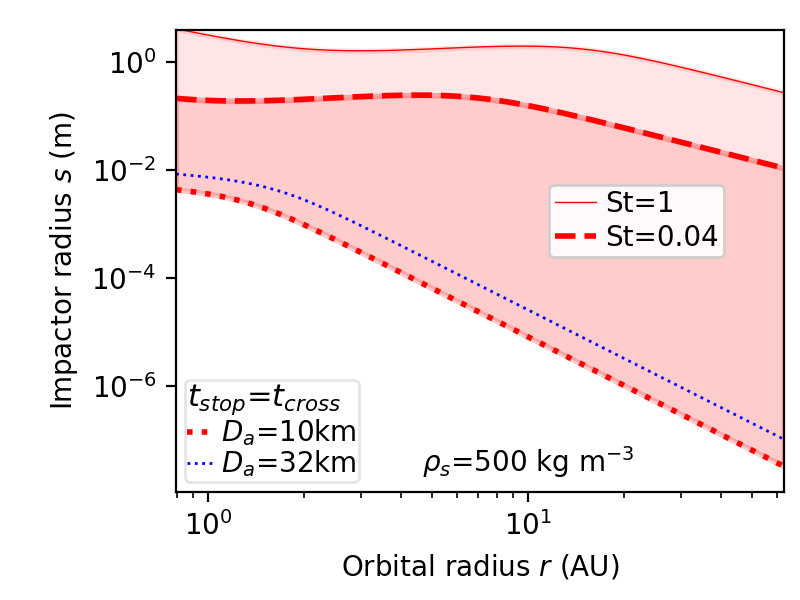}
    \caption{Range of particle sizes from a protostellar disk that can impact a planetesimal. The $x$ axis shows orbital radius and $y$ axis shows particle size. 
    The lower two dotted lines show where stopping time is equal to the time it takes the wind to cross a planetesimal of diameter $D_a $= 10 and 32 km.  Below these lines, particles are less likely
    to impact a planetesimal as they are carried away by the wind.  The top thick red dashed line show Stokes number $St=0.04$   which is the limiting value for particles that are active in the streaming instability \citep{Rucska_2023}.   The thin red line above it shows $St=1$. The bends in these two lines are due to the transition between Epstein and Stokes regime, with particles in the Stokes regime in the inner disk. In the outer solar system smaller dust particles can impact planetesimals. 
The pink shaded region shows particles of density $\rho_s = 500$ kg~m$^{-3}$ that can hit a $D_a=10$ km diameter object.  The region between $St=0.04$ and $St=1$ is shaded a lighter shade to 
    illustrate that some objects in this region could have been previously incorporated into planetesimals via streaming instability.  
    }
    \label{fig:sc}
\end{figure}

 
We consider planetesimals with escape velocity lower than the headwind 
velocity; $v_{esc,a} < u_{hw}$. 
A planetesimal should be larger than the mean free path in the gas $R_a > \lambda_g$,
and gas flow about a planetesimal has Reynolds number $Re_a >1$ (equation \ref{eqn:Re}).  
We refer to a crossing time as the time it takes the headwind at a velocity $u_{hw} $ to cross a planetesimal of diameter $D_a$, 
\begin{align}
t_{cross} = & \frac{D_a}{u_{hw}}  \nonumber \\
 =  &\ 26\ {\rm s} \left( \frac{r}{10\ {\rm AU} }\right)^0
 \left(\frac{D_a}{1 \ {\rm km}} \right) \tilde g_T^{-1}.\label{eqn:tcross}
\end{align}
Particles that have stopping time shorter than the crossing time, $t_{stop} < t_{cross}$, are 
less likely to hit the planetesimal due to aerodynamic deflection \citep{Visser_2016}. 
Particles large enough that $St>1$ are not as well coupled to the gas.  
For small planetesimals, (with wind velocity near or below the escape velocity from the planetesimal) a range of particles sizes  that are likely to hit the planetesimal is 
bounded by $t_{cross} \lesssim t_{stop} \lesssim \Omega_K^{-1}$  (e.g.,  
see section 7.2.2.3 by \citealt{Ormel_2017}).  
The radius of a particle that has stopping time equal to crossing time, $t_{stop}  = t_{cross}$,  we denote $s_{sc}$.   
For $D_a=10$ and 32 km diameter planetesimals this particle size is plotted in Figure \ref{fig:sc} as a function of radius $r$ from the Sun.   The critical particle size $s_{sc}$ is computed numerically by solving for a root of $t_{stop} - t_{cross} = 0$,  using the disk model described in section \ref{sec:disk}, a particle density $\rho_s = 500$ kg m$^{-3}$ and with acceleration due to drag from equation \ref{eqn:a_D}.   
At four different orbital radii, 
particle sizes corresponding to $t_{stop} = 3000$ s, computed in the same way, are listed in Table \ref{tab:tab}.   These values will be relevant in section \ref{sec:traj}.
 
For planetesimals large enough that the escape velocity exceeds the headwind velocity, $v_{esc} > u_{hw}$, 
of diameter a few hundred km,  
the collision probability for particles with
$ t_{cross}< t_{stop}< \Omega_K^{-1}$ is  higher than estimated geometrically from
the planetesimal cross sectional area 
due to gravitational focusing,  for example, see Figure 6 by \citet{Visser_2016}, but also see \citet{Lambrechts_2012,Ormel_2017}.      

If the streaming instability was responsible for planetesimal formation, $St=1$ and larger particles could have previously been incorporated into planetesimals.  Simulations show that particles small enough that $St<0.04$ do not clump into planetesimals via streaming instability  \citep{Rucska_2023} and so could remain in the disk where they later impact a planetesimal. 
In Figure \ref{fig:sc} we plot the particle size for $St = 0.04$ and $St=1$ to estimate the size of large particles (in the disk) that might impact a planetesimal.  Figure \ref{fig:sc} illustrates the range of sizes of particles, in the gaseous protostellar disk that are most likely to impact a small planetesimal.  The range of particle sizes is particularly wide in the outer solar system. 

\begin{figure}[ht]
    \centering
    \if \ispreprint1
    \includegraphics[width=3.5truein, trim = 0 0 0 0, clip]{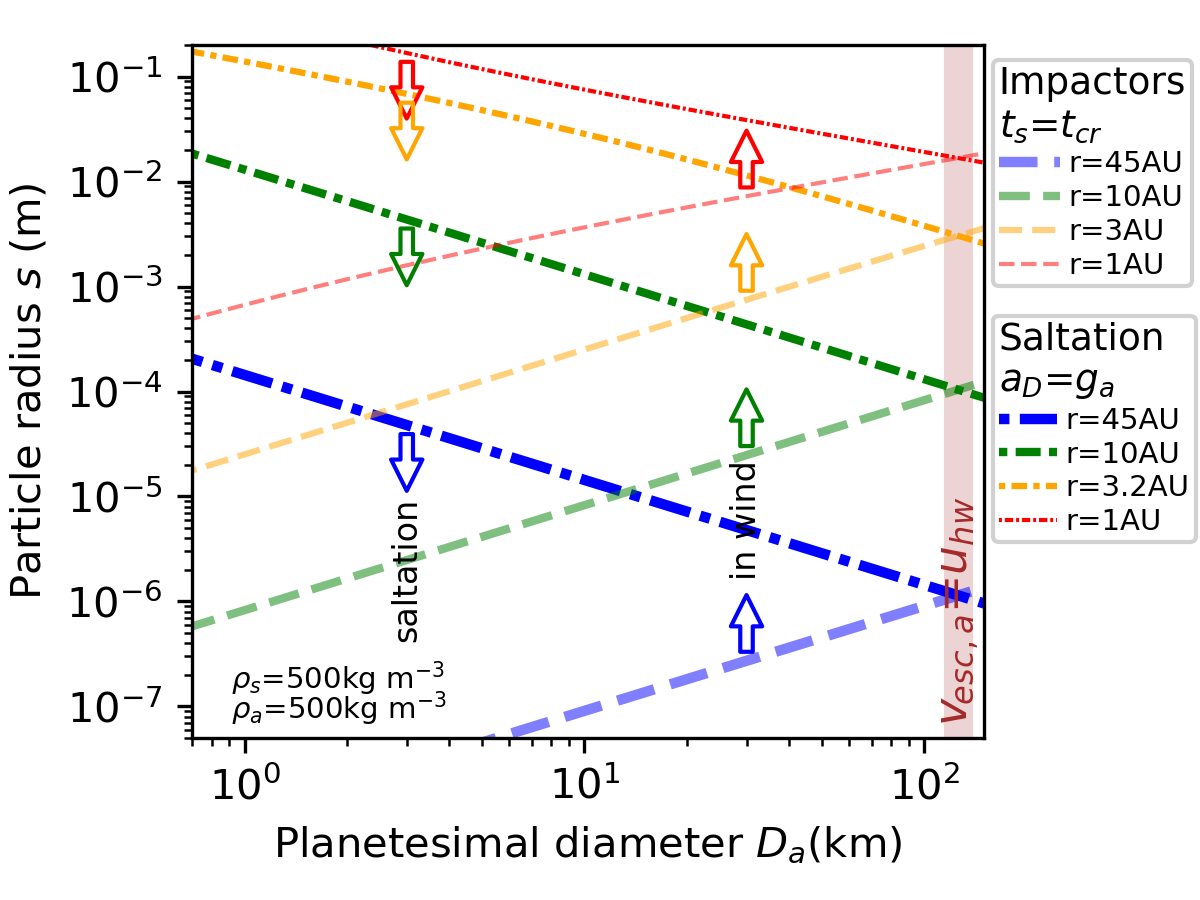}
    \else
    \includegraphics[width=4.5truein, trim = 0 0 0 0, clip]{Da_plot.png}
    \fi
    \caption{Saltation and impact regimes as a function of planetesimal diameter. 
The $y$ axis is particle radius in m and the $x$ axis is planetesimal diameter in km. 
The dashed lines give a lower limit on the size of wind particles that are likely to hit the planetesimal 
as the wind does not carry them away.  The dashed lines are 
where crossing time is equal to stopping time and are computed at orbital radii $r=$ 1, 3.2, 10, and 45 AU, 
in red, orange, green and blue, respectively, and in order of increasing line thickness.
The dot-dashed lines show limits for saltation of surface particles at the same orbital radii. 
Below the dashed lines, surface particles can be pushed by the wind. 
The planetesimal diameter where the escape velocity is equal to the wind velocity is shown 
with a wide vertical tan solid line on the right. 
This plot does not take into account gravitational focusing of impacts on larger planetesimals 
and consequently 
the $x$-axis is truncated at the tan solid line. 
Small particles that are accreted onto the surface are small enough that they can be pushed by 
the wind. 
   }
    \label{fig:Da_plot}
\end{figure}

In Figure \ref{fig:Da_plot} the minimum particle size that is not aerodynamically deflected (computed with 
stopping time equal to crossing time) is compared to 
 the minimum particle size allowing saltation
(computed with drag acceleration equal to surface gravity $a_D = g_a$).   
However, instead of plotting these quantities as a 
function of orbital radius, we plot them as a function of planetesimal diameter at 4 
different orbital radii.  The $x$-axis in Figure \ref{fig:Da_plot} is planetesimal diameter, so this figure is complimentary 
to Figure \ref{fig:codot} and \ref{fig:sc} where the horizontal axis  varies with orbital radius.   

Quantities in Figure \ref{fig:Da_plot} are computed with wind and surface particle density $\rho_s = 500 $ kg m$^{-3}$ and bulk planetesimal density  $\rho_a = \rho_s$, and our disk model. 
Dot-dashed lines show the saltation limits (particle sizes for which drag acceleration is
equal to gravitational acceleration $a_D = g_a$).  Dashed lines show particle sizes with 
stopping time equal to crossing time.  
The widths and colors of the lines depend upon the orbital radius. 
The arrows on the left show where saltation is allowed and the arrows on the right show where wind particles are not aerodynamically deflected and so are more likely to impact the planetesimal. 

If the planetesimal has escape velocity equal to the wind velocity; $v_{esc,a} =  u_{hw} $, 
the minimum particle size that allows saltation (with $a_D = g_a$) 
is equal to that with stopping equal to crossing time ($t_{cross} = t_{stop}$)  
This can be verified by manipulating equations \ref{eqn:assumption},  \ref{eqn:tstop}  and  \ref{eqn:tcross} 
and is why the pairs of dashed and dot-dashed lines on Figure \ref{fig:Da_plot}
 intersect on the right where the escape velocity equals the headwind speed.   
For small planetesimals
(with escape velocity below the wind velocity) the smallest particles that can be accreted
from the wind could be blown across the planetesimal surface by the wind.  
However, as shown in Figure \ref{fig:codot}, 
only in the inner solar system (inside a few AU) can the wind exceed cohesion, which is 
why we now discuss particles that are splashed off the surface due to impacts. 

\subsection{Crater ejecta caused by impacts from particles in a headwind}
\label{sec:ejecta}

The headwind velocity (as estimated in equation \ref{eqn:uhw}) exceeds the 10 m/s threshold, above which two colliding ice particles fail to stick together (\citealt{Gundlach_2015}; also see Figure 2 by \citealt{Kimura_2020}).  
The impact can cause a crater. In a low velocity regime, ejected material consists of splashed particles.  
Studies of ejecta from micrometeorites on airless bodies usually adopt a high, of order km/s, impact velocity  \citep{Zakharov_2022}.  
Impacts occurring at a headwind velocity of about 50 m/s are in a lower velocity regime. 
Recent compilations of crater dimensions suggest that scaling laws, based on dimensionless parameters, are remarkably good at matching crater properties, such as size and volume, over a wide range of impact and substrate properties, and including a low velocity regime \citep{Holsapple_1993,Housen_2011,Housen_2018,Celik_2022}.

For an impact on a planetesimal of density $\rho_a$ and gravitational acceleration 
$g_a$, dimensionless parameters used to characterize impacts 
\citep{Holsapple_1993,Housen_2011} are 
\begin{align}
\pi_2 & = \frac{g_a a_{pj}}{u_{pj}^2}  \label{eqn:pi2} \\
\pi_3 & = \frac{Y_a}{\rho_a u_{pj}^2}  \label{eqn:pi3} \\
\pi_4 & = \frac{\rho_a}{\rho_{pj}},  \label{eqn:pi4} 
\end{align}
where $a_{pj},u_{pj},\rho_{pj}$ are projectile radius, velocity, and density. 
The $\pi_2$ parameter is the inverse of the square of the Froude number. 
The $\pi_3$ parameter depends on the planetesimal's material strength $Y_a$. 
The projectile mass is assumed to be $m_{pj}= \frac{4\pi}{3} \rho_{pj} a_{pj}^3$. 
If crater growth is halted principally by the target strength, then crater formation is in the {\it strength} regime
and scaling laws are independent of $\pi_2$. Otherwise crater formation is said to be in the {\it gravity} regime and the scaling laws are independent of $\pi_3$.  

Craters are in the strength regime when 
\begin{align}
\pi_2 \lesssim \pi_3^{1 + \frac{\mu}{2}} \pi_4^\nu, \label{eqn:c_regime}
\end{align}
with exponents $\nu \sim 0.4$ and $\mu \sim 0.4$ for granular or porous materials  \citep{Housen_2011}.
We use the relation in equation \ref{eqn:c_regime} to place a constraint on the radius of the projectile below which the impact is in the strength regime; 
\begin{align}
    a_{pj} &\lesssim \left(\frac{Y_a}{\rho_a} \right)^{1 + \frac{\mu}{2} }\frac{\pi_4^\nu }{g_a u_{pj}^{\mu}}  \\
    & \sim 71\ {\rm m}\ \left(\frac{Y_a}{500~{\rm Pa}} \right)^{1.2} 
    \left( \frac{u_{pj}}{40~{\rm m/s}}\right)^{-0.4}
   \left( \frac{D_a}{10~{\rm km}} \right)^{-1}\nonumber \\
   & \ \ \ \ \times 
   \left(\frac{\rho_a}{1000~{\rm kg~m}^{-3}} \right)^{-2.2} \pi_4^{0.4}. \label{eqn:acrit}
\end{align} 
We have inserted a strength $Y_a =500$ Pa based on measurements of material strength in regolith \citep{Brisset_2022}.  
We expect the projectiles in the wind to be pebbles and smaller than a meter, so 
equation \ref{eqn:acrit} implies that craters formed from particles in the headwind would predominantly be in the strength regime. This is consistent with studies of rubble asteroids that take into account their cohesive strength \citep{Scheeres_2010}.  However, if the planetesimal is porous and low density, its material strength might be lower than 500 Pa and impacts could be near the transition between strength and gravity impact regimes.  


In the strength regime, scaling laws  
predict a crater radius $R_{cr}$ 
\begin{align}
R_{cr} = a_{pj} \left(\frac{4 \pi}{3}\right)^\frac{1}{3} H_2 
\pi_3^{-\frac{\mu}{2}}
\pi_4^{-\nu} \label{eqn:Rcr}
\end{align}
(from Table 1 by \citealt{Housen_2011}).
The radius depends on coefficient $H_2$ and exponents $\mu$, and $\nu$ which are measured from experiments in different substrate materials. 
We take coefficient and exponent values from column 5 for sand or column 7 
for weakly cohesive sand and fly ash from Table 3 by \citet{Housen_2011}.   
The coefficients and exponents are approximately 
\begin{align}
H_2 = 0.4, \ \ C_1 = 0.55, \ \  k = 0.3, \nonumber  \\ 
\ \ \mu = 0.4, \ \ \nu= 0.4,  \ \ n_1 = 1.2.  \label{eqn:cofs}
\end{align}
We included the additional coefficients  $C_1, k, n_1$ that will be used below. 
With these coefficients and with impact velocity set by the disk headwind 
from equation \ref{eqn:uhw}, $u_{pj} = u_{hw}$, the crater radius 
\begin{align}
R_{cr} \sim &\  3.2\ a_{pj} 
\left( \frac{Y_a}{500~{\rm Pa}} \right)^{-0.2}
\left( \frac{\rho_a}{1000~{\rm kg~m}^{-3}} \right)^{0.2} \nonumber \\
& \ \ \times 
\left( \frac{r}{10~{\rm AU}}\right)^0 
\pi_4^{-0.4} \tilde g_T^{0.4}. \label{eqn:Rcr2}
\end{align}
We find that craters on a planetesimal that are formed from impacts with 
particles in a headwind are only a few times larger than the headwind particles themselves. 
If the projectile density is lower than the planetesimal density, then $\pi_4>1$ and the crater radius would be smaller, as expected. 

\subsection{Erosion or accretion?}
\label{eqn:erode}

By comparing the velocity of ejecta to the escape velocity on a planetesimal, we determine whether
impacts from particles in a headwind would lead to an increase in mass or accretion,  or would lead to erosion, 
because so much ejecta escapes the planetesimal. 

Following \citet{Housen_2011}, 
we assume that a projectile acts like a point source when considering crater-related phenomena. 
The ejecta properties depend on 
 the horizontal distance from the site
of impact, $x$.  Between the projectile radius and the crater radius ($x$ ranging from about projectile radius $a_{pj}$ to the crater radius $R_{cr}$), the ejecta velocity $v_{ej}$ obeys a power law 
\begin{equation}
v_{ej}(x) \sim C_1 \! \left(\frac{x}{a_{pj}} \right)^{\!-\frac{1}{\mu}} \!\! \pi_4^{-\frac{\nu}{\mu}}  u_{pj} \ \ 
{\rm \  for\  }  n_1 a_{pj} \lesssim   x \lesssim R_{cr},
 \label{eqn:vejx}
\end{equation} 
(via Eqn.~14 by \citealt{Housen_2011}), with dimensionless coefficients $n_1$ and $C_1$.
This equation gives maximum and minimum ejection velocities at $x \sim n_1 a_{pj}$ and at $x \sim R_{cr}$, 
\begin{align}
v_{ej,max} &\sim C_1 \pi_4^{-\frac{\nu}{\mu}} n_1^{-\frac{1}{\mu} } u_{pj} \nonumber \\
v_{ej,min}& \sim C_1 \pi_4^{-\frac{\nu}{\mu}}   \left( \frac{a_{pj}}{R_{cr}} \right)^{\frac{1}{\mu}} u_{pj} \nonumber \\ & = 
C_1 \left( \frac{4\pi}{3} \right)^{-\frac{1}{3 \mu}} H_2^{- \frac{1}{\mu}} \pi_3^\frac{1}{2} u_{pj}.
\label{eqn:vej_lims}
\end{align}
Using coefficients listed in equation \ref{eqn:cofs} and density ratio $\pi_4 = 1$, the maximum ejection velocity 
(via equation \ref{eqn:vej_lims}) is $v_{ej,max} = 0.35~u_{pj}$. 

Inverting equation \ref{eqn:vejx} to estimate $x$ as a function of the ejection velocity 
\begin{equation}
x(v_{ej}) = a_{pj} \left( \frac{v_{ej}}{C_1 u_{pj}} \right)^{-\mu} \pi_4^{-\nu}.
\label{eqn:xvej}
\end{equation}
The integrated mass ejected at distances up to $x$ from site of impact  
\begin{align}
M(x) \sim k \rho_a \left( x^3 -  n_1^3 a_{pj}^3 \right)   \ \ \  {\rm \  for\  } n_1 a_{pj} \lesssim  x \lesssim R_{cr},
\label{eqn:Mx}
\end{align}
(Eqn.~18 by \citealt{Housen_2011}), with dimensionless coefficient $k$.
Inserting equation \ref{eqn:xvej} into equation \ref{eqn:Mx}
 gives total ejecta mass above velocity $v_{ej}$ 
\begin{align}
M(v_{ej}) \sim k \rho_a a_{pj}^3 
	\left( \left( \frac{v_{ej}}{C_1 u_{pj}}\right)^{-3\mu}\!\! \pi_4^{-3\nu} -n_1^3 \right),
\end{align}
for $v_{ej}$ within the range given by equations \ref{eqn:vej_lims}.
We compare this to the mass of the projectile 
\begin{align}
\frac{M(v_{ej}) }{m_{pj}} \sim \frac{3k}{4 \pi}   \pi_4
\left( \left( \frac{v_{ej}}{C_1 u_{pj}}\right)^{-3\mu}\!\! \pi_4^{-3\nu}  - n_1^3 \right). \label{eqn:Mvej}
\end{align} 
Note that equation \ref{eqn:Mvej} is independent of material strength $Y_a$. 

By setting ejecta velocity to the escape velocity on a planetesimal, $v_{ej} = v_{esc,a}$,  we estimate whether the impacts cause accretion or erosion on the planetesimal.   
If the fraction of ejecta above the escape velocity exceeds the projectile mass,  $M(v_{esc,a}) /m_{pj} > 1$, then the impact erodes the planetesimal,  otherwise it accretes mass.  
The dividing line occurs when $M(v_{esc,a}) /m_{pj} \sim 1$ which is at 
an escape velocity value that satisfies 
\begin{align}
v_{esc,a,crit}
&\sim   C_1 \left( \frac{3k}{4 \pi} \right)^\frac{1}{3\mu} \pi_4^\frac{1- 3 \nu}{3 \mu}u_{pj} \nonumber \\
 & \sim 0.061\ \pi_4^{-0.17} u_{pj} .  \label{eqn:vesc_b}
\end{align}
In the last step we have used the coefficients given by \citet{Housen_2011} and listed in equation 
\ref{eqn:cofs}.
In other words if
\begin{equation}
v_{esc,a} \gtrsim  v_{esc,a,crit} 
, \label{eqn:vesc}
\end{equation}
we expect impact from particles in the headwind to allow the planetesimal to accrete mass,  otherwise particles in the headwind would cause erosion. 
This expression is not sensitive to projectile mass or substrate material strength. 


The escape velocity of a spherical planetesimal with diameter $D_a$ is 
\begin{align}
    v_{esc,a} &= \sqrt{\frac{2 G M_a}{R_a}} 
     = \sqrt{ \frac{2 \pi G \rho_a}{3}}\ D_a. \label{eqn:vesc_a}
\end{align}   
Setting the escape velocity equal to the critical value  in equation \ref{eqn:vesc_b}, 
we estimate the diameter of a planetesimal below which impacts from particles in a headwind are erosional;
\begin{align}
D_{erode} &\approx  C_1 \left( \frac{3k}{4 \pi} \right)^\frac{1}{3\mu} \pi_4^\frac{1- 3 \nu}{3 \mu} 
\frac{u_{hw}}{\sqrt{ 2 \pi G \rho_a /3}}\nonumber \\
& \approx \frac{0.061\  \pi_4^{-0.17} u_{hw}}{\sqrt{ 2 \pi G \rho_a /3}}\nonumber \\
&\approx 6.3 \ {\rm km} \left( \frac{r}{10~{\rm AU}}\right)^0 \!
\left( \frac{\rho_a}{1000~{\rm kg~m}^{-3}}\right)^{-\frac{1}{2}} \nonumber \\
& \ \ \times \tilde g_T \pi_4^{-0.17}. \label{eqn:D_erode}
\end{align}
In the estimate for $D_{erode}$ we set the projectile velocity equal to the headwind velocity (neglecting
gravitational focusing) and used
Equation \ref{eqn:uhw} for the headwind velocity.  
This expression is independent of planetesimal material strength $Y_a$ and independent of the 
density of the gas disk. It is primarily dependent upon the impact speed of projectile particles.   

We find that impacts from headwind particles are not likely to erode the planetesimal if the planetesimal diameter exceeds about 8 km. 
The planetesimal size limit for erosion is independent of radius from the Sun because in our disk model the headwind velocity is independent of radius. 
If the protostellar disk has a gap or is cooler or has lower exponent for radial decay of disk density than we assumed in our disk model, then $u_{hw}$ could be lower than 47 m/s (estimated in equation \ref{eqn:headwind} and equation \ref{eqn:uhw}) gives a lower value of $D_{erode}$.  
During an epoch when streaming instability is active and yet the particle density is high, the relative speed between disk particles and a forming planetesimal could be lower than 50 m/s \citep{Johansen_2007}. 
If the particles within the headwind are fluffy (density $\rho_{pj}$ is low compared to the bulk density of the planetesimal surface), then parameter $\pi_4 = \rho_a/\rho_{pj} >1$ and the critical asteroid diameter below which erosion occurs could be below 8 km.

To derive the erosion size limit of equation \ref{eqn:D_erode}, we used crater and ejecta 
scaling laws.  Impacts from particles in a headwind are not 
expected to be at high velocity.  The planetesimal surface is granular and if the projectiles are smaller than surface particles, then the collision could be in a ballistic regime where a single particle is knocked off the surface.  Momentum conservation for the encounter between the two particles would give a low ejecta velocity for the larger particle.  If the size distribution of particles 
on the planetesimal surface is dominated by particles that are large compared to 
those in the wind that hit the surface, then equation \ref{eqn:D_erode} would overestimate the erosion limit. 
In other words, a planetesimal smaller than that in equation \ref{eqn:D_erode}, could withstand erosion if its surface lacked small grains.  

The estimate for $D_{erode}$ in equation \ref{eqn:D_erode} does not take into account planetesimal rotation. If an impact near the poles does not cause erosion, a similar velocity impact near the equator of a quickly rotating object could still produce ejecta above the escape velocity, causing erosion. 
The estimate does not take into account drag forces on the ejecta due to interaction with the disk,   collisions between ejecta and ejecta and wind particles, or heterogeneity in the substrate and surface topography. 
 
\subsection{Accretion and erosion rates}
\label{sec:erode}

For an impact of a headwind particle with mass $m_{pj}$, we use equation \ref{eqn:Mx} for the mass of ejecta at $x = R_{cr}$ and equation \ref{eqn:Rcr} for the radius of a crater to estimate the total mass ejected during crater formation 
\begin{align}
\frac{M_{ej} }{m_{pj} }&  \approx 
\frac{3k }{4\pi} \pi_4 \left(\frac{R_{cr}} {a_{pj}} \right)^3  = k \pi_4^{1-3\nu} \pi_3^{-\frac{3\mu}{2} } H_2^3 \label{eqn:Mcr1} \\
& \approx 2.4
\left( \frac{Y_a}{500~{\rm Pa}} \right)^{-0.6}
\left( \frac{\rho_a}{1000~{\rm kg~m}^{-3}} \right)^{0.6} \nonumber \\
& \ \ \ \ \ \ \ \times 
\left( \frac{r}{10~{\rm AU}}\right)^0 \pi_4^{-0.2}  \tilde g_T^{1.2}. \label{eqn:Mcr}
\end{align}
For this mass ratio estimate,  we have used the coefficients for crater ejecta scaling listed in equation \ref{eqn:cofs}.
Note that this mass ratio increases if the body has lower strength. 
Conveniently, in the strength regime for crater formation, the planetesimal diameter, setting the gravitational acceleration, is not relevant. 

For a setting where impacts from particles in the headwind lead to accretion,  we estimate 
the accretion rate onto the planetsimal from the mass flux of headwind particles.  
If impacts from the headwind lead to erosion, the erosion rate can be estimated from the particle mass flux and using the crater to projectile mass ratio in equation \ref{eqn:Mcr}.


We characterize the fraction of the protostellar disk mass in particles with the ratio $f_p \equiv \rho_p/\rho_g$, 
 where $\rho_{p}$ is the effective density in  the disk from particles, not the density of the particles themselves.  
Small particles may not hit a planetesimal due to aerodynamic deflection 
(\citealt{Visser_2016}, and as discussed in section \ref{sec:sc}) . 
We use a dimensionless factor $\xi_p $ to describe the mass fraction of disk particles that can impact a planetesimal.  The factor $\xi_p$ is the mass per unit time in disk particles that impact the planetesimal divided by the mass per unit time in disk particles that would have impacted the planetesimal in the absence of areodynamic deflection.  
 Neglecting gravitational focusing, $\xi_p <1$.    
The mass of particles impacting the planetesimal per unit area and time is 
\begin{align}
    F_m & \sim  f_p \xi_p \rho_g u_{hw} \label{eqn:Fm} \\
    &\approx 7 \times 10^{-9}\  
    {\rm kg~ m}^{-2}{\rm s}^{-1} \left( \frac{f_p}{10^{-2}} \right) 
\nonumber  \\
& \ \ \ \times \left(\frac{r}{10 \ {\rm AU} }\right)^{-\frac{11}{4}}
\xi_p \tilde g_\Sigma \tilde g_T^\frac{1}{2}
. \label{eqn:Fm1}
\end{align}
If impacts cause mass loss, we would multiply this mass flux by $M_{ej}/m_{pj}$ 
from equation \ref{eqn:Mcr} to estimate the erosion rate.  If impacts cause accretion, then the amount of material that is lofted above the surface per unit area and time via impacts can be estimated from $F_m$ multiplied by $M_{ej}/m_{pj}$. 

If the impacts lead to accretion, the mass flux in equation \ref{eqn:Fm} can be converted to a deposition rate by dividing by $4 \rho_{pj}$
with the factor of 4 taking into account the ratio of surface area to cross sectional area 
and $\rho_{pj}$ the density of the projectile particles, 
\begin{align}
    \frac{dh_{acc}}{dt} &\sim \frac{F_m}{4 \rho_s} \nonumber \\
    & \sim 0.11\ {\rm m~kyr}^{-1} 
    \left( \frac{f_p}{10^{-2}} \right) 
    \left(\frac{r}{10 \ {\rm AU} }\right)^{-\frac{11}{4}}
    \nonumber \\
& \ \ \ \times 
\left(\frac{\rho_{pj}}{500~ {\rm kg~m}^{-3}} \right)^{-1} \xi_p \tilde g_\Sigma \tilde g_T^\frac{1}{2}
. \label{eqn:h_acc}
\end{align}
If the impacts lead to erosion, then the mass flux should be multiplied by the escaping ejecta mass
fraction from equation \ref{eqn:Mcr} and divided by $4 \rho_a$, 
\begin{align}
    \frac{dh_{erode}}{dt} &\sim \frac{F_m}{4 \rho_a} \frac{M_{ej}}{m_{pj}}\nonumber \\
    & \sim 0.27\ {\rm m~kyr}^{-1} 
    \left( \frac{f_p}{10^{-2}} \right) 
    \left(\frac{r}{10 \ {\rm AU} }\right)^{-\frac{11}{4}}
    \nonumber \\
& \ \ \ \times 
\left( \frac{\rho_a}{1000~{\rm kg~m}^{-3}} \right)^{-0.4} \!\!
\left( \frac{Y_a}{500~{\rm Pa}} \right)^{-0.6} \nonumber \\
& \ \ \  \ \times  \xi_p 
 \pi_4^{-0.2}  \tilde g_T^{1.7} \tilde g_\Sigma 
 .  \label{eqn:h_erode}
\end{align}

For comparison, the lifetime of a protostellar disk is  2--10 Myr \citep{Williams_2011}.
Equation \ref{eqn:h_acc} and \ref{eqn:h_erode} imply that 
in the inner solar system where the disk density is higher,  accretion and erosion (depending upon
 planetesimal size) due to impacts 
 could be significant, with small planetesimals likely to erode and larger ones likely to accrete mass.  
For planetesimals with size near $D_{erode}$, the deposition rate 
(and the additional factor of  $M_{ej}/m_{pj}$) imply mixing between  
particles on the planetesimal surface and grains in the headwind. 
In the Transneptunian region ($>$ 30 AU),  erosion or accretion due to impacts
would not be significant unless 
there are epochs where the disk exhibits regions of high particle concentration (high values of $f_p$). 

\begin{figure}[!htbp]\centering
\includegraphics[width=3.3truein,trim = 1 13 5 5, clip]{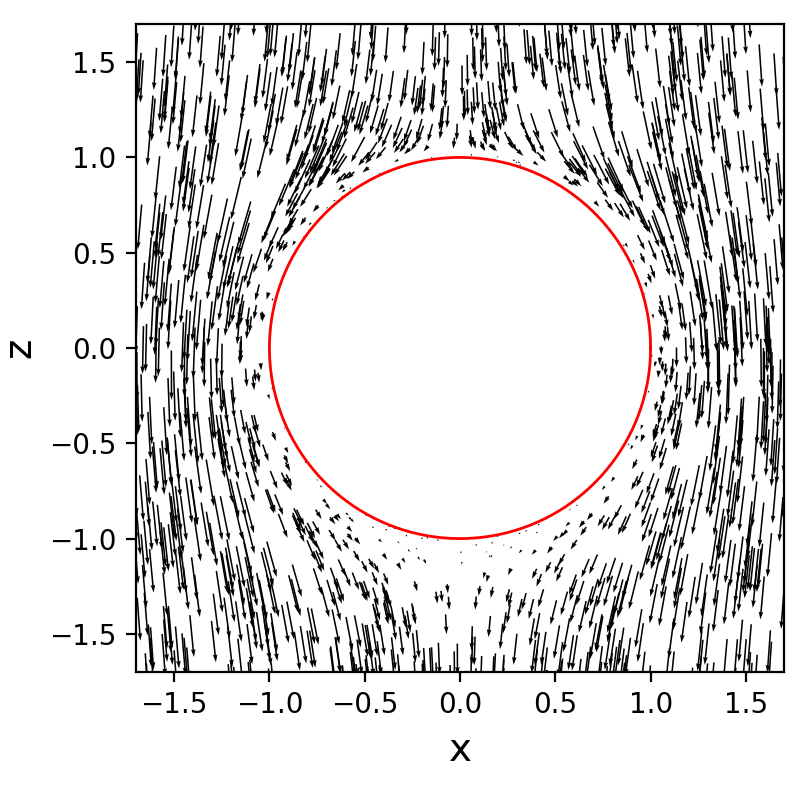}
\caption{Vector graph showing our simple model for flow of the disk gas about a spherical planetesimal. 
The flow vectors are given by a potential model that is modified near the
surface with a boundary layer with thickness  sensitive to Reynolds number.
The flow shown has Reynolds number $Re = 10$ and the boundary layer is particularly visible on the lower side.   We chose a low Reynolds number for this figure to better show the boundary layer. 
\label{fig:wind_illust}}
\end{figure}

\section{Trajectories of splash particles}
\label{sec:traj}

With integrations of splash particle trajectories, 
we explore how impacts generate ejecta that are transported across the surface or escape the planetesimal altogether.
Our ejecta particle integrations are similar to those of \citet{Larson_2021} for asteroid   
ejecta curtains, except impacts are low velocity and we include gas drag.  
For each impact site, we generate a splash (ejecta) particle velocity using velocity distribution described below in section \ref{sec:pej} that is consistent with scaling for crater ejecta developed by \citet{Housen_2011} and is discussed in section \ref{sec:ejecta}. 
The equation of motion for the splash particles  
is described in section \ref{sec:motion} and takes into account drag from a headwind, with flow about the planetesimal described in section \ref{sec:pot}. 

\subsection{Equation of motion for a splash particle in a wind}
\label{sec:motion}

Ignoring orbital motion about the Sun, we work in an inertial frame with 
 $+z$ pointing in the upwind direction and the $x,y$ plane
delineating the transition between windward and leeward sides. 
This coordinate system is illustrated in Figure \ref{fig:wind_illust}.
We define an impact parameter $w = \sqrt{x^2 + y^2}$. 

Assuming axi-symmetry  the wind velocity ${\bf u}(w,z)$ is a function of impact parameter $w$ and $z$. 
Upwind of the planetesimal, we assume a wind velocity $\lim_{z\to \infty} {\bf u}(w,z) = -u_{hw} \hat {\bf z}$. 
 
A splashed particle of radius $s$, has mass $m_s$,  position ${\bf r}_s$ (with respect to the center of the planetesimal), velocity ${\bf v}_s = \frac{d{\bf r}_s }{dt}$ and acceleration ${\bf a}_s = \frac{d{\bf v}_s }{dt} $ where time is $t$.
After a particle is splashed off the surface by an impact, its acceleration is 
\begin{align}
{\bf a}_s =  -\frac{GM_a}{r_s^3} {{\bf r}_s}   + \frac{{\bf F}_D}{m_s}, 
\label{eqn:acc}
\end{align} 
where the first term on the right is due to the gravity of the planetesimal, which is assumed to be a sphere, 
and ${\bf F}_D$ is the drag force from the wind.  The drag depends on the particle size, and the wind's velocity vector at position ${\bf r}_s$.   We ignore lift and a possible Magnus effect on the particle. 
In the Epstein and Stokes regimes, drag is proportional to velocity. 
The drag force   
\begin{align}
\frac{{\bf  F}_D}{m_s}  \approx  \frac{ ({\bf u} - {\bf v}_s)}{t_{stop}(s)}   ,
\end{align}
where the wind velocity is a function of position, ${\bf u}({\bf r}_s)$, 
and the stopping time is given by equation \ref{eqn:tstop}.
We divide equation \ref{eqn:acc} by the gravitational acceleration on the surface of planetesimal $g_a$, 
\begin{align}
\frac{{\bf a}_s}{g_a} &= - \left( \frac{R_a}{r_s} \right)^2 \frac{\hat {\bf r}_s }{R_a}
+ \frac{ ({\bf u} - {\bf v}_s) }{t_{stop}(s) g_a} .
\end{align}

We choose to work in gravitational units.
Distances are in units of planetesimal radius $R_a$, 
$\tilde{\bf r}_s = {\bf r}_s/R_a$,  and dimensionless time is $ \tau = \Omega_a t$ 
with $\Omega_a \equiv \sqrt{GM_a/R_a^3} = \sqrt{4 \pi G \rho_a/3}$.    
Note that $\Omega_a$ only depends on density $\rho_a$ as $M_a \propto R_a^3$. 
Velocity is in units of $v_a = \sqrt{GM_a/ R_a}$ and 
acceleration in units of $g_a = GM_a/R_a^2$.   In these dimensionless units the escape velocity is equal to $\sqrt{2}$.
The planetesimal spin vector we denote ${\boldsymbol \Omega}_{spin,a}$.  The associated dimensionless 
spin vector in gravitational units is
\begin{equation}
\tilde {\boldsymbol \omega}_a = \frac{{\boldsymbol \Omega}_{spin,a}}{\Omega_a} .
\end{equation}

With acceleration $\tilde {\bf a}_s  = {\bf a}_s/g_a$,   velocity 
$\tilde {\bf v}_s = {\bf v}_s/v_a$,  wind velocity $\tilde {\bf u} = {\bf u}/v_a$,  and stopping time 
$\tilde t_{s} = t_{stop}(s) \Omega_a$,
the equation of motion is 
\begin{align}
\frac{d^2 \tilde {\bf r}_s}{d\tau ^2}  = 
\tilde {\bf a}_s & = -\frac{ \tilde {\bf r}_s}{\tilde r_s^3} 
   + \frac{ ( \tilde {\bf u}(\tilde {\bf r}_s) - \tilde {\bf v}_s) }{\tilde t_{s} }.
   \label{eqn:eqm}
\end{align}
Equation \ref{eqn:eqm} is the equation of motion we integrate to create splash particle trajectories. 

\subsection{Wind flow}
\label{sec:pot}

The Reynolds number is greater than unity for flow of the gaseous disk about a planetesimal. However,  the gas mean free path could exceed the size of the splashed particles, putting them in a moderate Knudsen number regime with respect to the gas flow.  A boundary layer near the planetesimal surface should have structure dependent upon the mean free path and temperature \citep{Sharipov_2016}.  This gives a jump known as a slip velocity \citep{Maxwell_1879}, where the mean particle velocity is not equal to that of the surface within the boundary layer (see Chap 10 by \citealt{Sharipov_2016}). 
Because of its simple analytical form and following \citet{Visser_2016},  we adopt potential flow \citep{Batchelor_1967},  to describe the velocity of wind gas particles about a spherical planetesimal.  The advantages of a potential flow model are its simple analytical form and that there is flow next to the planetesimal surface, mimicking a slip velocity.  
However, a potential flow model has the disadvantage that it unrealistically gives no drag due to d'Alembert's paradox.  
We compromise by adopting a potential flow model, but modify the flow close to the planetesimal surface with a laminar boundary layer with thickness that depends on the Reynolds number.  



For a cylindrical coordinate system with cylindrical radius $w$ and $z$ coordinate as shown in Figure \ref{fig:wind_illust}, a velocity field for flow in the negative $z$ direction about a sphere with radius $R_a$ 
can be generated from an axisymmetric potential 
\begin{align}
\Phi(w,z) &=  -U z \left( 1 + \frac{R_a^3}{2 (w^2 + z^2)^\frac{3}{2}} \right).
\end{align}
The velocity components are found by taking the gradient of this potential 
\begin{align}
u_w & = \frac{\partial \Phi}{\partial w} = \frac{3 U}{2 } \frac{ w z R_a^3 }{(w^2 + z^2)^\frac{5}{2}}  \label{eqn:pot1}\\
u_z & =  \frac{\partial \Phi}{\partial z} = -U -  \frac{ U R_a^3}{2  (w^2 + z^2)^\frac{3}{2}}
\left(1 - \frac{3 z^2}{w^2 +z^2}  \right). \label{eqn:pot2}
\end{align}
Distant from the planetesimal, the wind velocity ${\bf u} = -U \hat{\bf z} =- u_{hw}\hat {\bf z}$.

In dimensionless units, the wind velocity as a function of position 
$\tilde {\bf r}_s = (\tilde x_s, \tilde y_s,\tilde  z_s)$  for the potential flow model (in equations \ref{eqn:pot1}, \ref{eqn:pot2})
\begin{align}
\tilde {\bf u}_{\rm pf}(\tilde {\bf r}_s) & = 
\tilde u_{hw} \Bigg[ \frac{3}{2} \frac{ \tilde z_s }{ \tilde r_s^5} \left(\tilde x_s \hat {\bf x}_s + \tilde y_s 
 \hat {\bf y}_s \right) \nonumber \\
 & \qquad\qquad - \left( 1 +   \frac{1}{2r_s^3 } \left( 1 -  \frac{3\tilde z^2}{\tilde r_s^2} \right) \right) \hat {\bf z}_s  \Bigg].
 \label{eqn:potflow}
\end{align}

We modify the potential flow model of equation \ref{eqn:potflow} by adding a laminar boundary layer based on Prandtl's quadratic scaling law. 
The width of the boundary layer is 
\begin{align}
\delta(\tilde x_b) = \sqrt{\frac{\tilde x_b}{Re}}, 
\end{align} 
where $\tilde x_b$ is the distance along the surface from the windward side's stagnation point and 
$Re$ is the Reynold's number of the flow about the planetesimal.
The distance to the surface of point $\tilde {\bf r}_s$ is 
$\tilde y_b = \left|\tilde {\bf r}_s -  \frac{\tilde  {\bf r}_s}{|\tilde {\bf r}_s|} \right|$.  Points within the boundary layer satisfy $\tilde y_b < \delta(\tilde x_b)$ where $\tilde x_b$ is integrated along the surface from the windward stagnation point to  $\hat {\bf r}_s$.  
Within the boundary layer, we smoothly vary the flow
\begin{align}
\tilde {\bf u}_{\delta}(\tilde {\bf r}_s) = \frac{\tilde y_b}{\delta(\tilde x_b)} \tilde {\bf u}_{\rm pf}(\tilde {\bf r}_s) + \left(1 - \frac{\tilde y_b}{\delta(\tilde x_b)} \right)\tilde {\boldsymbol \omega}_{a} \times   \tilde{\bf r}_s .
\label{eqn:udelta}
\end{align}
Here the function $\tilde {\bf u}_{pf}(\tilde {\bf r}_s)$ is given by equation \ref{eqn:potflow} and $\tilde {\boldsymbol\omega}_a$ is the planetesimal spin vector in gravitational units. 
The $y_b/\delta(x_b)$ factor is present in the self-similar Blasius boundary layer model. 
The term on the right of equation \ref{eqn:udelta} takes into into account rotation of the planetesimal surface. 

The full velocity field in dimensionless units is 
\begin{align}
\tilde {\bf u}( \tilde {\bf r}_s)  =
 \begin{cases}
	\tilde {\bf u}_{\rm pf} (\tilde{\bf  r}_s)  & {\rm for\ \ } \tilde y_b > \delta (\tilde x_b) \\
	\tilde {\bf u}_\delta (\tilde{\bf  r}_s) & {\rm otherwise} ,
\end{cases} \label{eqn:tildeu}
\end{align}
with $\tilde {\bf u}_{\rm pf}$ given by equation \ref{eqn:potflow} and $\tilde {\bf u}_\delta$
given by equation \ref{eqn:udelta}.

\subsection{Distribution of ejecta velocities}
\label{sec:pej}

We assume that impacts from particles in the wind are on the $+z$ hemisphere of the planetesimal and are uniformly distributed in cross sectional area.  We neglect gravitational focusing of the projectiles.  
For each simulated impact, we first generate a single splash particle at a randomly chosen impact point $\tilde {\bf r}_{s,init}$ on the planetesimal surface.   The impact site gives the initial position for a subsequently integrated splash particle trajectory. 

Each projectile has a velocity equal to that of the headwind velocity. 
In the frame rotating with the planetesimal, 
the projectile velocity 
\begin{align}
\tilde {\bf u}_{pj} = - \tilde u_{hw} \hat {\bf z} -  
 \tilde {\boldsymbol\omega }_a\times \tilde {\bf r}_{s,init}  .
\label{eqn:u_pj}
\end{align}
where $\tilde {\bf r}_{s,init}$ is the site of impact. 

By differentiating equation \ref{eqn:Mvej}, we estimate the velocity distribution of ejecta 
per unit mass 
\begin{align}
p(v_{ej})  dv_{ej}  \propto v_{ej}^{-3 \mu - 1} dv_{ej}  \label{eqn:pej}
\end{align}
The exponent is -2.2 for exponent $\mu=0.4$ from equation \ref{eqn:cofs}. 
Maximum and minimum ejection speeds are given in equation \ref{eqn:vej_lims}. By integrating equation \ref{eqn:cofs}, these limits normalize the distribution function. 
For each splash particle we choose an 
ejecta velocity magnitude with probability given by 
the power law distribution of \ref{eqn:pej}
and using maximum and minimum velocity limits, which are 
dependent upon the magnitude 
of the projectile velocity (via equation \ref{eqn:u_pj}) in the frame rotating with the planetesimal. 

We describe how we choose a velocity direction for each splash particle. 
Because our planetesimal is spherical and distances are normalized by planetesimal radius, the surface normal at the site of an impact $\hat {\bf n} = \tilde {\bf r}_{s,init}$. 
We define the downrange direction 
\begin{equation}
\hat {\bf d} =  \frac{ 
{\bf u}_{pj} - ( {\bf u}_{pj} \cdot \hat {\bf n} )\hat {\bf n} } 
{| 
{\bf u}_{pj} - ( {\bf u}_{pj} \cdot \hat {\bf n} )\hat {\bf n} |} . \label{eqn:dvec}
\end{equation}
The unit vector $\hat {\bf d}$ is computed by projecting 
the incoming projectile velocity into the plane tangent to the surface at the point of impact. 

The ejecta velocity $\tilde {\bf v}_{ej}$ (in the rotating frame)
has direction $\hat {\bf v}_{ej}$. 
We assume that the ejecta angle is independent of the impact angle, and 
is aligned with the downrange direction, 
\begin{align}
    \hat {\bf v}_{ej} = \sin \zeta_{ej} \hat {\bf n} + \cos \zeta_{ej} \hat {\bf d},  \label{eqn:hatvej}
\end{align}
where $\zeta_{ej}$ is an elevation angle. 
We assume a downrange direction, to mimic the strong azimuthal asymmetry of ejecta caused by oblique impacts into granular media \citep{Anderson_2003,Raducan_2022,Suo_2023}.
We assume an ejecta elevation angle of $\zeta_{ej} = 45^\circ$,  similar to those measured for oblique impacts in a granular medium \citep{Anderson_2003} and in simulations of oblique impacts \citep{Raducan_2022}.  
After choosing the magnitude of the ejecta velocity $\tilde v_{ej}$ from the power law distribution (equation \ref{eqn:pej}) and its direction with equation \ref{eqn:hatvej}, 
the ejecta velocity vector is transferred back into the inertial frame, giving initial splash particle velocity 
\begin{equation} 
\tilde {\bf v}_{s,init} = \tilde v_{ej} ( \sin \zeta_{ej} \hat {\bf n} + \cos \zeta_{ej} \hat {\bf d})  
 +  \tilde  {\boldsymbol \omega}_a \times \tilde {\bf r}_{s,init} .  \label{eqn:v_init}
 \end{equation}

\begin{figure}[!htbp]\centering
\includegraphics[width=3.3truein, trim = 5 10 20 30,clip]{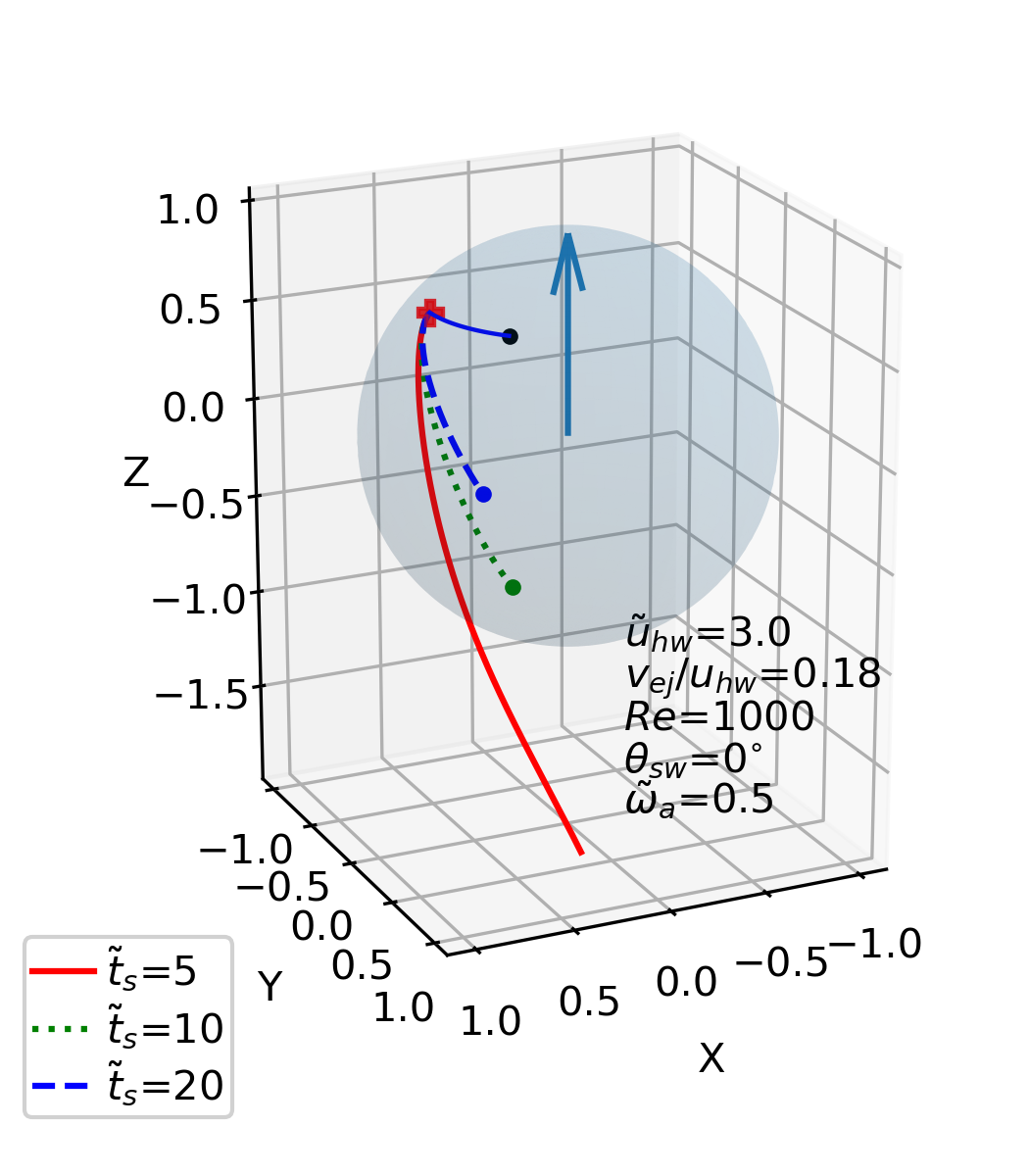}
\caption{Examples of the trajectories of integrated splashed particles.  Three trajectories with the same initial velocity but different stopping times are shown. 
The vertical vector shows the direction of planetesimal rotation.  The wind comes from the $+z$ direction.   The trajectories are computed with the same wind speed. 
The solid blue arc shows rotation of a point on the surface that rotates with the planetesimal. 
The arc shows how far a particle at the impact point moves compared to the splashed particle  with $\tilde t_s = 20 $ and shown with the dashed blue curve.   
The splash particle with the shortest stopping time, with $\tilde t_s = 5$ and shown with a red solid line,  was blown away by the wind and escaped the planetesimal. 
\label{fig:trajs}
  } 
\end{figure}

\subsection{Numerical integrations of splash particles}

Parameters needed to describe an integration of a single splash particle are the site of impact, giving the initial position of the splash particle on the windward side, the initial ejection velocity,  the wind velocity $\tilde u_{hw}$, the Reynold's number, setting the thickness of a boundary layer, 
the stopping time $\tilde t_s$, and 
the planetesimal spin vector $\tilde {\boldsymbol \omega}_{a}$. 
The orientation of the spin vector is described with 
the angle between planetesimal spin vector ${\boldsymbol \Omega}_{spin,a}$ and the direction from which the wind originates,  
$\theta_{sw} \in [0,\pi]$;
\begin{equation}
\cos \theta_{sw} = \frac{ \tilde {\boldsymbol \omega}_{a} \cdot \hat {\bf z} }{\tilde\omega_{a} }
. \end{equation} 
If $\theta_{sw} =0$ then the spin axis points upwind. 

The equation of motion,  equation \ref{eqn:eqm}, 
with wind velocity from equation \ref{eqn:tildeu}, is integrated using a leap-frog method.  In the absence of drag, the integrator is accurate to second order in the time step $d\tau$.
Integrations are terminated when the particle trajectory goes above the escape velocity or
when the particle returns to hit the planetesimal surface.  

Examples of three splash particle trajectories are shown in Figure \ref{fig:trajs} in a three dimensional plot.  In this figure the planetesimal surface is shown with the gray sphere. 
The three particle trajectories have the same initial position on the planetesimal surface and the same ejection velocity vector but different stopping times.  The parameters for the integrations are printed on the plot. 
The splash particles with shorter stopping times are more affected by the wind.  The particle shown with a red solid line has the shortest stopping time and it escaped the planetesimal.  A blue arc shows how far a surface particle initially at the same
position, moved due to rotation, before the $\tilde t_s = 20$ particle (shown with a dashed blue line) landed back on the surface.
A comparison between the landing position of the $\tilde t_s = 20$ particle trajectory and 
the end of the arc (shown with a black dot) illustrates how particles that are splashed off the surface, and do not escape,  are transported across the surface.   

\begin{figure}[!htbp]\centering
\if \ispreprint1
\else
\newgeometry{top=0.5cm, bottom=0.5cm}
\begin{adjustwidth}{-1.0cm}{-1.0cm} 
\fi
\includegraphics[width=3.3truein,trim = 10 0 10 0, clip]{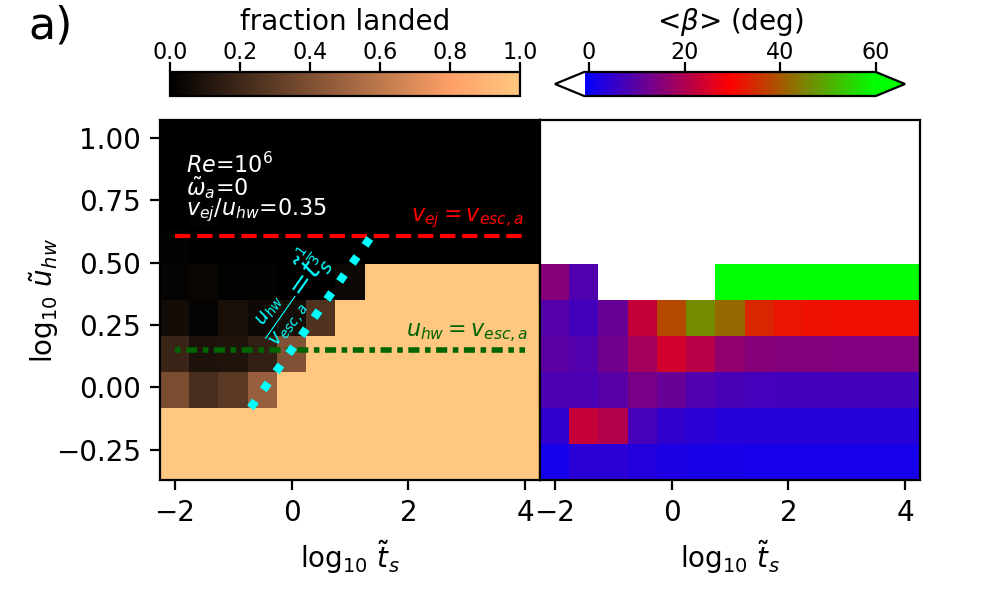}
\includegraphics[width=3.3truein,trim = 10 0 10 0, clip]{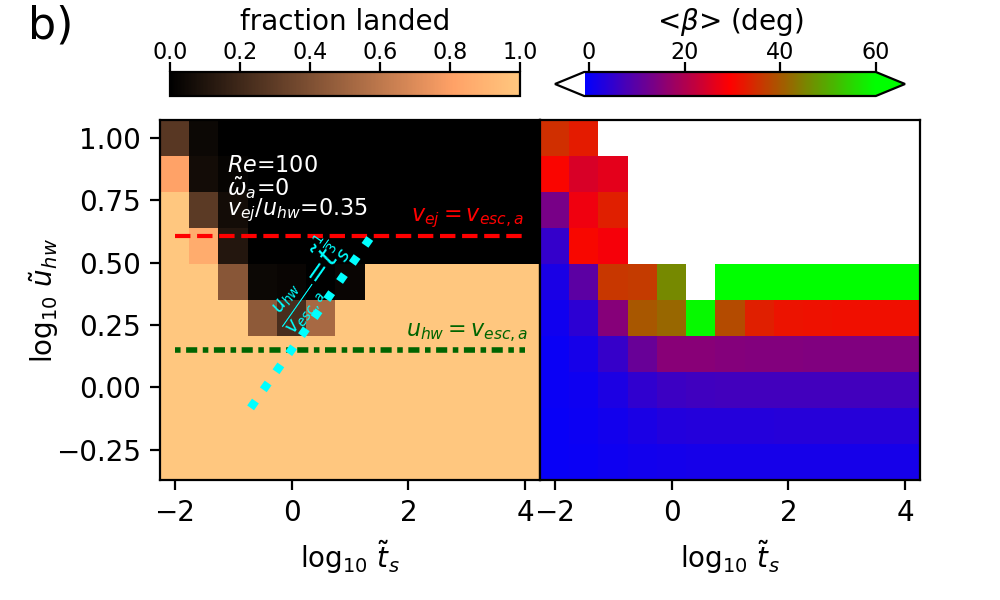}
\includegraphics[width=3.3truein,trim = 10 0 10 0, clip]{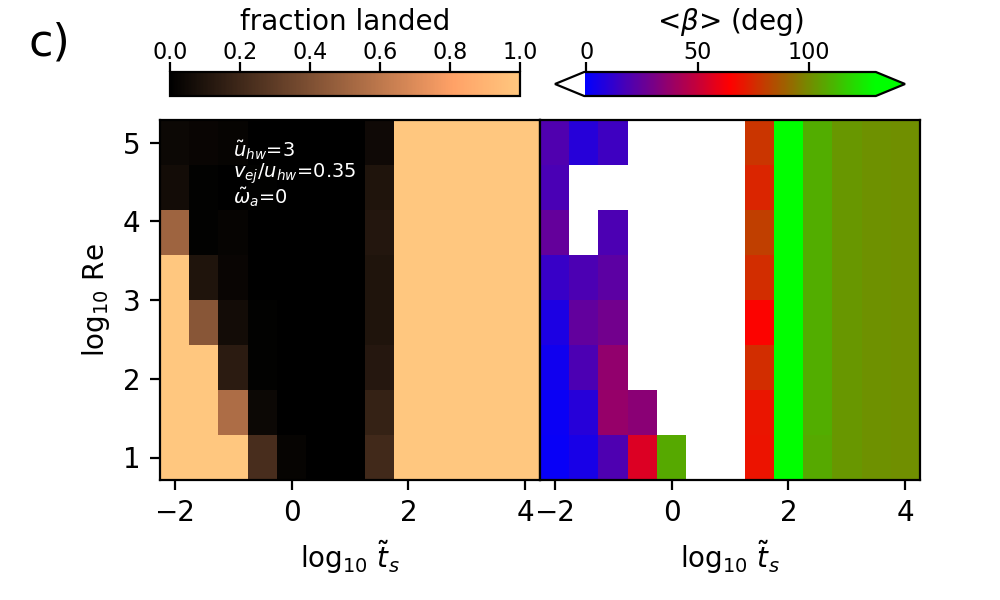}
\caption{Integrations of splashed particles caused by impacts from a headwind.  Ejecta particles feel drag from a headwind. In this set of integrations, the ejecta velocity is fixed, rather than chosen with a distribution function.  a) The left panel shows the fraction of particles that did not escape the planetesimal and the right panel shows the mean value of the angle travelled by the particle with respect to its initial position. 
Initial splash velocity is set to 0.35 $\tilde u_{hw}$ where $\tilde u_{hw}$ is the wind velocity
in gravitational units.   
Each pixel shows the fraction of particles that do not escape computed from 100 integrated particles. The $x$-axis shows the stopping time $\tilde t_s$ and the $y$-axis the wind speed, $\tilde u_{hw}$, both on log scales.  A boundary layer width depends on  
the Reynold's number of the flow, and $Re=10^6$.  The planetesimal is not rotating.  The dashed horizontal red line shows where
ejection velocity is equal to the escape velocity. 
b) Similar to a) except Reynolds number $Re=100$.
c) Similar to b) except we fix the wind velocity $\tilde u_{hw}=3$ and we vary the Reynolds number  on the $y$-axis.
A comparison between left and right panels shows that splash particles are transported larger distances across the surface when they are at a velocity that is just below the escape velocity. 
\label{fig:vesc_2D}}
\if \ispreprint1
\else
\end{adjustwidth}
\restoregeometry
\fi
\end{figure}

\begin{figure*}[!htbp]\centering
\if \ispreprint1
\else
\newgeometry{top=0.5cm, bottom=0.5cm}
\begin{adjustwidth}{-1.0cm}{-1.0cm} 
\fi
\includegraphics[width=3.4truein, trim = 10 0 10 0, clip]{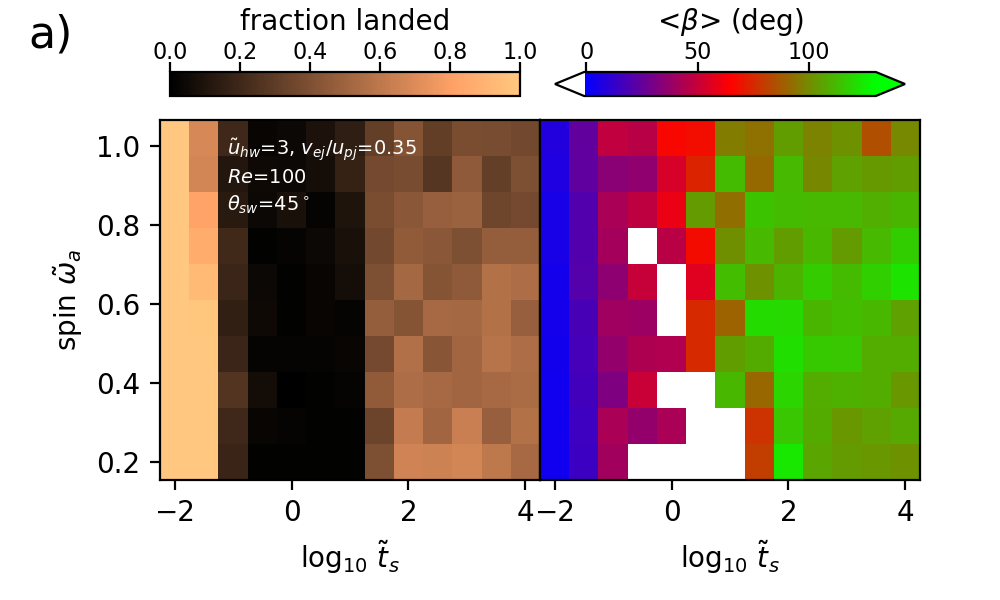}
\includegraphics[width=3.4truein,trim = 10 0 10 0, clip]{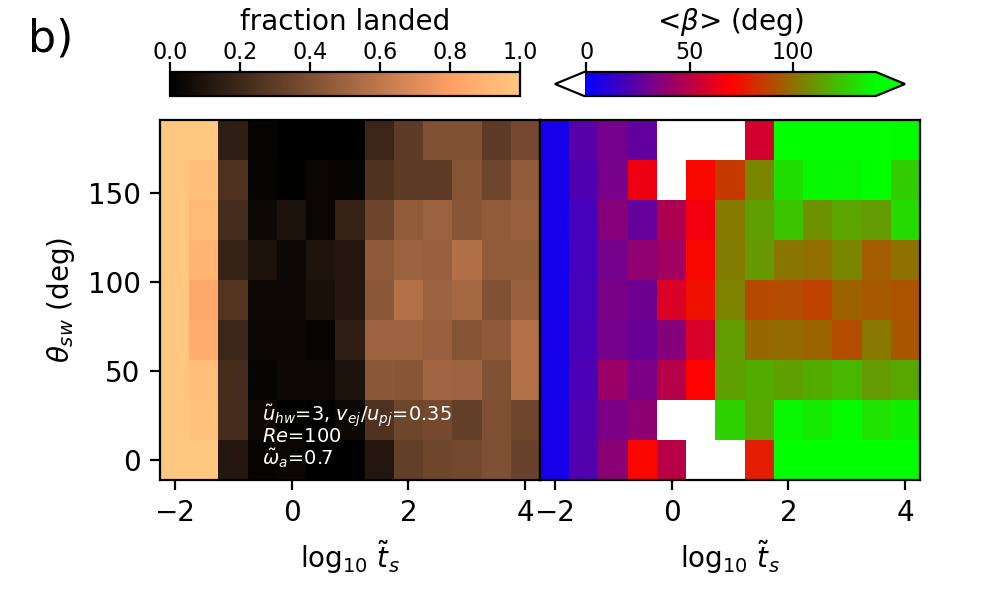}
\includegraphics[width=3.4truein,trim = 10 0 10 0, clip]{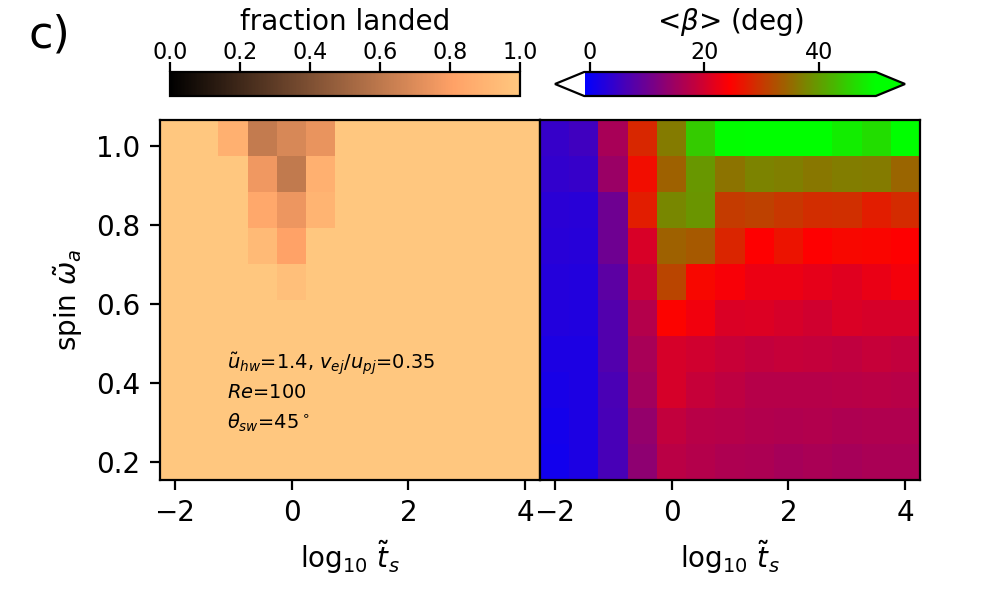}
\includegraphics[width=3.4truein,trim = 10 0 10 0, clip]{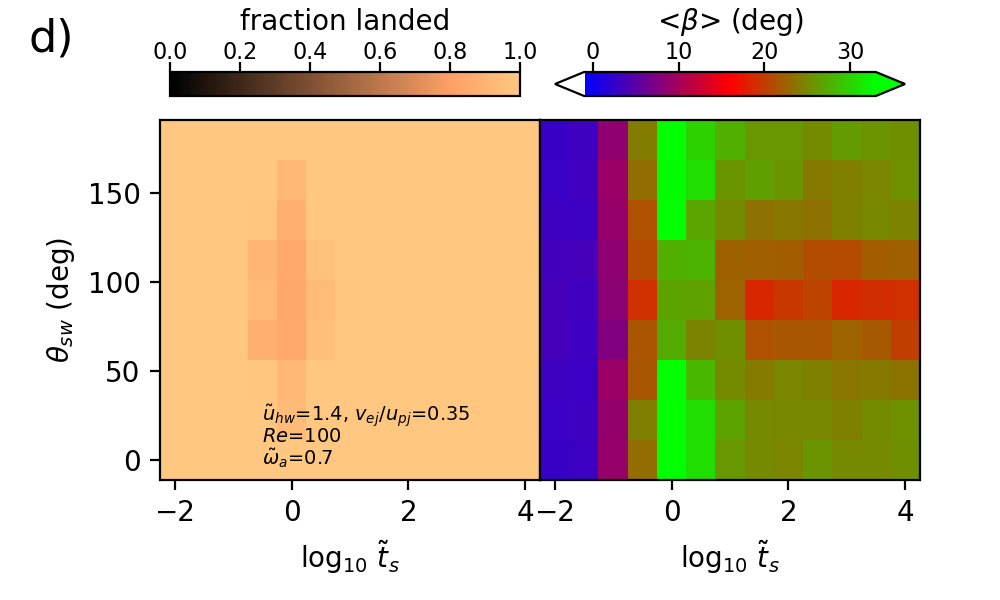}
\caption{Integrations of splashed particles with a specific ejection velocity. These figures are similar to 
those in Figure \ref{fig:vesc_2D} except we vary the planetesimal
spin rate and axis tilt.  In the frame moving rotating with the surface at the site of each impact, the ratio of ejecta velocity magnitude to projectile velocity is 0.35. 
a) The planetesimal spin rate is shown on the $y$-axis.   The axis tilt angle is fixed at $\theta_{sw} = 45^\circ$.
The wind velocity is $\tilde u_{hw} = 3$.
b) Similar to a) except the spin is set to $\tilde \omega_a = 0.7$ and on the $y$ axis we vary the spin axis angle $\theta_{ws}$. 
c) Similar to a) except $\tilde u_{hw}= 1.4$.
d) Similar to b) except $\tilde u_{hw}= 1.4$.
\label{fig:vesc_2D_spin}}
\if \ispreprint1
\else
\end{adjustwidth}
\restoregeometry
\fi
\end{figure*}

\subsection{Integrations of splash particles with a single ejection velocity}
\label{sec:vesc}

We carry out series of splash particle simulations with randomly chosen impact positions on the windward side of a spherical planetesimal,  but with a fixed value of the ratio of ejection velocity to wind speed $\tilde v_{ej}/\tilde u_{hw}$.   
We set the value for this ratio to 0.35, equal to the maximum ejection velocity given in equation \ref{eqn:vej_lims} using scaling coefficients 
from equation \ref{eqn:cofs}.   The ejecta direction is set via equations \ref{eqn:u_pj}, \ref{eqn:dvec}--\ref{eqn:v_init} with spin $\tilde \omega_a = 0$. 
For each set of parameters, 100 particles
are integrated and the fraction of particles that escaped are recorded. For each particle that did not escape and 
with origin at the center of the planetesimal, we compute 
the angle $\beta$ between the landing position and that of a particle that remained on the surface.   We compute the mean value $\langle \beta \rangle$ to show how far splashed particles travel across the surface.  The time step used for the integrations was $d\tau = 0.005$. 

Figure \ref{fig:vesc_2D} shows three series of splash particle integrations. 
On the $x$-axes we vary the dimensionless stopping time $\tilde t_s$.  On the $y$-axes we vary the wind velocity, $\tilde u_{hw}$ (Figures \ref{fig:vesc_2D}a,b), or the Reynolds number (Figures \ref{fig:vesc_2D}c). In figures \ref{fig:vesc_2D}a and b each pixel corresponds to 100 integrations at the same $\tilde u_{hw}$ and $\tilde t_s$. 
In figures \ref{fig:vesc_2D}c each pixel corresponds to 100 integrations at the same $Re$ and $\tilde t_s$. 
The integrations in Figure \ref{fig:vesc_2D}a and b are the same except they have Reynolds number $10^6$ and 100, respectively.  The left panels show the fraction of particles that did not escape and the right panel shows $\langle \beta\rangle$ for the non-escaping particles and for the same set of integrations. 
When the particles all escape, the color of a pixel is white in the right panel and black in the left panel. 
Because the ejection velocity is proportional to the projectile speed, a variation of wind 
speed in Figures \ref{fig:vesc_2D}a and b gives a variation in the ejection velocity.  
On Figures \ref{fig:vesc_2D}a and b, the horizontal red dashed lines show where the ejection velocity is equal to the escape velocity.   Particles tend to escape the planetesimal when the ejection velocity exceeds the escape velocity; $v_{ej} > v_{esc,a}$.

In the transition region on Figures \ref{fig:vesc_2D}a and b, where the wind velocity $u_{hw} > v_{esc}$ (the horizontal dot-dashed green lines)
and  the associated  ejection velocity $v_{ej} <v_{esc,a}$ (horizontal dashed red lines) we estimate
the transition between splash particles that escape and those that return to the surface
with cyan dotted lines. In this transition region,   the cyan dotted lines 
delineate where particles the integrations suggest that particles tend to escape 
\begin{align}
\tilde t_s \lesssim  
\left (\frac{u_{hw}}{v_{esc,a}}\right)^3 .
\label{eqn:tscyan}
\end{align}

Our splash particle integrations suggest that particles tend to escape 
if they have stopping time shorter than a value that scales with the cube of the wind velocity. 
A time $t_B \equiv \frac{GM_a}{u_{hw}^3}$,  was used by  \citet{Lambrechts_2012} to characterize accretion of particles in a wind onto a planetesimal of mass $M_a$. This timescale was called a Bondi time, even though it is not dependent upon the sound speed in the gas. 
The scaling with the cube of the velocity in equation \ref{eqn:tscyan} suggests that the splash particle dynamics is similar in some ways to the dynamics of disk particles  that can accrete onto a planetesimal.  We note that 
the integrations of Figures \ref{fig:vesc_2D}a and b were carried out at a fixed
ratio of $v_{ej}/u_{hw}$ so equation \ref{eqn:tscyan} does not imply that particles are less likely to escape
at higher ejection velocity.  

In Figure \ref{fig:vesc_2D}c, the Reynolds number is varied and wind velocity is fixed at $\tilde u_{hw} = 3$.   As $\log_{10} 3 \approx 0.48$, 
this wind velocity is just below the dashed red line in Figures \ref{fig:vesc_2D}a and b.  A comparison between Figure \ref{fig:vesc_2D}a, b and c shows that a thicker boundary layer, corresponding to lower Reynolds number $Re$, aids in preventing small particles, with short stopping times, from escaping. 
For particles with $\tilde t_s >1 $, 
there is sensitivity to stopping time with larger particles less likely to feel the wind and escape. 

In Figure \ref{fig:vesc_2D_spin} we show a series  of integrations that are similar to those shown in Figure \ref{fig:vesc_2D}, however in these we vary the planetesimal spin rate and axis angle.  In Figure 
\ref{fig:vesc_2D_spin}a and b, the wind velocity is $u_{hw} =3$, high enough that ejecta velocity is near the escape velocity. 
In Figure  \ref{fig:vesc_2D_spin}c and d, $u_{hw} =1.4$, corresponding to a higher mass planetesimal where the ejecta velocity is lower than the escape velocity.
Because the planetesimal is rotating, the projectile velocity $u_{pj}$ in the rotating frame is sensitive to the impact location, the spin angular rotation rate and spin axis orientation.  For these integrations,  the ejecta velocity magnitude in the rotating frame is set to $v_{ej} = 0.35 u_{pj}$.  Ejecta angle for integrations shown in Figure \ref{fig:vesc_2D_spin} are computed using equations  \ref{eqn:u_pj}, \ref{eqn:dvec}--\ref{eqn:v_init}. 

In Figure \ref{fig:vesc_2D_spin}a, c, the spin rate is varied and
in Figure \ref{fig:vesc_2D_spin}b, d, the spin axis angle (with respect to wind direction) is varied.  As was evident in Figure \ref{fig:vesc_2D},  transport across the surface, as seen in larger values of $\langle \beta \rangle$, is enhanced when particles are more likely to escape.  
Figure \ref{fig:vesc_2D_spin}a and c show that planetesimal spin can increase the fraction of particles that escape.  This is expected as the planetesimal rotation adds a component to the ejecta velocities. 
Figure \ref{fig:vesc_2D_spin}c and d show that
the wind can cause significant transport across the surface even if splash particles are unlikely to escape.  Figures \ref{fig:vesc_2D_spin}b and d
show that if the spin axis is parallel to the wind direction, particles are more likely to escape and transport across the surface is enhanced. This is because the ejecta velocity in the rotating frame is always higher than the wind speed due to rotation if 
the spin axis angle $\theta_{sw} = 0$ or $\pi$, whereas if  $\theta_{sw} = \pi/2$, the spin can increase or decrease the projectile velocity in the rotating frame. 

\subsection{Splash particle integrations using an ejecta velocity distribution}
\label{sec:nesc}

In the previous section we discussed splash particle integrations with parameters given in dimensionless gravitational units and with a single ejection velocity or a specific ratio of ejection velocity to impact velocity in the rotating frame.   In this section we use physical units to set the parameters for the integrations and the ejecta velocities are drawn from the velocity distribution described in section \ref{sec:pej}.   
We carry out 3 series of splash particle integrations.  For each series of integrations, we choose a set of fiducial values, listed in Table \ref{tab:fiducial},  and then vary one of them in addition to the dimensionless stopping time for each set of integrations.  For each set of parameters, we integrate 100 splash particles, and as described in section \ref{sec:vesc}, record the fraction of particles that escape and for those that do not escape, the mean change in angle $\langle \beta \rangle$ which characterizes how far splash particles are transported across the surface. 

Physical parameters for the fiducial model are 
chosen to be similar to the physical properties of the Transneptunian object (486958) Arrokoth.  
We set the headwind velocity to $u_{hw} = 50$ m/s (approximately the value of equation \ref{eqn:uhw} from our circumstellar disk model).
The planetesimal diameter for the fiducial model is set to $D_a =20$ km, and its density to 250 kg~m$^{-3}$, similar to the values estimated for Arrokoth \citep{Keane_2022}. 
Planetesimal diameter and density are used to convert physical units for velocity and stopping time into dimensionless gravitational units. 
A low strength value $Y_a = 250$ Pa is chosen based on plausible ranges for the cohesion strength of a granular system  \citep{Sanchez_2014,Brisset_2022}. 
The Reynolds number is chosen to be 100, similar to that estimated for a 20 km diameter planetesimal that is at 45 AU (see Table \ref{tab:tab}). 
The particle size corresponding to $\tilde t_s\sim 1$ at different orbital radii can be estimated from the values listed in Table \ref{tab:tab} for a stopping time of 3000 s. 
The minimum and maximum ejection velocity for the ejecta velocity distribution
are set from equations \ref{eqn:vej_lims}
using a projectile velocity of $u_{pj} = u_{hw}$, dimensionless density ratio parameter $\pi_4 =1$, the strength $Y_a$, and equation \ref{eqn:Rcr2} giving the crater radius. 
Arrokoth's physical properties are discussed in more detail in section \ref{sec:apps} below.

The resulting integrations are shown in Figure \ref{fig:nesc_2D}, which is similar to Figure \ref{fig:vesc_2D}.  The left panels show the fraction of particles that do not escape and the right panels show the mean angular difference between landing position and that of a nearly surface particle that remained on the surface, for particles that did not escape.  In Figure \ref{fig:nesc_2D}a, we carry out sets of integrations with different planetesimal diameter, shown on the $y$-axis.   In Figure \ref{fig:nesc_2D}b, integrations have different head wind speeds $u_{hw}$ and in Figure \ref{fig:nesc_2D}c, integrations have different material strengths $Y_a$. 

In Figures \ref{fig:nesc_2D}a and b we show where the maximum ejection velocity equals the escape velocity with horizontal green dashed lines.  For these integrations the maximum ejection velocity is about 1/3 that of the wind speed. 
Particles can escape at planetesimal diameter below the dashed green line on Figure \ref{fig:nesc_2D}a 
because some ejecta particles are above the escape velocity.
Similarly, particles can escape at winds speeds above the dashed green line on Figure \ref{fig:nesc_2D}b because some ejecta particles are above the escape velocity.

In Figures \ref{fig:nesc_2D}a and c, we show with horizontal cyan dashed lines the planetesimal diameter giving a minimum ejection speed equal to the escape velocity. This is computed using Equation \ref{eqn:vej_lims} for the maximum and minimum ejection velocities.  For planetesimal diameter below the cyan dashed lines on Figure \ref{fig:nesc_2D}a, all ejecta should escape. 
For material strength above this line on Figure 
\ref{fig:nesc_2D}c all ejecta should escape,  except the smallest particles at low stopping $\tilde t_s$ which are trapped in the boundary layer. 

\begin{table}[htbp]\centering
\caption{Fiducial parameters for splash particle integrations of Figure \ref{fig:nesc_2D} \label{tab:fiducial}}
\begin{tabular}{llllllll}
\hline 
Parameter  & Symbol  & value  \\
\hline
Dimensionless  stopping time & $\tilde t_s$ &  [0.01, 3160]   \\
Planetesimal  diameter & $D_a$       &   20  km     \\
Planetesimal density  & $\rho_a$   & 250 kg m$^{-3}$ \\
Gravitational velocity & $v_a$       & 2.6 m/s  \\
Gravitational time-scale  & $\Omega_a^{-1}$ & 3782 s \\
Escape velocity & $v_{a,esc}$       & 3.7 m/s  \\
Planetesimal  strength & $ Y_a$     & 250 Pa  \\
Dimensionless parameter & $\pi_3$    & $4.5 \times 10^{-4}$ \\
Dimensionless parameter & $\pi_4$    & 1\\
Reynolds number & $Re$        & $100$    \\ 
Planetesimal  spin & $\tilde \omega_{a}$ & 0   \\
Wind speed & $u_{hw}$   & 50  m/s  \\
Dimensionless wind velocity & $\tilde u_{hw}$ & 17.8 \\ 
Maximum ejection velocity & $\tilde v_{ej,max}$  & 6.6 \\
Minimum ejection velocity & $\tilde v_{ej,min}$  &  0.62 \\
Time step & $d \tau$ & 0.005 \\
\hline
\end{tabular}
{\\ Notes:   Numbers in brackets for $\tilde t_s$ show the range of values used for series of integrations.  The interval for the stopping time is 0.5 in the log based 10. 
}
\end{table}

For  the integrations  with a single ejection velocity we found that splash particles tended
to escape with stopping time below a line proportional to $u_{hw}^3$, as given in 
equation \ref{eqn:tscyan}.  For the integrations with ejecta velocity drawn from a distribution,
we update the location of this approximate boundary.  The splash particle integrations suggest that 
splash particles are likely to escape if 
\begin{align}
\tilde t_s     \lesssim  \frac{1}{2} \left(\frac{u_{hw}}{v_{esc,a}}\right)^{3} .
\label{eqn:ts_boundary}
 \end{align}
 In Figure \ref{fig:nesc_2D}a and b, this approximate boundary is shown with a blue dotted line. 
This line is not derived or taken from the literature, but chosen to approximately mark where there is a change in behavior in our integrations. 
Particles that are large enough that they lie to the right of this line are less likely to escape. 
The $u_{hw}^3$ scaling is consistent with that of equation \ref{eqn:tscyan} but with a different constant of proportionality.  Here the constant of proportionality is sensitive to the ejecta velocity distribution, whereas equation \ref{eqn:tscyan} delineated the likelihood of particle escape for integrations with a single ejection  velocity.  

In Figure \ref{fig:nesc_2D}c, the strength parameter $Y_a$ is varied,  affecting the minimum ejection velocity. 
The splash particle integrations suggest that  particles with stopping time  
\begin{align}
  t_s \lesssim   \left(\frac{Y_a}{22\ {\rm Pa}} \right)^2  ,
    \label{eqn:Yaboundary}
\end{align}
are more likely to escape.  This transition is shown with a 
a dot-dashed light blue line in Figure \ref{fig:nesc_2D}c.  
This line is not derived or taken from the literature, but chosen to approximately mark where there is a change in behavior in our integrations.
The dependence on $Y_a$ probably arises because 
the minimum ejection velocity $v_{ej,min} \propto Y_a^{1/2}$ in equation \ref{eqn:vej_lims} through its dependence on the dimensionless $\pi_3$ parameter. 

The stopping time limit of equation \ref{eqn:ts_boundary} implies that splash particles could escape, depending upon their size,  even if the total mass of ejecta is low enough that the planetesimal is above the erosional limit estimated with $D_{erode}$ in equation \ref{eqn:D_erode}.   
We compute the diameter of a planetesimal for which the headwind velocity in our disk model is equal to the escape velocity; $u_{hw} = v_{esc,a}$,
\begin{align}
D_{a,vw}  & =   \frac{u_{hw}}{\sqrt{2 \pi G \rho_a/3}} \nonumber \\
& = 102\ {\rm km} \left( \frac{\rho_a}{1000\ {\rm kg~m}^{-3}} \right)^{\!\!-\frac{1}{2}}\! \tilde g_T .
\label{eqn:Davw}
\end{align}
On the surface of a planetesimal in the range $D_{erode} \lesssim D_a \lesssim D_{a,vw}$, small splashed  particles could escape.  On these bodies, if the impacting particle flux is high, there could be an evolution 
of the particle size distribution.
Erosion would take place until the surface is dominated by larger particles, above the limit in equation \ref{eqn:ts_boundary}.
 
\begin{figure}[!htbp]\centering
\if \ispreprint1
\else
\newgeometry{top=0.5cm, bottom=0.5cm}
\begin{adjustwidth}{-1.0cm}{-1.0cm} 
\fi
\includegraphics[width=3.5truein,trim = 10 0 10 0, clip]{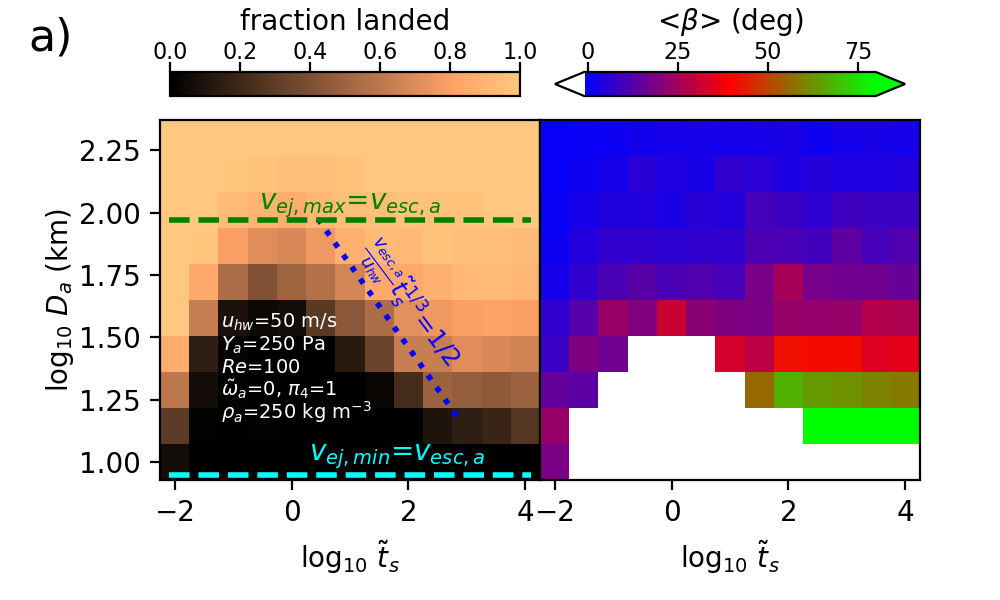}
\includegraphics[width=3.5truein,trim = 10 0 10 0, clip]{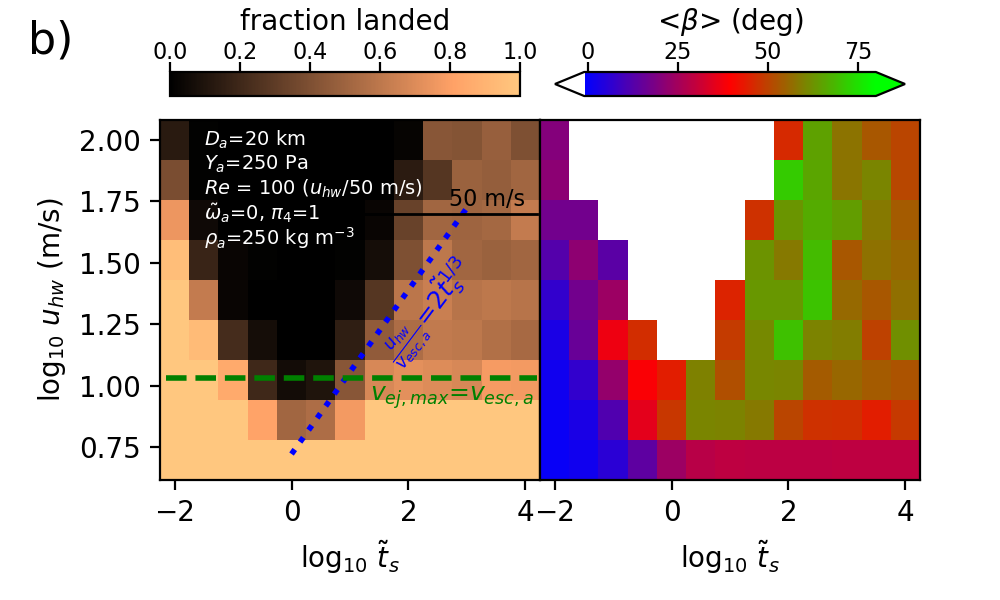}
\includegraphics[width=3.5truein,trim = 10 0 10 0, clip]{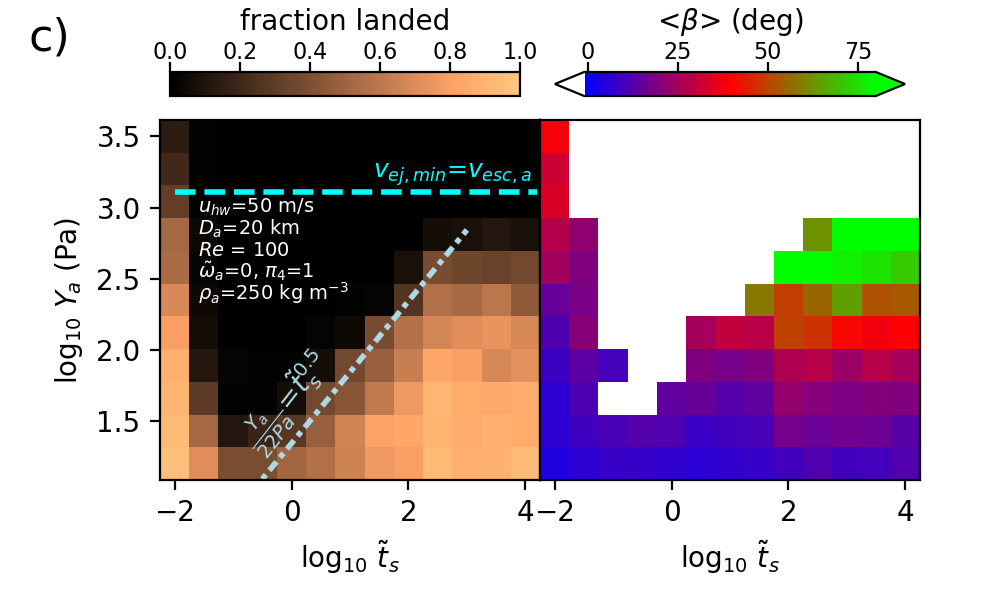}
\caption{Series of integrations based on physical parameters for a Transneptunian object like Arrokoth.  The integrations are similar to those shown in Figure \ref{fig:vesc_2D}, except both impact positions and ejecta velocity are drawn from distributions.  The left panels show the fraction of integrated splash particles that do not escape. 
The right panels show the mean angular distance across the surface travelled by splash particles that do not escape. 
The $x$ axes show dimensionless stopping time $\tilde t_s$.
Each pixel shows a series of 100 integrations with the same parameters. In each subfigure a different parameter is varied from the fiducial parameter set listed in Table \ref{tab:fiducial}. 
a) On the $y$-axis we show planetesimal diameter which is used to set the maximum and minimum velocities of the ejecta velocity distribution.  Green horizontal dashed line show the planetesimal diameter with escape velocity equal to the maximum ejection velocity.  Cyan horizontal dashed line shows planetesimal diameter with escape velocity equal to the minimum ejection velocity.  The blue dotted line is equation \ref{eqn:ts_boundary}.
b) Similar to a) except wind velocity is varied. 
c) Similar to a) except strength $Y_a$ is varied.
The light-blue dot-dashed line is equation \ref{eqn:Yaboundary}.
\label{fig:nesc_2D}}
\if \ispreprint1
\else
\end{adjustwidth}
\restoregeometry
\fi
\end{figure}

\subsection{Splash particle redistribution}
\label{sec:redistribution}

\begin{figure}[!ht]
    \centering
\includegraphics[width=3truein,trim = 0 15 0 0, clip]{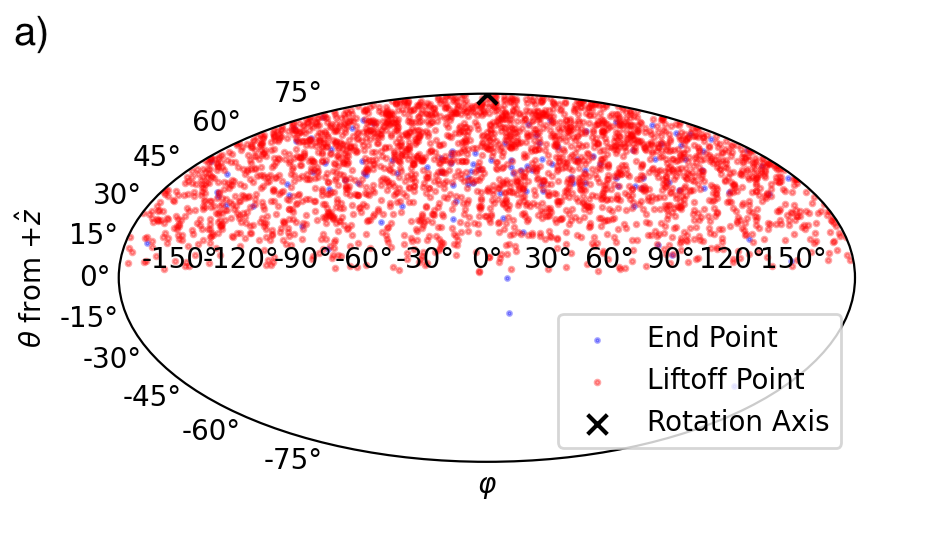}
\includegraphics[width=3truein,trim = 0 15 0 0, clip]{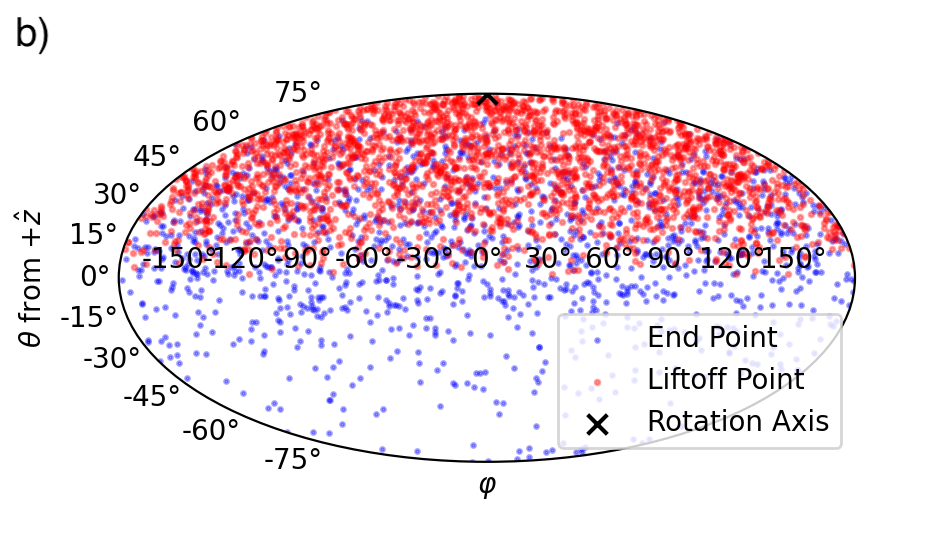}
    \caption{Initial and landing positions of splash particles.    a) Initial particle positions are shown in red and landing positions are shown in blue as a function of longitude and colatitude for particles with stopping time $\tilde t_s = 10$.   Impacts occur on the top hemisphere. Ejecta particles are integrated for a 20~km diameter spherical body assuming no rotation, with fiducial parameters listed in Table \ref{tab:fiducial}. 
    Many splashed particles escaped.
    b) Similar to a) but for stopping time $\tilde t_s = 100$.
    Ejecta particles can land distant from their impact sites.    The fate of splash particles is sensitive to their size, with larger particles less likely to be swept away by the wind.  }
    \label{fig:mollweide_Redistribution}
\end{figure}

\begin{figure}[!ht]
    \centering
\includegraphics[width=3truein,trim = 0 0 50 40, clip]{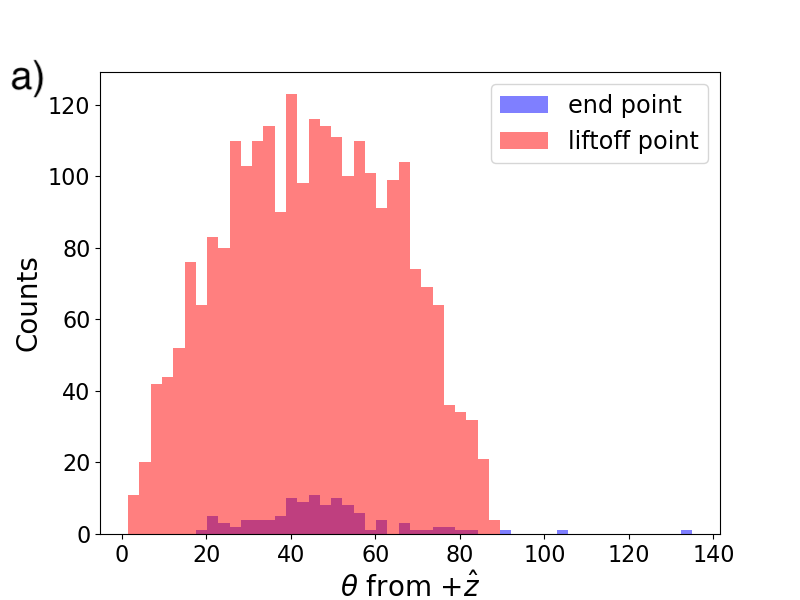}
\includegraphics[width=3truein,trim = 0 0 50 40, clip]{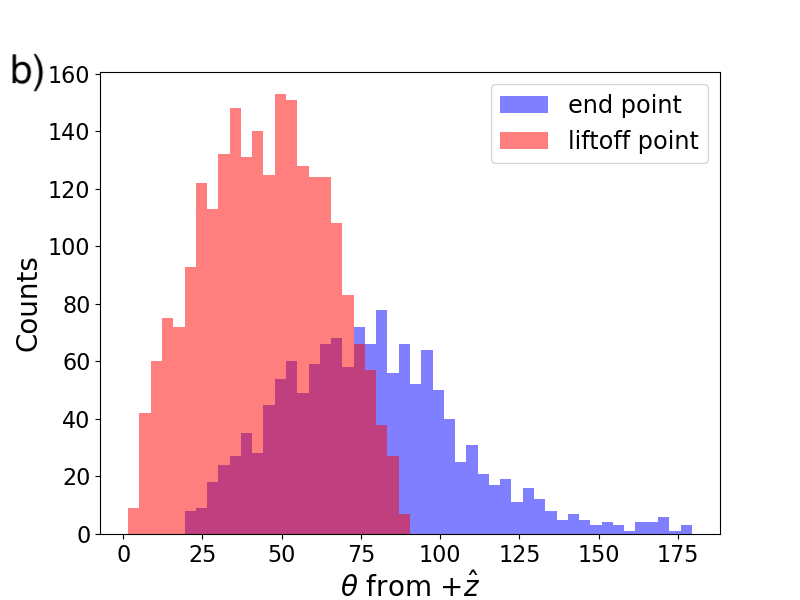}
    \caption{Colatitude distributions of splashed particle initial and landing positions are shown in red and blue, respectively.  
    The integrations shown are the same as in Figure \ref{fig:mollweide_Redistribution}.  a)  Particles with stopping $\tilde t_s = 10$. b) Same as a) but for for particles with stopping time of 100.  
    Particles are transported from the windward hemisphere to the leeward one.  
    }
    \label{fig:dTheta_Redistribution}
\end{figure}

We use the integrations discussed in sections \ref{sec:vesc} and \ref{sec:nesc} to look at the difference between initial and final positions of splashed particles.  
In Figure \ref{fig:mollweide_Redistribution} we show  
the initial and final positions of 2500 integrated splashed particles with stopping times of $\tilde t_s = 10$ and 100.  
The red points show initial positions, which are restricted to the windward side.  Blue points show landing positions for particles that did not escape. The larger particles with longer stopping times shown in Figure \ref{fig:mollweide_Redistribution}b land all over the body, including on most regions of the leeward side. 
In Figure \ref{fig:dTheta_Redistribution} we show colatitude histograms for the same integrations. 

Figure \ref{fig:dTheta_Redistribution} shows the same information as Figure \ref{fig:mollweide_Redistribution}, however it better illustrates the net motion of approximately 40$^\circ$ in the polar coordinate  $\theta$.  This view is complimentary to the mean angular difference $\langle \beta \rangle $  
shown in our previous Figures \ref{fig:vesc_2D} and \ref{fig:nesc_2D}.

\subsection{Trends with planetesimal size}

\begin{figure}[!ht]\centering
\if \ispreprint1
\includegraphics[width=3.5truein, trim = 0 0 0 0,clip]{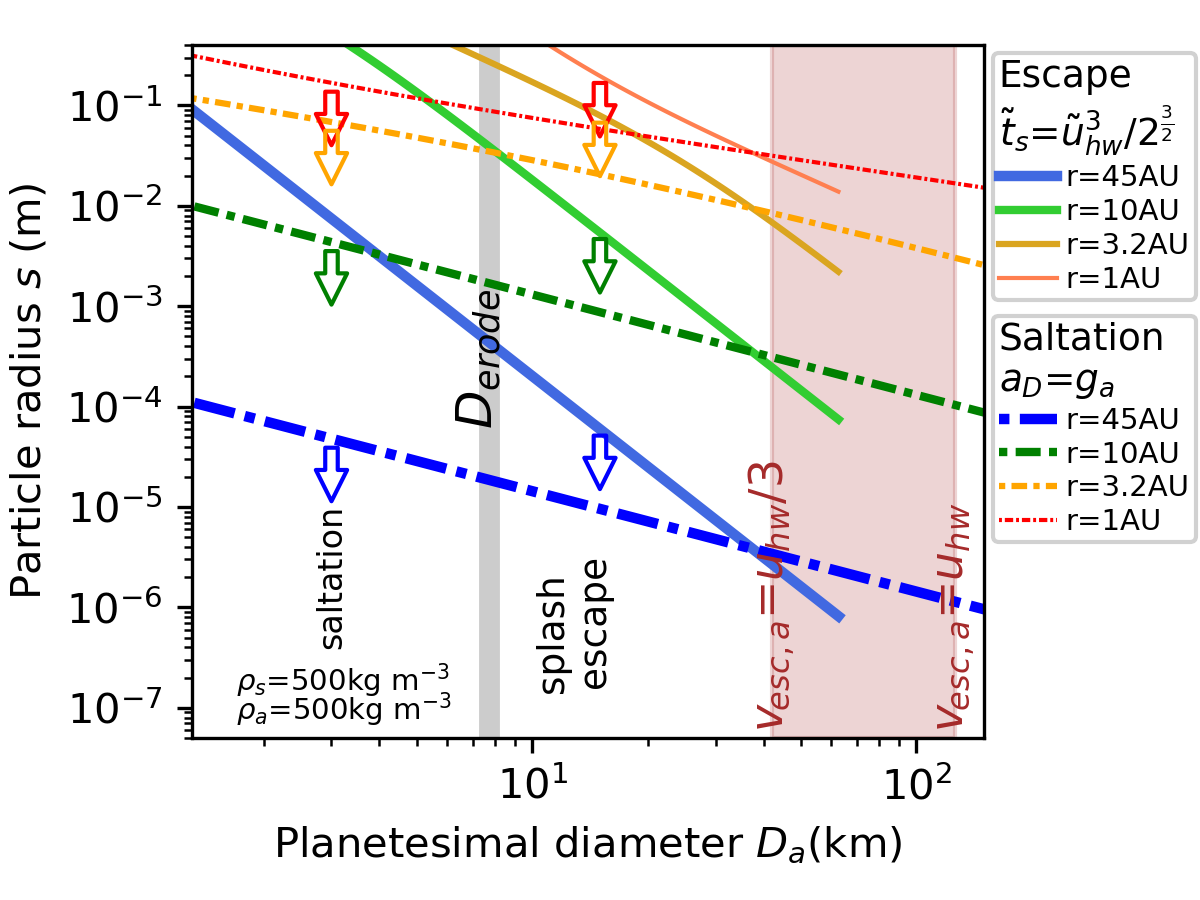}
\else
\includegraphics[width=4.5truein, trim = 0 0 0 0,clip]{Da_plot2.png}
\fi
\caption{Saltation and splash particle regimes as a function of planetesimal diameter.   
The $y$ axis is  particle radius in m and the $x$ axis is planetesimal diameter in km. 
Splash particles are assumed to be ejecta from impacts caused by headwind particles at 
the headwind velocity of disk model;  38 m/s.
The solid lines shows the maximum splash particle that is likely to be carried away by the wind
and is estimated via equation \ref{eqn:ts_boundary} at orbital radii $r=$ 1, 3.2, 10, and 45 AU, 
in red, orange, green and blue, respectively, and in order of increasing line thickness.
The dot-dashed lines show limits for saltation at the same orbital radii. 
A splash particle below one of the solid lines is likely to escape the planetesimal.  
Below the saltation lines, surface particles can be pushed by the wind. 
The thick vertical gray line shows the erosion limit $D_{erode}$ (equation \ref{eqn:D_erode}).  To the left of this line, impacts from headwind particles cause erosion.  
The wide tan rectangle shows the range of planetesimal
sizes where $u_{hw}/3 < v_{esc,a} < u_{hw} $.   Inside and to the right of this rectangle, few ejecta particles would escape.  
Between the vertical grey bar and the red rectangle, splash particles tend to be  
redistributed across the surface.  Particles that are below the solid lines are more likely to escape, 
so there could be a slow increase in the mean surface particle size. 
Particles that can saltate can also escape if they are splashed off the surface by an impact. 
If the wind contains small particles (see Figure \ref{fig:Da_plot}) then the surface can 
 harbor small particles, that are blown across the surface and then could eventually escape. 
Within the red rectangle, splash particles do not tend to escape.  In this regime, 
particles on the surface can be blown by the wind as long as they are below 
the dot-dashed lines.  
Quantities are computed with splash and surface particle density $\rho_s = 500 $ kg m$^{-3}$
and bulk planetesimal density  $\rho_a = 500 $ kg m$^{-3}$. 
 \label{fig:splash_Da}
}
\end{figure}

Splash particles are likely to escape if they are smaller than the limit given by the inequality of equation \ref{eqn:ts_boundary}, suggested by our splash particle integrations.
In Figure \ref{fig:splash_Da}, we compare this limit to the maximum particle allowing saltation
by the headwind.  
Figure \ref{fig:splash_Da} shows regimes for saltation, erosion and transport of splash particles from impacts from particles in a wind with head velocity about 40 km/s and the protostellar disk model of section \ref{sec:disk}.  
The $x$-axis in Figure \ref{fig:splash_Da} is planetesimal diameter, so this figure is complimentary 
to Figure \ref{fig:codot} and \ref{fig:sc} where the horizontal axis is sensitive to orbital radius.
Figure \ref{fig:splash_Da} is similar to Figure \ref{fig:Da_plot} which shows the minimum
 size of impactors in the wind 
  (those with stopping time equal to the planetesimal crossing time). 
In Figure \ref{fig:splash_Da}, we compute quantities for a particle density and bulk planetesimal density of 
$\rho_s = \rho_a = 500 $ kg m$^{-3}$. 

In Figure \ref{fig:splash_Da}, dot-dashed and solid 
lines are shown at 4 different orbital radii, $r=$ 1, 3.2, 10, and 45 AU, 
in red, orange, green and blue, respectively,  and with line thickness increasing with $r$.    
The dot-dashed lines show limits for saltation.  
Saltation limits are computed by solving for the particle size at which acceleration by drag equals surface acceleration,  $a_D = g_a$,  using equation \ref{eqn:a_D} for $a_D$ and the protostellar disk model of section \ref{sec:disk}.
The solid lines shows the maximum splash particle that is likely to be carried away by the wind
and these are estimated via equation \ref{eqn:ts_boundary}. 


In Figure \ref{fig:splash_Da}, with a thick vertical gray line
we plot the erosion limiting size of equation \ref{eqn:D_erode}). 
For planetesimal diameter $D_a > D_{a,vw}$ (with escape velocity equal to the headwind velocity; equation \ref{eqn:Davw}), splash particles of any size are unlikely to escape the planetesimal.  The right side of the large tan rectangle on Figure \ref{fig:splash_Da} shows this limit.  The left side of the same rectangle corresponds to a diameter giving escape velocity equal to $u_{hw}/3$ which is the maximum ejection velocity estimated from ejecta scaling models (from equation \ref{eqn:vej_lims}).   Thus splash particles 
are unlikely to escape for planetesimals within the tan rectangle. 

On the left side of Figure \ref{fig:splash_Da}, impacts from particles in the wind cause erosion on a small planetesimal, though if the surface is covered by particles with sizes lying above the solid lines, the erosion rate can be reduced. 

In between the vertical gray line and tan rectangle on Figure \ref{fig:splash_Da}, impacts from particles in the wind cause some splash particles, those with higher ejecta velocities,  to escape the planetesimal.  
Particles below the solid lines would be most likely to escape, leading to evolution of the size distribution 
of particles on the surface.   The saltation lines are below the escape lines, so splash particles that are large 
enough that they would not escape, would not saltate.  As the wind can contain particles
below the saltation line (see figure \ref{fig:Da_plot}) small particles could be replenished via accretion  
and these could be transported by the wind. 

Within and to the right of the tan rectangle on Figure \ref{fig:splash_Da}, splashed particles are unlikely to escape.    We have not taken into account gravitational focusing when 
computing the impact velocity, so the $x$ axis on Figure \ref{fig:splash_Da} is truncated where the escape velocity is equal to the headwind velocity. 
Above this planetesimal diameter, wind particles would impact the planetesimal near the escape velocity, 
and splash particles would still be unlikely to escape because the maximum ejecta velocity should
be lower than the escape velocity.  Within the tan rectangle, particles that are splashed
of the surface could be blown across the surface by the wind.  With impacts overcoming cohesion, 
small splashed particles (those below a dot-dashed line) can saltate.


As mentioned in section \ref{sec:gusts}, the planetesimal could experience a higher or lower wind velocity if there are perturbations in the disk, and 
a higher headwind velocity would facilitate erosion, transport and saltation.  

\subsection{Caveats/Discussion}

We discuss physical affects that could be incorporated into an improved model for splash particle dynamics.   

We have neglected the dynamics of the wind particles as they approach the planetesimal.  
For larger bodies with escape velocity above the wind velocity, $ v_{esc,a}>u_{hw}$,   the impact velocity should be larger than $u_{hw}$ (as assumed in our simulations) due to gravitational focusing (e.g, \citealt{Ormel_2010,Lambrechts_2012,Visser_2016,Ormel_2017}).  
Future studies could more accurately model the trajectories of the impacting wind particles.  
Larger planetesimals could harbor an atmosphere which could be disturbed by the wind
and could increase drag near the planetesimal surface.  Integrations of both impacting and splash particle trajectories could take into account a planetesimal atmosphere. 
  
We used a crater ejecta scaling approach to estimate mass and velocity of splashed particles. However studies relevant for aeolian processes on Earth \citep{Beladjine_2007,Kok_2012}
adopt similar but different descriptions for the ejecta mass and velocity distributions.   Experimental studies, particularly for oblique impacts, might better constrain the parameters needed for crater ejecta scaling in this setting. 

Impacts with larger (greater than a meter) objects are not discussed but would also lead to mixing and surface modification.  The orbital eccentricities of planetesimal and wind particles would affect the 
impact speeds \citep{Huang_2023}. 
The role of the size distribution of impacting particles and that for particles on the surface particles could be explored.  
We assumed a bulk strength for the substrate, however cohesion between individual surface particles and between impactor and surface particles may be relevant for predicting splash particle dynamics. 

In our splash particle integrations, initial conditions for ejecta direction in the frame rotating with the surface were in the downrange direction (with respect to the impacting headwind projectile velocity), at $45^\circ$ from the surface normal,  and  independent of impact angle. 
Future studies could adopt more realistic ejecta angle, velocity and mass distributions.   

We assumed a spherical planetesimal. Future studies could consider non-spherical bodies and explore the role of surface topography in affecting ejecta distributions and splash particle dynamics. 

As the planetesimal orbits the Sun, the direction of the wind (in an inertial frame) would rotate.   Also the planetesimal spin axis can precess.  
Some latitudes or regions might be more likely to accumulate particles. Future studies could take into account planetesimal rotation and orbital motion and variations in headwind direction to better predict which regions would be most likely to accumulate splashed material. 

Models for flow about planetesimal could be improved to better model the boundary layer near the surface,  lift,  rarefied gas dynamics (for flows in the outer solar system), eddies and turbulence that could be present in the flows. 
On Earth and other planets the wind velocity as a function of height is often described with a logarithmic profile which is a semi-empirical relationship describing the vertical distribution of horizontal mean wind speeds (e.g., \citealt{Kok_2012}). Perhaps analogous models could be constructed for the boundary layers for flows about planetesimals. 


For our splash particle trajectory integrations we only included a drag force from the wind, 
however,  lift near the surface is also likely to be present, (see discussion in the supplements 
by \citet{Gunn_2022} and \citealt{Loth_2008b,Luo_2016}). 
If the splash particles
are spinning, then there could also be a tangential force on 
the particle.  This force, akin to the Magnus effect,  can be present even in a rarefied gas dynamics regime at Knudsen number greater than 1 \citep{Taguchi_2022}.

Disk structure evolves (e.g., \citealt{Lenz_2020,Estrada_2022}).  
Disks can have gaps.  The disk has vertical structure with varying distribution of particles as a function of height above the disk and this too evolves prior to 
the evaporation of the disk.  Future time dependent models could be developed to explore the
the role of disk evolution on planetesimal surface modification processes.

Secondary ejecta have been neglected in our model, however,  secondary ejecta, could be above the cohesion limit, particularly if the mass of a splash particle exceeds that of the surface particle that it hits.   Ricochets are likely at low velocity \citep{Wright_2020b,Suo_2023}. Ejecta particles would move further across the surface if they can bounce off the surface.   


\begin{figure}[!htbp]
    \centering
    \if \ispreprint1 
    \includegraphics[width=3.5truein,trim = 0 0 0 10 ,clip]
    {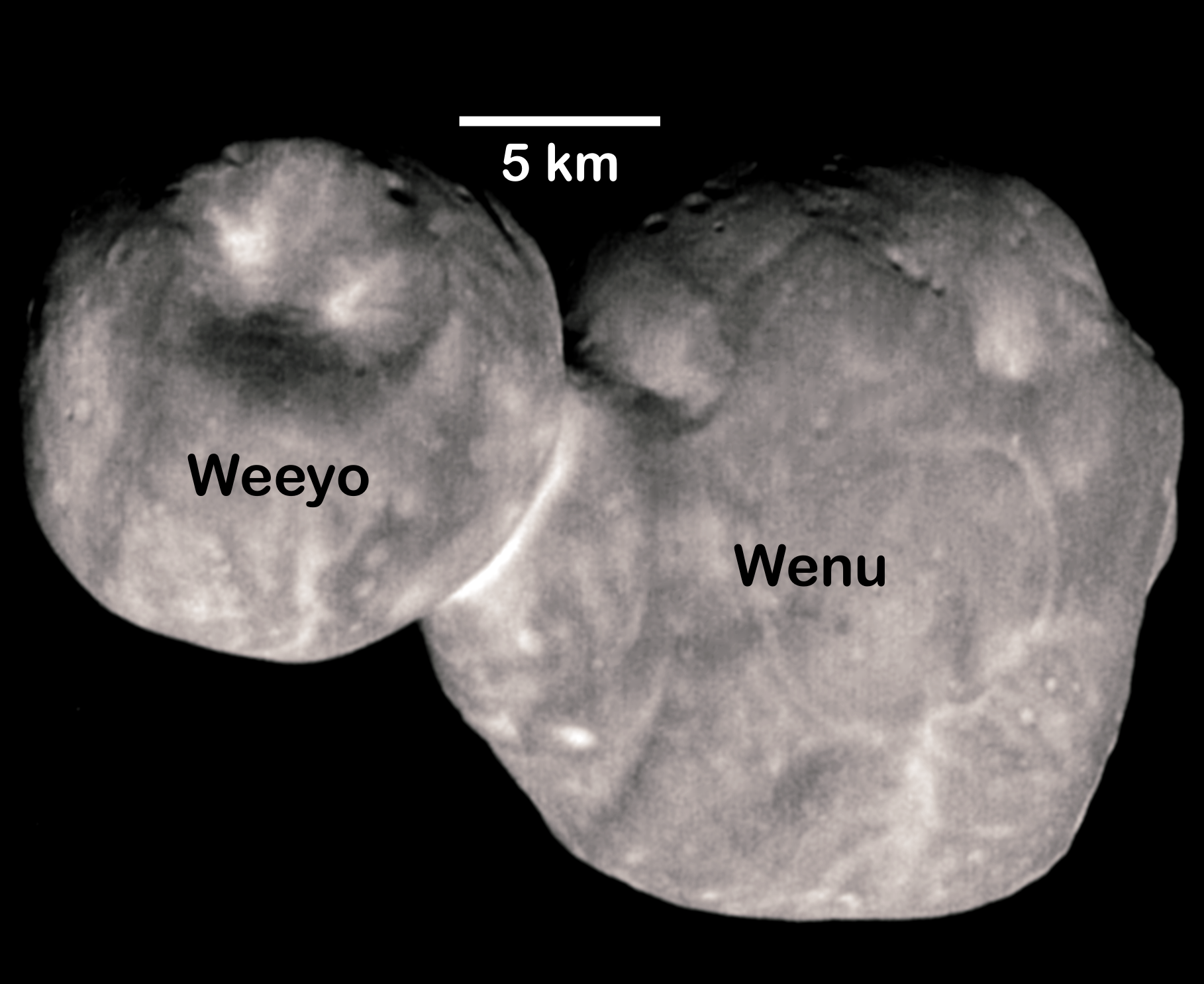}
    \else
    \includegraphics[width=4.5truein,trim = 0 0 0 10 ,clip]
    {ca06_v5_l10b.png}
    \fi
    \caption{The CA06 LORRI Mosaic of Arrokoth.  The undulating, smooth terrain on Wenu is interpreted in terms of mounds that coallesced at low speed to form Arrkoth \citep{Stern_2023}.  The smooth terrain of Arrokoth is quite unlike those of imaged Jupiter family comets \citep{Spencer_2020}.  The large crater on Weeyo is called the Sky Crater. }
    \label{fig:Arr}
\end{figure}

\section{Winds on Arrokoth}  
\label{sec:apps}

(486958) Arrokoth (formerly 2014 MU$_{69}$) is a bilobate cold classical Transneptunian object, which most likely formed in the outer Solar system, near its current semi-major axis of 44.6 AU.  
Due to its quiescence and isolation, Arrokoth is described as primordial and pristine compared
to other bodies in the Solar system \citep{McKinnon_2020}.
Arrokoth's gravitational surface slope distribution suggests that Arrokoth is a remarkably low-density body, 155--600 kg m$^{-3}$ \citep{Keane_2022}, implying substantial porosity 
$\gtrsim 75\%$,  assuming icy composition \citep{Grundy_2020,Keane_2022}.
Arrokoth was likely created by the merger of two progenitor transneptunian objects,
at low relative velocity \citep{Marohnic_2021, Mao_2021,Lyra_2021}.

Images from the New Horizons New Horizons' LOng Range Reconnaissance Imager (LORRI) have been combined into a mosaic called CA06 (see  \citealt{Stern_2023}) that is comprised of a stack of 9 sequentially observed registered images and is shown in Figure \ref{fig:Arr}. Amongst Arrokoth's most striking features are the smooth and undulating terrain present on its larger lobe (or head),  also called Wenu  \citep{Stern_2019,Spencer_2020,McKinnon_2020,McKinnon_2022,Stern_2023}.  Arrokoth's surface lacks cliffs, chasms, and perched boulders, and it is remarkably smooth in comparison to Jupiter family comets (see Figure S3 by \citet{Spencer_2020} contrasting these objects). Inspired by Arrokoth's unusual morphology, we explore the possibility that interactions with the protosolar disk could have smoothed its surface. 

Detection of methanol ice (CH$_3$OH) \citep{Grundy_2020} suggests that formation of Arrokoth took place in an optically thick disk, shielded from the luminous young stellar object that was then the Sun \citep{Lisse_2021}.  This suggests that Arrokoth's building blocks spent time within a gaseous disk after it formed. 
After the evaporation of the protosolar nebula, sublimation processes could have altered Arrokoth's surface  \citep{Lisse_2021,Steckloff_2021,Zhao_2021}.
Although sublimation was likely too weak to significantly alter Arrokoth's spin state,
between 10 to 100 Myr after formation, sublimation could transport sufficient mass to erode topographic features on length scales of 10 -- 100 m \citep{Steckloff_2021} or flatten its two lobes \citep{Zhao_2021}.
Thereafter, Arrokoth would be quiescent, with volatile production rates insufficient to drive either significant shape changes, mass transport or torque \citep{Steckloff_2021}.   

Impact generated seismic shaking can cause crater erasure and 
downslope movements of a loose surface layer \citep{Richardson_2004,Thomas_2005}. However, seismic attenuation in granular systems is rapid,  suggesting that erosion is localized, even following a large impact \citep{Quillen_2022,Sanchez_2022,Suo_2023}.   

Wenu's morphology suggests that Wenu was formed from smaller spheroidal building blocks \citep{Spencer_2020,Stern_2023}.
Weeyo (the smaller lobe) does not display distinct units of rolling topography near the Sky crater, possibly as a result of resurfacing caused by the impact event that created this crater \citep{Stern_2019}.
\cite{Spencer_2020} suggested that
sublimation and associated erosion taking place due to sublimation over the past 4.5 billion years (Ga)
would not explain the smooth terrain seen at the $\gtrsim $ 30 meter scale of the New Horizons imaging resolution at closest approach on the larger lobe.  

Would sublimation on Arrokoth have roughened or smoothed its surface?
Sublimation of snow on Earth can roughen a surface if darker dust particles heat or shade underlying snow \citep{Betterton_2001}.
Outgassing can cause comet surface material to fragment and crumble, leading to a surface covered in boulders and pebbles \citep{Vincent_2017}.  Outgassing can also generate deposits of fine grained material that can later develop ripples \citep{ElMaarry_2017}.
The erosion rate on comets (due to outgassing) is sensitive to porosity, dust to ice mass ratio and composition \citep{Benseguane_2022}, so heterogeneity in the surface material properties and composition could cause variations in topography.
However, based on the relative erosion between plateaus and valley bottoms, sublimation is predicted to erode, rather than amplify, pits at the surface of comet 67P/Churyumov-Gerasimenko  \citep{Benseguane_2022}.
Comparison of images of Jupiter Family Comet nuclei to Arrokoth (see Figure 8 and S3 by \citealt{Spencer_2020}) shows that Arrokoth's undulating, smooth and lobate substructures are unlike the geography of comets as Arrokoth lacks boulder-strewn surfaces and cliffs.  
During its most active period, the erosion rate on Arrokoth is estimated to be an order of magnitude lower than that on comet 67P/Churyumov-Gerasimenko \citep{Steckloff_2021}.
\citet{Steckloff_2021} proposed that Arrokoth's higher escape speed and weaker sublimation pressure would not allow grains in a sublimation flow to escape. 
So while \citet{Spencer_2020} suggested that Arrokoth's smooth regions might not be explained via erosion associated with sublimation, \citet{Steckloff_2021} suggested the opposite.  

Color and shape variations of Wenu's surface have been interpreted in terms of mounds that coalesced at low relative speeds ($< 0.5$ m/s) to form Wenu \citep{Stern_2023}. 
Taking the surface areas for the 9 mounds identified on Wenu and listed in Table 1 by \citet{Stern_2023}, but excluding those that are viewed obliquely, we compute the diameter of circles with equivalent surface area.  The mean value of these area-equivalent diameters is about 4.6 km. Soft-sphere simulations suggest that Wenu could have been formed from a gentle collapse of a rotating cloud of a few dozen spherical 5 km diameter progenitors \citep{Stern_2023}.  Subunits may have been soft enough that they deformed upon merging \citep{Spencer_2020, Jutzi_2015}.   It is possible that the subunits were originally smooth 
due to accretion.  For example, smooth ridges on some of Saturn's Moon's have been 
interpreted in terms of accretion of streams of fine particulates \citep{Quillen_2021}. 
Alternatively Arrokoth's subunits could have eroded prior to merging and been smoothed during erosion. 

Figure \ref{fig:codot} shows that in the green circle and at an orbital radius of 45 AU, a disk wind would not overcome cohesion on the surface.  However, as is true on Earth, impacts from particles entrained in a wind can splash small particles off the surface that would not directly be lofted by the wind
\citep{Bagnold_1941,Kok_2012}.  

To estimate the velocity distribution of ejecta from impacts, we need an estimate for the substrate material strength.
The bounce of the Philae lander on comet 67P/ Churyumov-Gerasimenko implies that its primitive ice has a very low compressive strength, less than 12 Pa \citep{ORourke_2020}.  However this value is likely too low to be a bulk value on this comet, based on its steep cliffs,  \citep{Groussin_2015}.   
Compressive stress on the core of Arrokoth and tensile stress on its neck imply that its strength must exceed a few tens of Pa \citep{McKinnon_2022}.
Based on comparison to porous solids \citep{Housen_2018}, compressive strength values of 25-100 kPa are plausible for Arrokoth \citep{McKinnon_2022}. 
Based on their spin values, rubble asteroids have cohesion strength at most a few Pa \citep{Sanchez_2014}.  Laboratory studies of regolith give a strength of about 500 Pa \citep{Brisset_2022}.
The crush strength of snow is a few kPa \citep{Huang_2013}.
If Arrokoth is granular rather than a porous solid, its strength could be lower than a few kPa. 
In summary, there are lower limits on the compressive, tensile and cohesive strength of Arrokoth \citep{McKinnon_2022},  however, its bulk strength, particularly if it is a heterogeneous granular system, is difficult to exactly pin down, though the range of 25 to a few kPa is plausible \citep{McKinnon_2022}.

Figure \ref{fig:sc} shows that at 45 AU particles within a protostellar disk that have radius between about 0.1~m and 1~$\mu$m could impact a planetesimal with the size of Arrokoth.  The range of particle sizes that can impact a planetesimal in the outer solar system is large, suggesting that the parameter  describing the mass fraction of disk particles that would impact the surface $\xi_p \sim 1$.  
However, Arrokoth itself is near the maximum diameter where impacts from particles in a wind could be erosional rather than accreting (as discussed in \ref{sec:erode}).
\citet{Stern_2023} suggest that Wenu's smooth terrain is due to
the smooth surfaces of smaller building blocks. Thus we should also consider the possibility that wind driven processes smoothed these smaller 5 km diameter mounds, in addition to smoothing Arrokoth itself.  As Arrokoth is near the erosion limit for impacts at a headwind velocity of about 50 m/s, its building blocks would be below this limit. 


\subsection{Conditions for wind related particle transport on Arrokoth's building blocks}

What conditions would allow wind-driven impacts to redistribute material on Arrokoth's building blocks? To answer this question, we assume that the building blocks are the mounds identified by \citet{Stern_2023} which have a diameter of about $D_{mound} =$ 5 km. We adopt a density of $\rho_a = 250$ kg m$^{-3}$ based on estimates of Arrokoth's density.  With this mean density and diameter, the escape velocity from a mound is $v_{esc,mound} \approx 1 $ m/s.  This is similar to the minimum escape speed from Arrokoth (shown in Figure 11 by \citealt{Keane_2022}) and about 3 times lower than the guaranteed escape speed (shown in 
Figure 12 by \citealt{Keane_2022}). 

Rather than adopt a particular headwind speed based on a disk model, we estimate the range of headwind speeds that would be consistent with sufficient particle transport to smooth the surface of Arrokoth's mounds but not cause extreme levels of erosion.  We then estimate the required particle density in the disk.

Figures \ref{fig:vesc_2D}a, b, \ref{fig:nesc_2D}a and b illustrate that particles are not transported very far across the surface when the wind speed is too low compared to the escape velocity on the planetesimal $v_{esc,a}$. 
The wind speed sets the speed of particles hitting the planetesimal and so the maximum ejection speed for splashed particles. For splash particle transport we estimate  
\begin{align} 
u_{hw} \gtrsim 3 v_{esc,a}. \label{eqn:lim1}
\end{align} 
The factor of 3 arises from equation \ref{eqn:vej_lims} giving the maximum ejection velocity $v_{ej,max}$. The limit is consistent with the green dashed horizontal lines shown in Figure \ref{fig:nesc_2D}a and b.

We estimate an upper limit for the wind speed $u_{hw}$ from the erosion limit estimated in equations \ref{eqn:vesc_b} and \ref{eqn:vesc} giving 
\begin{align}
    u_{hw} \lesssim 16~ v_{esc,a}. \label{eqn:lim2}
\end{align} 
The factor of 16 arises from inverting equation \ref{eqn:vesc_b} for the critical escape velocity.
For wind speeds above this velocity, we expect that escaping ejecta mass exceeds that accreting onto the surface.  

We use $u_{hw,mound}$ to refer to a typical value for the headwind velocity for a particle rich wind that would have smoothed the surface of Arrokoth's building blocks due to redistribution of splashed particles. 
For an escape velocity of 1 m/s for Arrokoth's 5 km diameter building blocks, the limits of equations \ref{eqn:lim1}, \ref{eqn:lim2} give a range 
\begin{align}
    3~ {\rm m/s} \lesssim u_{hw,mound} \lesssim 16~ {\rm m/s}. \label{eqn:umound}
\end{align}
With the same assumed density of 250 kg m$^{-3}$, a spherical planetesimal with a 20 km diameter, similar to Arrokoth itself, has an escape velocity about 4 times higher and this would give a range of headwind velocities $12 \lesssim u_{hw} \lesssim 64 $ m/s.   


A disk with a lower temperature or different density profile (with orbital radius) could have a lower headwind velocity.  However, the mounds that later coalesced into Arrokoth could have been born within clumps during streaming instability \citep{Nesvorny_2021} or within a turbulent concentration of particles \citep{Hartlep_2020}.  The simulations by \citet{Nesvorny_2021}
exhibited a variety of phenomena for clumps embedded within particle rich filaments that are present during 
streaming instability, including disruption of spinning clumps and ejection of mass after formation of a binary.  
 If the clumps or concentrations have a high mass density of particles compared to gas, then the gas is dragged by the particles and the velocity of the gas with respect to forming planetesimals could be lower than the mean headwind velocity of the unperturbed disk \citep{Johansen_2007}.  \cite{Johansen_2007} reported that ``The particles that make up a clump continuously leak out downstream to the radial drift flow and are replaced with new particles drifting in from upstream.''  They also saw variations of order unity in the radial drift rates of clumps within their simulations.   
%
If Arrokoth's clump's experienced a range of headwind velocities while streaming instability took place, 
then its clumps could have experienced periods of erosion (at higher wind velocities) as well as 
periods of accretion (at lower wind velocities).   Epiodes of erosion and accretion  could also have smoothed  
the surface of Arrokoth's sub-units. 

The rate that particles hit the surface is given by the mass of particles per unit area and time hitting the surface via equation \ref{eqn:Fm}.
Using a wind velocity $u_{hw,mound}$ and a time $\Delta t$ for the epoch where the particle flux is high, the depth of material redistributed is 
\begin{align}
h  
& \sim  \frac{ f_p \xi_p \rho_g u_{hw,mound} \Delta t}{4 \rho_a}. \label{eqn:hhhh}
\end{align}
We have neglected the ratio of ejecta to projectile mass as it depends on material strength and is of order unity (see equation \ref{eqn:Mcr}). 
The dimensionless parameter $\xi_p$ encompasses uncertainty in the fraction of disk particles that impact the surface due to aerodynamic deflection, whereas $f_p$ denotes the fraction of disk mass that is in particles.   It could also take into account the fraction of mass in small particles that stick to the surface via cohesion \citep{Gundlach_2015} rather than splash particles off the surface. 

\begin{figure}[!ht]
    \centering
    \includegraphics[width = 3.75truein,trim = 0 0 0 0,clip]{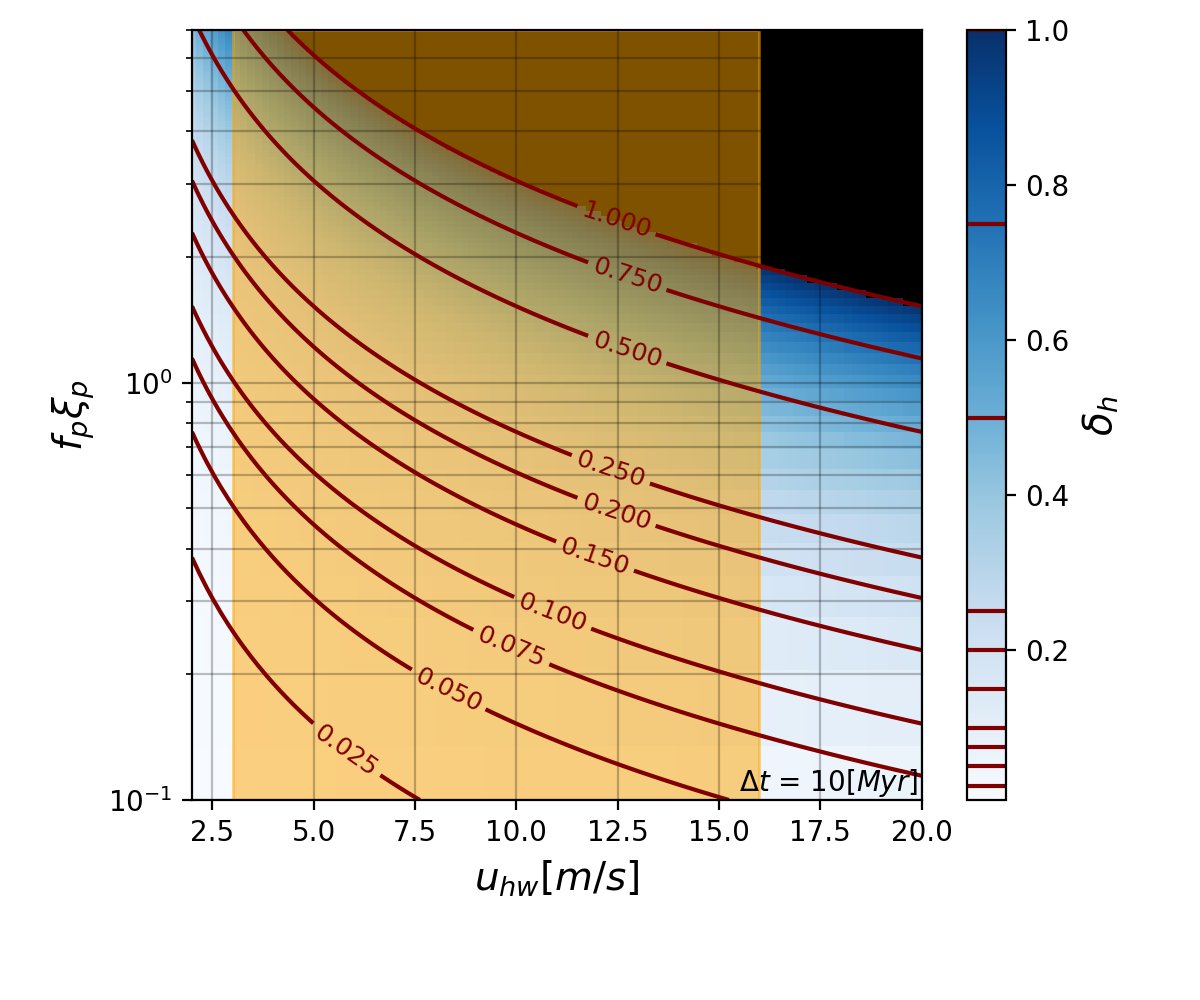}
    \caption{Redistribution depth ratio $\delta_h$ as a function of the headwind velocity, $u_{hw}$, and the particle-to-gas density ratio times an efficiency factor, $f_p \xi_p$, for a time interval $\Delta t = 10$ Myr.   Contours show the redistribution depth ratio $\delta_h$ (from equation \ref{eqn:delta_h}). The shaded orange region indicates the headwind velocity range of Eq.~\ref{eqn:umound} estimated for Arrokoth's mounds.  Smoothing of Arrokoth's mounds would require a particle rich disk for an extended length of time that seems incompatible with models for the protosolar disk. }
    \label{fig:Redistribution_Depth}
\end{figure}

We describe the smoothness of Arrokoth's mounds in terms of the ratio of the depth of splashed material to the planetesimal radius  
\begin{align}
   \delta_h & \equiv  \frac{h}{R_{mound}} \nonumber \\
& \sim f_p \xi_p \Delta t \frac{ \rho_g u_{hw,mound} }{2 \rho_a D_{mound}}. \label{eqn:delta_h} 
\end{align}
Equivalently, 
\begin{align}
f_p \xi_p \Delta t \sim 
\frac{\delta_h 2 \rho_{a,A} D_{mound}}{\rho_g u_{hw,mound}}.
\end{align}
We assume that the disk has midplane gas density $\rho_g$ equal to that of our disk model at 45 AU   
(from Table \ref{tab:tab}). This gives 
\begin{align}
f_p \xi_p \Delta t 
& \sim  10^6\ {\rm yr} \left( \frac{\delta_h}{0.035}  \right)
\left(\frac{u_{hw,mound}}{10~ {\rm m/s}} \right)^{\!\!-1}\!\! \left( \frac{D_{mound} }{5\ {\rm km}}\right)  \nonumber \\
& \ \ \ \ \times 
\left(\frac{\rho_{a}}{250~ {\rm kg~m}^{-3} } \right) \left(\frac{\rho_g} {2.6\! \times\! 10^{-10}~ {\rm kg~m}^{-3} } \right)^{\!\!-1},  \label{eqn:fpxi}
\end{align}
where we have used the middle of the range from equation \ref{eqn:umound} for the headwind velocity, $u_{hw,mound}$. The ratio $\delta_h = 0.035$ corresponds to about 90 m on a 5 km diameter mound which is about 3 times the $\approx $ 30 m resolution scale of the New Horizons imaging in Figure \ref{fig:Arr}.
With $\Delta t = 10$ Myr, a characteristic lifetime for the protosolar disk and wind velocity 10 m/s we would require $f_p \xi_p \sim 1/10$ for a ratio of $\delta_h \sim 3\%$.  This implies that the particle density must be similar in size to the gas density for an extended period of time.    Even if particles settle, a particle to gas density near 1 is unrealistically high for the mean value in the protosolar disk. The large required value of $f_p \xi_p$ to see meaningful surface particle redistribution for this duration of time is illustrated in Figure \ref{fig:Redistribution_Depth} which shows contours of $\delta_h$ (from equation \ref{eqn:delta_h}) as a function of headwind velocity and $f_p \xi_p$ at  $\Delta t = 10$ Myr,  with planetesimal density $\rho_a = 250$ kg m$^{-3}$, $D_{mound} = 5$ km and midplane gas density $\rho_g$  from the disk model at 45 AU from Table \ref{tab:tab}.  Arrokoth's mounds and Arrokoth itself probably cannot be smoothed via headwind associated impacts in a non-clumpy protostellar disk during the lifetime of the disk. 

\begin{figure}[!hb]
    \centering
\includegraphics[width=3.75truein,trim = 0 0 0 0,clip]{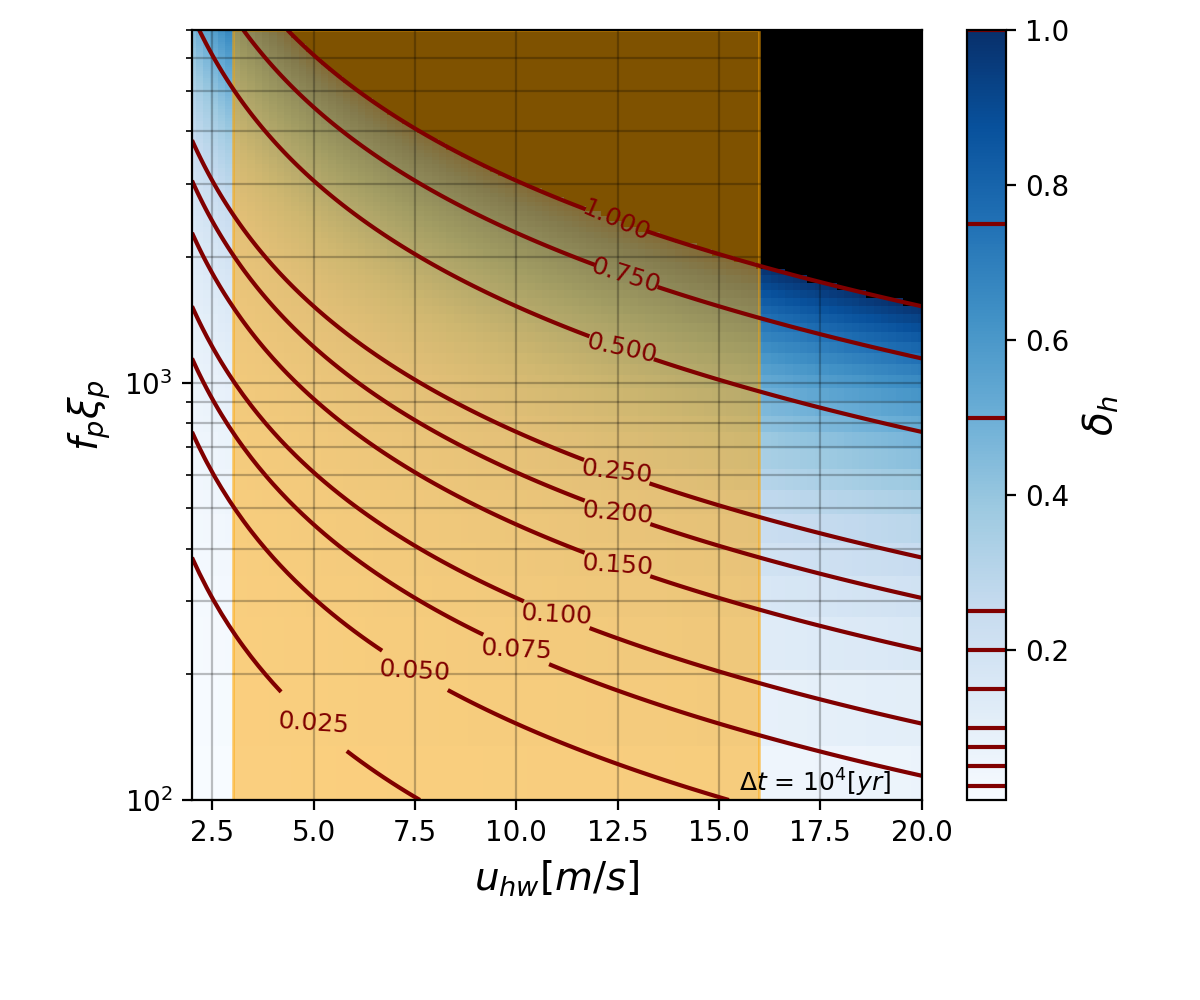}
    \caption{Similar to Figure \ref{fig:Redistribution_Depth} except showing $\delta_h$ for a duration $\Delta t = 10^4$ year. Smoothing of Arrokoth's mounds would require conditions for particle flux and wind velocity compatible with an epoch of streaming instability. }
\label{fig:Redistribution_Depth_SI_Duration}
\end{figure}

During streaming instability, clumps form with high particle concentrations \citep{Johansen_2007}.  Simulations by \citet{Li_2021} show clumps with particle to gas density ratio $f_p \sim 10^1$ to $10^3$ (see their figure 12). Peak values in the simulations by \citet{Rucska_2023} were similar in size but they reported $\Sigma_p/\Sigma_g$.
Because dust scale height is expected to be lower than the gas scale height, the ratio of midplane densities should exceed the ratio of surface densities $f_p = \rho_p/\rho_g > \Sigma_p/\Sigma_g$.  
A range of growth rates for the streaming instability are possible, ranging from 30 to 2000 in units of the Keplerian orbital angular rotation rate $\Omega_K$ (see Figure 8 \citealt{Li_2021}). This range of growth periods correspond to about $10^3$ to $10^4$ years at 45 AU.  
\citet{Simon_2022} speculate that in marginally stable systems,  planetesimal formation could take longer and be less efficient than those with short growth times. 
If we take $\Delta t = 10^3$ years,  equation \ref{eqn:fpxi} gives $f_p \xi_p \sim 10^3$ which is similar to the peak values for the particle to gas density seen in streaming instability simulations. Figure \ref{fig:Redistribution_Depth_SI_Duration} is similar to Figure \ref{fig:Redistribution_Depth} but using $\Delta t = 10^4$ years. 
Smoothing of Arrokoth's mounds by headwind associated impacts and splashed particle redistribution is compatible with their formation within clumps during an epoch of streaming instability
and subsequent evolution while embedded in particle rich filaments while streaming instability was active.  


\section{Summary and Discussion}

Using estimates for the drag force from a headwind from the protosolar nebula on an planetesimal and a protostellar disk model, we estimate the maximum size of a particle that might be lofted from the surface. 
In the inner solar system, gravity and cohesion can be overcome by the drag force from a protostellar disk headwind, allowing mm-sized particles on a planetesimal to be lofted off the surface.
However in the outer solar system, we estimate that cohesion is too strong for the wind to directly pull particles from a planetesimal's surface. 

As has been proposed to account for aeolian features on Pluto \citep{Tefler_2018}, we consider alternative
mechanisms for pulling particles off of the planetesimal's surface.  
Particles within the headwind can impact the planetesimal,
splashing particles off its surface. 


For particles within a disk that impact a planetesimal, 
we estimate the velocity distribution of ejecta using crater scaling laws, following \citet{Housen_2011}.  
To determine whether these impacts cause erosion, we estimate the mass in ejecta that is above the escape velocity. 
Using an estimate for the headwind velocity, we estimate that impacts from headwind particles are erosional for planetesimals with diameter smaller than about 8 km.  A similar estimate gives a total mass in ejecta about 3 times the projectile mass, though this is dependent upon the material strength of the planetesimal surface.  The protostellar disk gas density and an estimate for the fraction of mass in particles gives an estimate the erosion rate, if the planetesimal diameter is smaller than the 8 km erosion limit,  or the accretion rate from the wind, otherwise. 
If splashed particles do not tend to escape the planetesimal, then the accretion rate times the ratio of ejecta to projectile mass gives an estimate for the depth of material transported across the surface as splashed particles. The erosion or accretion rates are higher in the inner solar system where the density of the disk is higher. 


With a powerlaw ejecta velocity distributions typical of impact craters (e.g., \citealt{Housen_2011}) and taking into account drag from a wind, we integrate trajectories of particles splashed off the surface due to impacts from headwind particles.  We find that splash particles are most likely to be transported across the surface when the wind velocity, setting the ejecta velocity, is within about an order of magnitude of the  escape velocity of the planetesimal. 
The fate of splashed particles is sensitive to the planetesimal material strength, which affects the ejecta velocity distribution, the particle stopping time, which determines the coupling between particle and wind, and the planetesimal spin rotation rate and spin axis angle.
Large particles on the surface are less likely to escape when impacted by headwind particles, and this could cause evolution of the surface size distribution.  For planetesimals small enough that some ejecta particles can escape, low velocity impacts from particles in a headwind could mix material and exchange angular momentum between the forming planetesimal and the disk reservoir.  

Motivated by the smooth undulating terrain on the larger lobe of the pristine transneptunian object
Arrokoth, we consider the possibility that winds may have smoothed or eroded the surface of its approximately 5 km diameter building blocks, described as mounds by \citet{Stern_2023}.   We consider requirements on the wind velocity and particle flux that would allow transport of splashed particles across the surfaces of these building blocks and would not cause high levels of erosion. 
We explore a scenario where particles within a disk wind impact the surface of a mound, splashing particles from the surface.
The splashed particles land distant from the impact site, so they would blanket the surface, smoothing its terrain.   This wind related smoothing scenario is an alternative to erosion mechanisms
that involve impact generated seismic motions  \citep{Richardson_2004,Thomas_2005}  or sublimation  \citep{Steckloff_2021}.   
Assuming that the smooth terrain is caused by redistribution of surface particles across a depth of a few percent of the diameter of a 5-km primordial building block,  a headwind speed of approximately 10 m/s is required for the wind related smoothing scenario.  This speed is lower than that estimated in the disk, however  low headwind speeds could be present on particle concentrations during epochs of streaming instability.  We estimate that a mass density in particles a few hundred times that of the gas disk and present for about $10^4$ years could cause a few percent depth to be redistributed. The duration, high particle density and low headwind velocity are plausible values for clumps during streaming instability in the outer solar system. If Arrokoth's mounds formed during streaming instability, their smooth terrain suggests the physical conditions that were present in the clumps in which they formed.  Qualitatively, Arrokoth's building blocks could have been sandblasted by icy particles from the snowstorm in which they were born.  



While cohesion prevents a disk wind from lofting particles off a planetesimal surface in the outer solar system, the disk gas density is high enough to blow particles across the surface of 10 to 100 km diameter planetesimals in the inner solar system. 
There may have been mass exchange and mixing between planetesimal surfaces and a reservoir of particles residing in the disk. 
Because saltation and the probability of splash particle escape are particle size dependent, interactions between planetesimal surfaces and headwinds could affect the 
size distribution of surface particles on planetesimals, 
in addition to causing erosion   \citep{Rozner_2020,Demirci_2020,Demirci_2020b,Cedenblad_2021}. 
Interactions between particle rich headwinds and planetesimals are likely to cause a variety of interesting phenomena which could be the focus of future studies. 

The largest uncertainties in our models for splash particle dynamics are caused by uncertainty in the lift force in a rarefied gas dynamics regime near a surface and the difficulty of modeling the gas flow around a planetesimal.  
Future studies could explore the role of more realistic flows
by taking into account structure in the boundary layer, velocity shear in the wind and gravitational focusing. 

\vskip 0.1 truein
{Acknowledgements:}

This material is based upon work supported by NASA grants  80NSSC21K0143 and 80NSSC21K1184. 
We thank Hesam Askari and Paul Sanchez for helpful discussions.  We thank Alan Stern for helpful correspondence. 
We thank Tod Lauer for making available to us the CA06 mosaic of LORRI images of Arrokoth. 

Scripts and code used in this manuscript are available at \url{https://github.com/aquillen/Arrokoth}.

\bibliographystyle{elsarticle-harv}
\bibliography{Arrokoth}

\if \ispreprint1
\appendix
\section{Nomenclature}

Symbols used in the manuscript are summarized in the following tables. 

\if \ispreprint1
\else
\newgeometry{top=0.5cm, bottom=0.5cm}
\fi

\setcounter{table}{0} 
\begin{table}[tp] \centering
\caption{Nomenclature}
\begin{tabular}{llll}
\hline
$G$ & Gravitational constant \\
$k_B$ & Boltzmann constant \\
$M_\odot$ & Solar mass \\
\hline
$r$   & Orbital radius from the Sun \\
$z$  & Distance above disk midplane \\
$v_K $ &  Keplerian velocity, $ \sqrt{GM_\odot/r}$\\
$\Omega_K $ &   Keplerian angular velocity, $ v_K/r$ \\
\hline
$\rho_g$ & Protosolar disk gas density at midplane \\
$\Sigma_g$ & Gas surface density of protosolar disk \\
$P_g$ & Gas pressure in midplane of protosolar disk \\
$T_g$  & Temperature of gas in protosolar disk \\
$c_g$ & Isothermal sound speed in protosolar disk \\
$v_{\rm th}$ & Thermal velocity \\
$h_g$ & Vertical scale height of gas in protosolar disk \\
$\lambda_g$ & Mean free path in gas \\
$\eta_g$ &  Pressure parameter (equation \ref{eqn:eta_g}) \\
$\alpha_g$ & Alpha parameter for a turbulent disk \\
$\bar m$ & Mean molecular mass \\
$\nu_g$  & Kinematic gas viscosity \\
$\sigma_{H2}$ & Collision cross section for molecular hydrogen \\
$g_\Sigma,g_T$ & Coefficients describing disk model \\
$\tilde g_\Sigma, \tilde g_T $ & Ratio of coefficients w.r.t. to adopted values \\
$u$  & Gas velocity \\
$u_{hw}$ & Headwind velocity \\
$u_{turb}$ & Turbulent eddy velocity, $\sqrt{\alpha_g} c_g$  \\
\hline
$s$  & Grain radius, particle in disk or splash particle \\
$\rho_s$  & Grain density \\
$t_{stop}$ & Stopping time \\
$St$ & Stokes number, dimensionless stopping time \\
 & \qquad $St = t_{stop}\Omega_K$ \\
${\bf F}_D$ & Drag force \\
$a_D$ & Acceleration on a particle from drag \\
\hline
$R_a$ & Radius of planetesimal \\
$D_a$ & Diameter of planetesimal \\
$M_a$ & Mass of planetesimal \\
$\rho_a$ & Density of planetesimal \\
$g_a$ & Surface gravitational acceleration,  $GM_a/R_a^2$ \\
$v_{esc,a}$ & Escape velocity, $  \sqrt{2GM_a/R_a}$ \\
$v_a$     & Gravitational velocity $\sqrt{GM_a/R_a}$ \\
$\Omega_a$ & Inverse gravitational timescale  $v_a/R_a$\\
$Y_a$ & Material strength \\
$ {\boldsymbol \Omega}_{spin,a}$  & Spin vector of planetesimal  \\
$t_{cross}$ & Crossing time, $D_a/u_{hw}$\\
\hline
$F_{coh}$   & Cohesive pull-off force \\
${\cal B}$   & Parameter describing surface roughness \\
$\gamma_c$  & Cohesive contact energy per unit area \\
\hline
\end{tabular}
\end{table}

\setcounter{table}{0} 

\begin{table}[!t]\centering
\caption{Nomenclature-continued}
\begin{tabular}{llll}
\hline
$\rho_{pj}$  & Projectile density  \\
$a_{pj}$  & Projectile radius \\
$m_{pj}$  & Projectile mass \\
$u_{pj}$ & Projectile velocity \\
$v_{ej}$ & Ejecta velocity \\
$R_{cr}$ & Crater radius \\
$M()$ & Mass ejecta function  \\
$M_{ej}$ & Total ejecta mass \\
$x$ & Distance to impact site \\
$C_1,H_2,k,n_1$ & Coefficients for crater scaling \\
$\mu,\nu$ & Exponents for crater scaling \\
\hline
$Re$ & Reynolds number \\
$M\!a$ & Mach number \\
$Fr$ & Froude number \\
$K\!n$ & Knudsen number \\
$\pi_2,\pi_3,\pi_4$ & Dimensionless crater scaling parameters \\
\hline 
$U$ & Wind velocity distant from planetesimal \\
$t$  & Time \\
$w$  & Impact parameter \\
$\tilde {\bf r}_s$  & Particle position in units of $R_a$ \\
$\tilde {\bf v}_s$  & Particle velocity in units of $v_a$ \\
$\tilde {\bf a}_s$  & Particle acceleration in units of $g_a$ \\
$\tilde {\bf u}$   & Wind velocity in units of $v_a$ \\
$\tilde {\bf u}_{\rm pf}$  & Flow field of potential flow \\
$\tilde {\bf u}_\delta$  & Flow field within boundary layer \\
$\tau$   & Dimensionless time, $\tau = t \Omega_a$ \\
$\tilde t_s$   & Dimensionless stopping time; $t_{stop} \Omega_a$ \\
$\tilde {\boldsymbol \omega}_{a}$  & Planetesimal spin vector in units of $\Omega_a$ \\
$\theta_{sw}$  & Angle between spin axis and wind direction \\
$\delta$  & Boundary layer thickness \\
$\tilde x_b$   & Distance from stagnation point along surface \\
$\tilde y_b$   & Distance from surface  \\ 
$\tilde {\bf u}_{pj}$ & Projectile velocity in frame \\ & \ \ \ rotating with the surface \\
$\tilde {\bf v}_{ej}$ & Ejecta velocity in frame rotating \\ & \ \ with the surface \\
$\hat {\bf d}$  & Unit vector pointing downrange \\
& \ \ \ with respect to the impact velocity \\
$\hat {\bf n}$ & Surface normal unit vector \\
$\zeta_{ej}$ & Elevation angle of ejecta velocity \\
$\beta$ & Angle describing how far a splash particle \\
 &  \ \ moves with respect to the surface \\
 \hline
\end{tabular} 
\end{table}

\setcounter{table}{0} 
\begin{table}[!ht]\centering
\caption{Nomenclature-continued}
\begin{tabular}{llll}
\hline
$D_{erode}$  & Diameter of a planetesimal with \\ 
& \ \ a critical erosion rate \\
$D_{a,vw}$    & Diameter of a planetesimal with \\
 & \ \ \ $v_{esc,a}  = u_{hw}$ \\
$f_p$  & Mean mass density in disk particles  \\ & \ \ divided by gas density \\
$\xi_p$  & Mass fraction of particles that can  \\ & \ \ impact a planetesimal \\
$F_m$  & Particle flux \\
$h$ & Depth \\
$\frac{d h_{erode}}{dt}, \frac{dh_{acc}}{dt}$ & Accretion and erosion rates (depth/time) \\
$\delta_h$  & Depth of redistributed material divided  \\ & \ \ \ by planetesimal radius \\
\hline
\end{tabular}
\end{table}

\fi

\if \ispreprint 1 
\else 
\restoregeometry
\fi

\end{document}